\documentclass[iop]{emulateapj}

\usepackage{graphics}
\usepackage{epsfig}
\usepackage{natbib}
\newcommand{\kms}{km~s\ensuremath{^{-1}}}
\newcommand{\msun}{$M_{\odot}$}

\newcommand{\mstar}{$M_{\ast}$}
\newcommand{\mdust}{$M_{dust}$}
\newcommand{\tdust}{$T_{dust}$}

\newcommand{\tdep}{$t_{dep}({\rm H_2})$}
 
\newcommand{\fgas}{$f_{gas}$} 
\newcommand{\xco}{$\alpha_{CO}$}
\newcommand{\arc}{8:00arc}

\shorttitle{Herschel/IRAM study of $z>2$ lensed galaxies}
\shortauthors{Saintonge et al.}

\begin{document}

\title{Validation of the equilibrium model for galaxy evolution to $z\sim3$ through molecular gas and dust observations of lensed star-forming galaxies\footnotemark[1]}

\author{Am\'elie Saintonge\altaffilmark{2}, Dieter Lutz\altaffilmark{2}, Reinhard Genzel\altaffilmark{2},  Benjamin Magnelli\altaffilmark{3}, Raanan Nordon\altaffilmark{4},  Linda J. Tacconi\altaffilmark{2},  Andrew J. Baker\altaffilmark{5}, 
 Kaushala Bandara\altaffilmark{6,2},   Stefano Berta\altaffilmark{2}, Natascha M. F\"orster Schreiber\altaffilmark{2}, \\Albrecht Poglitsch\altaffilmark{2}, Eckhard Sturm\altaffilmark{2}, Eva Wuyts\altaffilmark{2} \& Stijn Wuyts\altaffilmark{2}}

\footnotetext[1]{Based on observations carried out with the IRAM Plateau de Bure Interferometer. IRAM is supported by INSU/CNRS (France), MPG (Germany) and IGN (Spain). Based also on observations from {\it Herschel}, an ESA space observatory with science instruments provided by European-led Principal Investigator consortia and with important participation from NASA. }

\altaffiltext{2}{Max-Planck Institut f\"ur extraterrestrische Physik, 85741 Garching, Germany}
\altaffiltext{3}{Argelander-Institut f\"ur Astronomy, Universit\"at Bonn, 53121 Bonn, Germany}
\altaffiltext{4}{School of Physics and Astronomy, The Raymond and Beverly Sackler Faculty of Exact Sciences, Tel-Aviv University, Tel-Aviv 69978, Israel}
\altaffiltext{5}{Department of Physics and Astronomy, Rutgers, The State University of New Jersey, Piscataway, NJ 08854-8019 USA}
\altaffiltext{6}{Department of Physics and Astronomy, University of Victoria, Victoria, BC V8P 5C2, Canada}

%===================================
\begin{abstract}
We combine IRAM Plateau de Bure Interferometer and {\it Herschel} PACS and SPIRE measurements to study the dust and gas contents of high-redshift star forming galaxies.  We present new observations for a sample of 17 lensed galaxies at $z=1.4-3.1$, which allow us to directly probe the cold ISM of normal star-forming galaxies with stellar masses of $\sim10^{10}$\msun, a regime otherwise not (yet) accessible by individual detections in {\it Herschel} and molecular gas studies.  The lensed galaxies are combined with reference samples of sub-millimeter and normal $z\sim1-2$ star-forming galaxies with similar far-infrared photometry to study the gas and dust properties of galaxies in the SFR-\mstar-redshift parameter space.  The mean gas depletion timescale of main sequence galaxies at $z>2$ is measured to be only $\sim450$Myr, a factor of $\sim1.5$ ($\sim5$) shorter than at $z=1$ ($z=0$), in agreement with a $(1+z)^{-1}$ scaling.  The mean gas mass fraction at $z=2.8$ is $40\pm15\%$ (44\% after incompleteness correction), suggesting a flattening or even a reversal of the trend of increasing gas fractions with redshift recently observed up to $z\sim2$.  The depletion timescale and gas fractions of the $z>2$ normal star-forming galaxies can be explained under the ``equilibrium model" for galaxy evolution, in which the gas reservoir of galaxies is the primary driver of the redshift evolution of specific star formation rates.  Due to their high star formation efficiencies and low metallicities, the $z>2$ lensed galaxies have warm dust despite being located on the star formation main sequence. At fixed metallicity, they also have a gas-to-dust ratio 1.7 times larger than observed locally when using the same standard techniques, suggesting that applying the local calibration of the $\delta_{GDR}$-metallicity relation to infer the molecular gas mass of high redshift galaxies may lead to systematic differences with CO-based estimates.
\end{abstract}

\keywords{galaxies: evolution -- galaxies: high-redshift -- infrared: ISM -- ISM: dust -- ISM: molecules}

%===================================
\section{Introduction}
\label{intro}

Most star-forming galaxies follow a relation between their stellar masses and star formation rates that has been well characterized up to $z\sim2.5$ \citep[e.g.][]{noeske07,salim07,elbaz11,whitaker12}.  The very existence and tightness of this relation suggests that these galaxies live in a state of equilibrium where their ability to form stars is regulated by the availability of gas and the amount of material they return to the circum-galactic medium through outflows \citep[e.g.][]{genel08,bouche10,dave11,dave12,krumholz12,lilly13}.  Simultaneously, it downplays the importance of galaxy mergers in the global star formation budget of the Universe \citep{robaina09,rodighiero11,kaviraj13} and highlights the influence of secular processes with longer duty cycles such as gas accretion, bar formation and bulge growth \citep{genzel08}.  We refer to this general framework as the ``equilibrium model" for galaxy evolution, the formalism of which is detailed in recent work \citep[e.g.][]{dave12,lilly13}. 

This paradigm was ushered in by a combination of large imaging surveys, detailed kinematics studies, molecular gas measurements, and theoretical efforts.  For example, large optical and infrared surveys have contributed by allowing for the accurate measurement of stellar masses and star formation rates in very large galaxy samples, often supported by significant spectroscopic observing campaigns. Recently, {\it Herschel} observations in the wavelength range of 70-500$\mu$m with the PACS and SPIRE instruments have further contributed, by providing direct calorimetric SFRs that have been used to improve and recalibrate other indicators \citep[e.g.][]{nordon10,nordon12,elbaz10,wuyts11sfr}.  

At the same time, near-infrared integral field spectroscopy measurements at $z\sim2$ have started revealing that a significant fraction of high redshift galaxies are rotation-dominated discs \citep[e.g.][]{forster06,wright07,genzel06,genzel08,forster09,shapiro08,jones10b,gnerucci11,wisnioski11,epinat12,newman13}.  These observations convincingly demonstrate that the high SFRs measured in these galaxies are generally not caused by major mergers, as was previously assumed by analogy with the local ULIRGs which have comparable SFRs and are all major mergers \citep{sanders96,veilleux02}.  Instead, the high SFRs of $z\sim2$ galaxies are caused by high molecular gas fractions, well above the $5-10$\% typically observed at $z=0$ \citep{tacconi10,daddi10,tacconi13}.

Some of the strongest direct evidence in favor of the equilibrium model above indeed comes from molecular gas observations.  In the local Universe, with the exception of ULIRGSs, it has now been directly observed that the location of a galaxy in the SFR-\mstar\ plane is mostly determined by its supply of molecular gas, with variations in star formation efficiency playing a second order role \citep{saintonge12}.  Similar conclusions have also been reached based on high-redshift galaxy samples, whether directly using CO data \citep{tacconi13}, or indirectly using far-infrared photometry to estimate gas masses \citep{magdis12a}.  Also, outflows of molecular material, which are an important element in setting the balance between gas and star formation in the models, have now been directly observed in a range of objects \citep[e.g.][]{sturm11}.  But mostly, it is now possible to detect CO line emission in normal star-forming galaxies at $z>1$ \citep[e.g.][]{tacconi10,tacconi13,daddi10}.  These observations convincingly show that the rapid decline in the specific SFR of galaxies since $z\sim2$ can be  explained by the measured gas fractions and a slowly varying depletion timescale \citep[$t_{dep}\propto(1+z)^{-1}$,][]{tacconi13}. 

Out to $z\sim1.5$ it is possible to detect far-infrared emission and CO lines in individual objects with current instrumentation.  However, even in the deepest {\it Herschel} fields and with long integrations at the IRAM PdBI, it is still not possible to measure directly the dust and gas contents of individual normal star-forming galaxies with masses $\sim10^{10}$\msun\ at $z\gtrsim2$.  A proven way to study molecular gas in high redshift galaxies with lower masses and lower SFRs is instead to target objects that are gravitationally lensed \citep{baker04,coppin07,danielson11}.  Samples of lensed star-forming galaxies have for example also been used to extend to higher redshifts and lower stellar masses studies of the kinematics and stellar populations of disc galaxies, of the mass-metallicity relation \citep[e.g.][]{jones10b,richard11,wuyts12,wuyts12b}, and of the origin of metallicity gradients in galaxy discs \citep[][]{yuan11,jones13}.   In this paper, we explore the relation between dust, gas and star formation at $z=2-3$ using a sample of 17 UV-bright lensed galaxies targeted for deep {\it Herschel} PACS and SPIRE observations and having SFRs and stellar masses characteristic of main-sequence objects at their redshifts. We analyze the results of these observations in the context of the equilibrium model.  In addition, we present new IRAM PdBI observations that more than double the number of published lensed galaxies with CO line measurements. 

After we describe the sample in \S \ref{sample}, the multi-wavelength observations are presented in \S \ref{data}, including the new Herschel and IRAM PdBI observations.  In \S \ref{quantities}, we describe how key quantities such as stellar masses, star formation rates, dust masses, dust temperatures, gas masses and metallicities were calculated homogeneously.   The key results of this study are presented in \S \ref{results} and summarized in \S \ref{summary}; in short, we find that the $z>2$ lensed galaxies have low dust and gas masses, but high dust temperatures as a consequence of an efficient conversion of their gas into stars.  Gas mass fractions and depletion times follow a redshift evolution out to $z=3$ that is consistent with the expected scaling relations under the equilibrium model.    

All rest-frame and derived quantities in this work assume a \citet{chabrier03} IMF, and a cosmology with $H_0=70$\kms\ Mpc$^{-1}$, $\Omega_m=0.3$ and $\Omega_{\Lambda}=0.7$.  All molecular gas masses ($M_{H2}$) and derived quantities such as gas fractions and depletion times, presented and plotted include a factor of 1.36 to account for the presence of helium.
\\
\\

%===================================
\section{Sample}
\label{sample}

\subsection{Lensed galaxies sample}

The main sample consists of 17 lensed galaxies that were selected for deep {\it Herschel} PACS/SPIRE imaging (details of these observations are presented in \S \ref{herscheldata}).  These galaxies, selected from the literature, have been discovered as bright blue arcs of conspicuous morphology and then spectroscopically confirmed to be high redshift lensed objects.  As shown in \S \ref{samplebias}, their intrinsic properties are similar to UV-selected Lyman Break Galaxies (LBGs), or their $z\sim2$ BX/BM analogs, but the observed fluxes are strongly amplified.  The sources come both from the traditional method of searching for galaxies lensed by massive clusters, and from the recent searches for bright blue arcs in the Sloan Digital Sky Survey (SDSS), with lenses that typically are individual luminous red galaxies \citep[e.g.][]{allam07}.  In addition, we also consider the submm-identified Eyelash galaxy \citep{swinbank10}, as it is located within the field of view of our {\it Herschel|} observations of the Cosmic Eye.

The sample is by no means a complete census of UV-bright galaxies, but instead samples some of the best-known objects with rich multi-wavelength observations, including near-infrared imaging and spectroscopy and millimeter continuum and line measurements.  In Table \ref{Sampletab} the basic properties of the sample are given (coordinates, redshifts and amplification factors).  The galaxies have redshifts in the range $2<z<3$, with the exception of four galaxies with $z\sim1.5$ (median redshift is 2.3), and most have stellar masses in the range $9.5<\log M_{\ast}/M_{\odot}<11.0$ (median, $1.6\times10^{10}$\msun), as will be presented in \S \ref{mstarsection}. 

\begin{deluxetable*}{lccclcl}
\tabletypesize{\small}
\tablewidth{0pt}
\tablecaption{Lensed galaxy sample \label{Sampletab}}
\tablehead{
\colhead{Name} & \colhead{$\alpha_{\rm J2000}$} & \colhead{$\delta_{\rm J2000}$} & \colhead{$z_{spec}$} & \colhead{$z$ ref.} &
\colhead{Magnification} & \colhead{Magnif. ref.}  }
\startdata
       \arc & 00h22m40.9s & +14d31m14.0s & 2.730     & \citet{allam07} &  $12.3^{+15}_{-3.6}$         & \citet{allam07} \\ 
      J0712 & 07h12m17.5s & +59d32m16.3s & 2.646    & \citet{jones10b} &  27.5$\pm$8.1 & \citet{richard11} \\ 
      J0744 & 07h44m47.8s & +39d27m25.7s & 2.209    & \citet{jones10b} &  16.0$\pm$2.6 & \citet{richard11} \\ 
      J0900 & 09h00m03.3s & +22d34m07.6s & 2.032     & \citet{diehl09} &   4.8$\pm$1.0\tablenotemark{a}          & \citet{bian10} \\ 
      J0901 & 09h01m22.4s & +18d14m32.3s & 2.256     & \citet{diehl09} &   8.0$\pm$1.6\tablenotemark{a}        & \citet{fadely10} \\ 
      J1133 & 11h33m14.2s & +50d08m39.5s & 1.544      & \citet{sand04} &  14.0$\pm$2.0     & \citet{livermore12} \\ 
      J1137 & 11h37m40.4s & +49d36m35.8s & 1.411      & \citet{kubo09} &  17.0$\pm$3.0       & this work (appendix A)\\ 
  Horseshoe & 11h48m33.1s & +19d30m03.2s & 2.379 & \citet{belokurov07} &  30.0$\pm$5.0     & \citet{belokurov07} \\ 
      J1149 & 11h49m35.2s & +22d23m45.9s & 1.491     & \citet{smith09} &  23.0$\pm$4.6\tablenotemark{a}         & \citet{smith09} \\ 
      Clone & 12h06m02.1s & +51d42m29.5s & 2.001       & \citet{lin09} &  28.1$\pm$1.4        & \citet{jones10a} \\       
      J1226 & 12h26m51.3s & +21d52m20.3s & 2.925     & \citet{wuyts12} &  40.0$\pm$8.0\tablenotemark{a}       & \citet{koester10} \\ 
      J1343 & 13h43m34.0s & +41d55m09.0s & 2.092     & \citet{diehl09} &  40.0$\pm$8.0\tablenotemark{a}        & \citet{wuyts12b} \\ 
      J1441 & 14h41m49.4s & +14d41m21.3s & 1.433   & \citet{pettini10} &   4.5$\pm$1.5       & \citet{pettini10} \\ 
        cB58 & 15h14m22.2s & +36d36m25.2s & 2.729   & \citet{teplitz00} &  31.8$\pm$8.0         & \citet{seitz98} \\ 
      J1527 & 15h27m45.2s & +06d52m19.2s & 2.762     & \citet{wuyts12} &  15.0$\pm$3.0\tablenotemark{a}       & \citet{koester10} \\ 
              Eye & 21h35m12.7s & -01d01m42.9s & 3.074     & \citet{smail07} &  30.0$\pm$3.3   & \citet{dye07} \\ 
    Eyelash & 21h35m11.6s & -01d02m52.0s & 2.326  & \citet{swinbank10} &  32.5$\pm$4.5      & \citet{swinbank10} 
\enddata
\tablenotetext{a}{Assuming a 20\% uncertainty on the magnification, see \S \ref{lensmodel}.}
\end{deluxetable*}

\subsubsection{Is the lensed galaxies sample representative of the $z>2$ star-forming population?}
\label{samplebias}
As outlined in the introduction, the power of gravitational lensing is used here to extend the galaxy population where direct measurements of both molecular gas and dust masses can be performed.  In particular, most galaxies in the sample have a stellar mass of $\sim10^{10}$\msun\ and modest SFRs, as determined from IR and UV photometry, of $\sim50$ \msun\ yr$^{-1}$, given their median redshift of $z=2.4$.   We stress that the only way to directly detect the far-infrared emission of such galaxies across multiple {\it Herschel} bands is to target systems that are gravitationally lensed.  Even in the deepest PACS and SPIRE blank fields, this mass/redshift/SFR regime is barely accessible through stacking of tens of galaxies \citep[][]{reddy12,magnelli13}.  Therefore, it is impossible to directly compare the FIR properties of our lensed galaxies to those of a well-matched, un-lensed reference sample.  

Such a comparison is however desirable.  Since the lensed galaxies were selected visually based on their rest-frame UV light, there is a possibility that they represent a biased sub-sample of the high-redshift star-forming galaxy population.  For example, we could expect the selection technique to preferentially isolate objects that are particularly UV-bright and therefore dust- and metal-poor.  

To answer this question, we create a control sample of un-lensed galaxies, matching on mass, redshift, and IR luminosity, where the latter is obtained either from PACS or MIPS 24$\mu$m photometry when available, or else derived indirectly from the optical/UV photometry.   
We then compare the UV-to-IR ratio between the lensed and control samples.  More exactly, for each lensed galaxy, we extract a control galaxy from the GOODS fields catalog  \citep{wuyts11}, within 0.2 dex in $M_{\ast}$, 0.2 dex in $L_{IR}$ and 0.2 in $z$.  To check whether the lensed galaxies are less extincted than the control sample, the distributions of $\log({\rm SFR}_{IR}/{\rm SFR}_{UV})$ are compared.   As described in \S \ref{sfrsection}, for the lensed galaxies, we calculate SFR$_{IR}$ from the 160$\mu$m fluxes and SFR$_{UV}$ from the B- or V-band photometry.  For the GOODS sample, \citet{wuyts11sfr} used a ladder of star formation indicators to infer the respective contributions of obscured and un-obscured star-formation, even in the absence of flux measurements at FIR wavelengths, therefore also allowing a measurement of  $\log({\rm SFR}_{IR}/{\rm SFR}_{UV})$.    A KS test gives a probability $>90\%$ that the distributions of $\log({\rm SFR}_{IR}/{\rm SFR}_{UV})$ in the lensed and control samples are representative of the same parent population. 

\begin{figure}
\epsscale{1.1}
\plotone{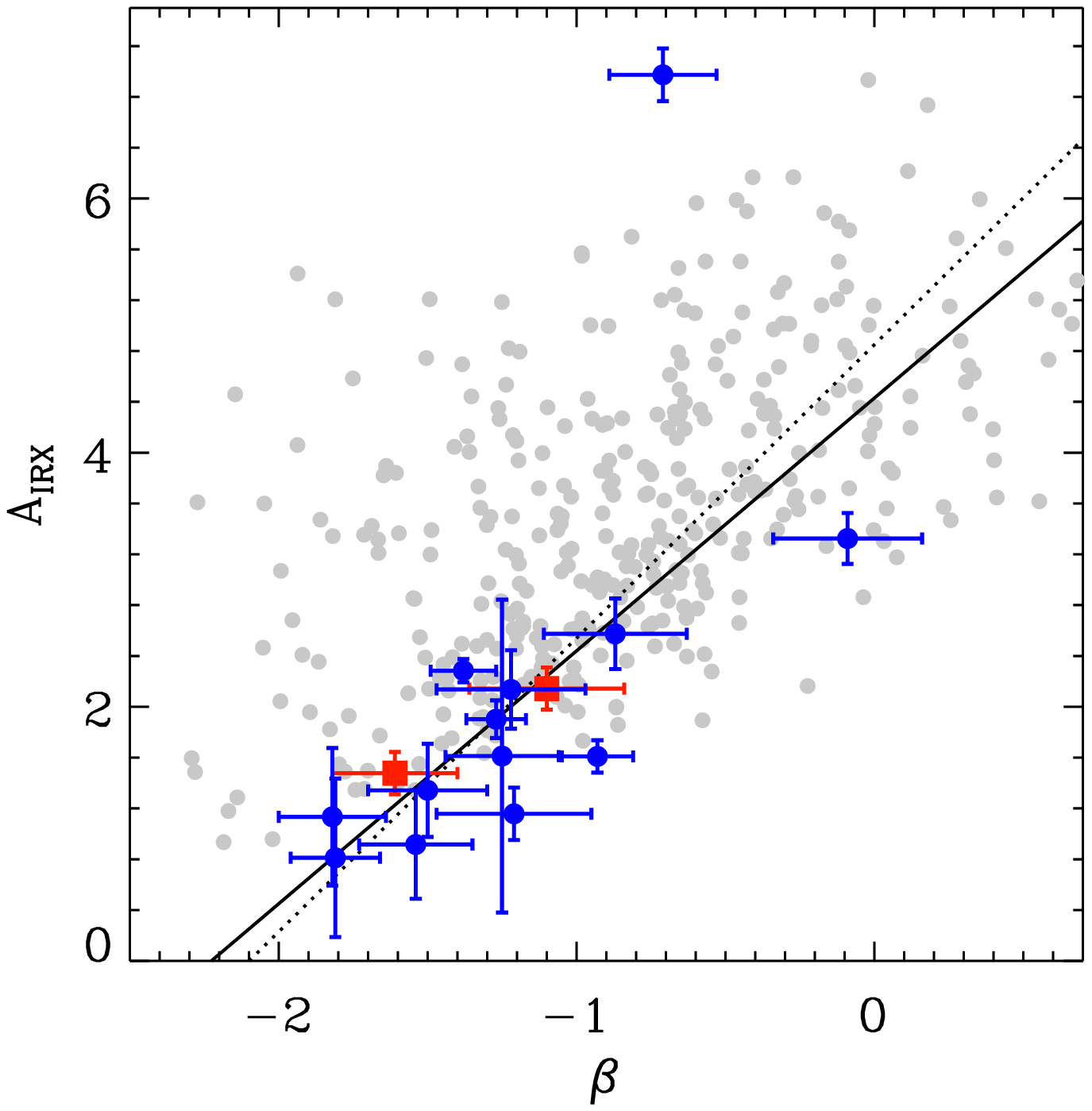}
\caption{Relation between the effective UV attenuation ($A_{IRX}$) and the UV spectral slope ($\beta$), showing the lensed galaxies (blue circles) and stacked points from \citet{reddy12} for UV-selected BX/BM galaxies in two bins of $\beta$ (red squares).  The $\beta$ values for the lensed galaxies are computed from HST photometry, or taken from \citet{richard11} and \citet{wuyts12}.  As a comparison, the relations of \citet{calzetti00} (dotted line), \citet{meurer99} (solid line) are shown, as well as the IR-selected sample of \citet{nordon13} (gray circles).   The lensed galaxy in this figure with the very high value of $A_{IRX}$ is the Eyelash. \label{irxbeta}}
\end{figure}

As a second test, we study the UV slope $\beta$ of the lensed galaxies.  In Figure \ref{irxbeta} the lensed galaxies are shown in the $A_{IRX}-\beta$ plane, where $A_{IRX}$ is the effective UV attenuation, 
\begin{equation}
A_{IRX}=2.5\log \left ( \frac{{\rm SFR}_{IR}}{{\rm SFR}_{UV}} +1 \right ). 
\end{equation}
The UV slope is measured from HST optical photometry using bands corresponding to rest-frame wavelengths redwards of Lyman $\alpha$ .  The observed values are interpolated to derive the magnitudes at 1600\AA\ and 2800\AA\ (rest-frame) which are then used to compute $\beta$.  As a comparison, we use again the GOODS fields sample for which \citet{nordon13} have derived values of the UV slope $\beta$.  This reference sample is shown in Fig. \ref{irxbeta}  as gray circles.  As these are PACS-detected galaxies only in the GOODS fields, the sample is biased towards large values of $A_{IRX}$.  To circumvent the necessity of detecting individual objects in several Herschel bands, \citet{reddy12} used stacks of UV-selected galaxies to probe galaxies with typically lower masses and SFRs.  Their results are shown in Fig. \ref{irxbeta} for two different bins of $\beta$.  The lensed galaxies are located in the same region of the $A_{IRX}-\beta$ as the BX/BM galaxies of \citet{reddy12}, and are also seen to follow the \citet{meurer99} extinction law (solid line).  The only exception is the Eyelash, which is an outlier with high UV attenuation as expected since it is the only submillimeter-selected galaxy in the sample. Using the sample of \citet{nordon13} (both Herschel-detected and undetected galaxies) we can also define a control sample for each of our lensed galaxies by matching on \mstar, SFR and redshift.  In 10 of the 12 galaxies where we have sufficient data to conduct this experiment the median of the $\beta$ distribution in the control sample agrees with the measured value of $\beta$ for the lensed galaxy within its uncertainty.  The two exceptions are J0900 and J1527 where measured values of $\beta$ are -1.8 while the control samples suggest -0.6 and -1.2, respectively. We also note that while previous studies of the Cosmic Eye and cB58 suggested that these objects were better represented by an SMC extinction law than by the \citet{meurer99} relation \citep[see e.g.][]{wuyts12}, the new Herschel data revise the IR luminosities upward and find these galaxies in agreement with the Meurer/Calzetti relation.  

The lack of significant differences in either the effective UV attenuation ($A_{IRX}$) or the UV slope ($\beta$) between the lensed galaxies of this study and their respective matched control samples suggests that they are not a biased subsample of the underlying population.  We therefore proceed under the assumption that they are representative of the bulk of the star-forming galaxy population at their specific redshifts and modest masses.

\subsubsection{Does differential lensing introduce biases in the measurements?}
\label{difflens}

In Table \ref{Sampletab}, we give the magnification factors used to correct measured quantities for the lensing effect.  The magnification factors are derived from optical imaging (typically from HST), and represent the average factor across the discs of these extended galaxies.  However, in reality, different regions of the galaxies may be lensed differentially, with regions falling on or near caustics being more strongly magnified. Therefore, components of the galaxy having different spatial extents and compactness may not be magnified uniformly.  This effect is known as differential lensing \citep[e.g.][]{blandford92}.  

In practice, this could impact our analysis since both warm dust and high$-J$ CO transitions are typically emitted in denser, more compact regions than cold dust or CO (1-0). Both the far-infrared SED and the CO spectral line energy distribution (SLED) might be biased, and standard calibrations to measure dust masses and temperatures, as well as excitation corrections to extrapolate total gas masses from a (3-2) line flux, may not apply \citep{downes95,serjeant12}. For the specific case of galaxy lenses selected in large area, flux-limited submm/FIR surveys, there is a bias towards galaxies with very compact dusty star-forming regions, as these are most likely to benefit from the strongest magnifications \citep{hezaveh12}.  With the exception of the Eyelash, our lenses are not submm/FIR-identified, but rather selected from cluster fields and SDSS imaging.   In the discovery papers, SDSS galaxy lenses were typically identified visually as blue arcs. This includes the arc length, driven by the basic geometric configuration of lens and background object, as well as flux. They will therefore not suffer the specific biases discussed by \citet{hezaveh12}, but the rest-UV emission used to derive the lensing model may still differ in extent or centroid from that at longer wavelengths.

It is difficult to directly assess the impact of differential lensing, as the exercise requires detailed magnification maps, and high resolution imaging of the different components (optical/UV continuum, FIR continuum, CO line emission,...).  The effect has therefore been so far quantified only based on simulations and very few, special galaxies.  In particular, most of the literature on this concentrates on the special case of submm- or FIR-selected galaxies such as the Eyelash \citep[e.g.][]{hezaveh12,serjeant12,fu12,wardlow13}.  As can be seen in Figure \ref{irxbeta}, the Eyelash is a very special object, the only one in our sample to be submm-selected, and any conclusions drawn from such extreme dusty systems may not apply to all normal star-forming galaxies.  We can however use these studies to get a sense of the amplitude of the issue in the most extreme cases. 

Using a submillimeter galaxy as a model, \citet{serjeant12} estimate the impact of differential lensing on FIR-selected galaxy samples.  They find that even when comparing quantities measured from rest-frame optical/NIR wavelengths (e.g. stellar masses) and FIR observations (e.g. SFRs), the median differential magnification ratio is $\sim0.8$  with small dispersion.  Therefore even in these cases, differential lensing does not affect the position of a galaxies in the SFR-mass plane significantly.  \citet{serjeant12} find a stronger effect of differential lensing on the CO SLED, but the effect manifests itself mostly for transitions with $J_{upper}>4$.  Indeed, for the few galaxies in our sample where CO(1-0)  measurements are also available, a typical (3-2)/(1-0) line ratio of $\sim 0.7$ is retrieved \citep{riechers10,danielson11}, similar to what is measured in unlensed galaxies \citep[e.g.][]{harris10,ivison11,bothwell13}.   A common explanation for this ubiquitous value of the (3-2)/(1-0) line ratio is that the lines are emitted from the same moderately excited component of the ISM, but \citet{harris10} argue that it may instead reflect a generic feature of multi-component star-forming ISMs, in which the different lines are emitted by different but well mixed, optically thick and thermalized components with different but characteristic filling factors. 

Since our {\it Herschel} and IRAM observations are mostly unresolved, it is not possible to directly assess the impact of differential lensing on our measurements.  Generally, this effect has not yet been extensively studied for normal star-forming high-redshift discs (as opposed to submm- or FIR-selected galaxies).  However, based on the arguments above and the hypothesis that the FIR continuum and the CO line emission originate from similar physical regions \citep[see however][]{fu12}, we can conclude that if occurring, differential lensing is unlikely to significantly affect the results of this study.  It may be biasing our gas masses high if the CO SLEDs are affected, but by no more than 40\%, comparing our adopted excitation correction $R_{13}=2.0$ and the value of 1.4 indicated by the observations of the Eye, Eyelash and cB58 \citep[][see also \S \ref{MH2section}]{riechers10,danielson11}.  

Similarly, differential lensing could be biasing the inferred dust properties.  If the more compact, hotter regions of the galaxies are more strongly lensed than the diffuse component, the inferred dust temperatures could be too high \citep[although][show that the effect can be reversed in systems with lower magnification factors of $\mu \sim 10$]{hezaveh12}.  This would result in under-estimated dust masses as well.  However, the fact that we do not observe a strong effect on the CO SLED argues against a strong effect, but we cannot rule out the possibility of a small bias and include this caveat in the discussion in \S \ref{evolution}.

\subsection{Comparison samples}
We add to the $z\sim2$ lensed galaxy sample additional objects from the literature to serve as a comparison point. First, we use a compilation of 16 galaxies in the GOODS-North field for which CO line fluxes as well as deep Herschel photometry are available \citep{magnellisaintonge}.  These galaxies are located mostly at $1.0<z<1.5$ and sample well the SFR-stellar mass plane.  We also use the compilation of submillimeter galaxies (SMGs) from \citet{magnelli12}, which expand the sample toward higher SFRs at fixed stellar mass and redshift. All the SMGs from the \citet{magnelli12} compilation are used in the analysis when only dust measurements are involved, and the subset of these also found in  \citet{bothwell13} whenever CO measurements are also needed.  All the galaxies in the comparison sample have {\it Herschel} PACS and SPIRE photometry, which was processed identically as for the lensed galaxy sample in order to derive infrared luminosities, SFRs, dust temperatures and dust masses (see details in \S \ref{FIRproducts}).

\begin{figure*}
\epsscale{1.2}
\plotone{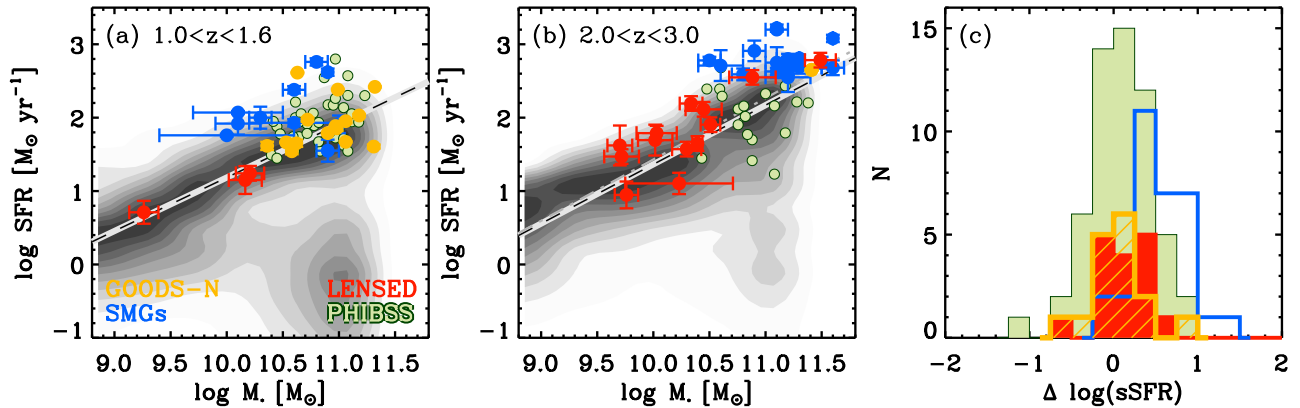}
\caption{Distribution of the sample in the SFR-\mstar\ plane for the redshift ranges of (a) $1.0<z<1.6$ and (b) $2.0<z<3.0$.  The lensed galaxy sample is represented in red, while the GOODS-N sample is in orange and the SMG compilation in blue (see sample description in \S \ref{sample}).  Additionally, we show the position in this plane of the PHIBSS galaxies \citep{tacconi13}, which are used in the analysis in \S \ref{evolution}. The gray contours represent the distribution of galaxies from the GOODS and EGS fields, with SFRs and stellar masses on the same calibration scheme as our main galaxy sample \citep{wuyts11}.  The star formation main sequence (dashed line) is derived at each redshift from this large reference sample, and given by eq. \ref{eqMS}.  In panel (b), we differentiate between the main sequence at $z=2.0-2.5$ (dashed line) and at $z=2.5-3.0$ (dashed-dotted line). In panel (c) we show the distribution of offsets from the star formation main sequence for all samples (same color scheme as in panels a and b), defined at fixed stellar mass as $\Delta \log(sSFR)=\log{\rm (SFR)}-\log{\rm (SFR)_{MS}}$.  \label{SFRmass}}
\end{figure*}

The lensed and comparison galaxy samples are shown in Figure \ref{SFRmass} in the SFR-\mstar\ plane for two redshift intervals, $1.0<z<1.6$ and $2.0<z<3.0$, using the GOODS and EGS catalogs of \citet{wuyts11} to provide a reference and define the main sequence (MS).  These are the samples onto which our measurements of stellar mass and star formation rate are calibrated (sections \ref{mstarsection} and \ref{sfrsection}), making them the best matched reference catalogs.   The MS we derive from these is 
\begin{equation}
\log {\rm SFR}_{MS} = a + b \log (M_{\ast}/M_{\odot}), 
\label{eqMS}
\end{equation}
where $a=[-6.102,-6.704,-6.923]$ and $b=[0.728,0.807,0.834]$ for the three redshift intervals of $[1.0-1.6]$, $[2.0-2.5]$ and $[2.5-3.0]$, respectively.

When discussing the molecular gas properties of the lensed galaxies, we also use additional references.  In particular, we make use of the data from the COLD GASS survey \citep{COLDGASS1}, which includes CO(1-0) measurements for a representative sample of 365 SDSS-selected galaxies at $0.025<z<0.050$ with $M_{\ast}>10^{10}$\msun.  For high redshift galaxies, the PHIBSS sample \citep{tacconi13} is by far the largest compilation of CO measurements for normal star-forming galaxies, and is the perfect reference to study e.g. the redshift evolution of the gas contents of galaxies (see section \ref{evolution}).

%===================================
\section{Data}
\label{data}

\subsection{Herschel photometry}
\label{herscheldata}

\begin{deluxetable*}{lcccccccc}
\tabletypesize{\small}
\tablewidth{0pt}
\tablecaption{PACS, SPIRE and MAMBO photometry \label{FIRtab}}
\tablehead{
\colhead{Name} & \colhead{$S_{70}$} & \colhead{$S_{100}$} & \colhead{$S_{160}$} & \colhead{$S_{250}$} &
\colhead{$S_{350}$} & \colhead{$S_{500}$}  & \colhead{$S_{1200}$}  \\
 & \colhead{mJy} & \colhead{mJy} & \colhead{mJy} & \colhead{mJy} & \colhead{mJy} & \colhead{mJy} & \colhead{mJy} }
\startdata
        \arc & 
  15.1$\pm$1.0 & 
  36.8$\pm$1.5 & 
  63.9$\pm$2.6 & 
  91.3$\pm$3.6 & 
  78.5$\pm$4.2 & 
  47.8$\pm$4.0 & 
   2.9$\pm$0.5 \\ 
       J0712 & 
   0.1$\pm$1.0 & 
   0.1$\pm$1.8 & 
   1.2$\pm$2.9 & 
   9.1$\pm$2.9 & 
   3.8$\pm$3.4 & 
   4.7$\pm$3.1 & 
    \nodata \\ 
       J0744 & 
   0.6$\pm$2.0 & 
   2.2$\pm$4.1 & 
   1.4$\pm$5.5 & 
   0.4$\pm$3.4 & 
   3.9$\pm$3.6 & 
   4.0$\pm$3.3 & 
    \nodata \\ 
       J0900 & 
   0.8$\pm$1.0 & 
   9.2$\pm$1.6 & 
  11.1$\pm$3.6 & 
   7.3$\pm$3.5 & 
   7.2$\pm$3.4 & 
   2.1$\pm$3.2 & 
    \nodata \\ 
       J0901 & 
  18.9$\pm$1.0 & 
  45.5$\pm$2.2 & 
 119.3$\pm$5.4 & 
 226.4$\pm$7.0 & 
 276.1$\pm$6.2 & 
 208.7$\pm$6.0 & 
  15.3$\pm$1.2\tablenotemark{a} \\ 
       J1133 & 
   1.0$\pm$1.2 & 
   0.6$\pm$1.6 & 
   4.9$\pm$2.3 & 
   9.1$\pm$3.7 & 
   6.9$\pm$3.6 & 
   6.1$\pm$3.1 & 
    \nodata \\ 
       J1137 & 
   8.6$\pm$1.0 & 
  18.5$\pm$1.0 & 
  23.0$\pm$2.4 & 
  19.3$\pm$4.5 & 
   9.9$\pm$3.4 & 
   0.9$\pm$4.0 & 
    \nodata \\ 
       Horseshoe & 
   2.2$\pm$0.9 & 
   6.7$\pm$1.1 & 
  15.2$\pm$1.3 & 
  24.8$\pm$3.5 & 
  16.2$\pm$4.4 & 
   9.3$\pm$3.7 & 
    \nodata \\ 
       J1149 & 
   0.5$\pm$0.8 & 
   2.2$\pm$1.1 & 
   4.9$\pm$4.0 & 
  10.6$\pm$3.5 & 
   8.0$\pm$3.5 & 
   3.7$\pm$3.5 & 
    \nodata \\ 
           Clone & 
   9.7$\pm$1.2 & 
  20.6$\pm$1.3 & 
  32.2$\pm$3.1 & 
  32.0$\pm$2.9 & 
  20.8$\pm$4.3 & 
  13.4$\pm$3.9 & 
    \nodata \\ 
       J1226 & 
   1.5$\pm$0.9 & 
   5.1$\pm$0.9 & 
   8.6$\pm$2.3 & 
   9.7$\pm$5.3 & 
   3.7$\pm$3.3 & 
   3.1$\pm$5.0 & 
    \nodata \\ 
       J1343 & 
   1.3$\pm$0.8 & 
   2.2$\pm$1.0 & 
   4.3$\pm$1.5 & 
   2.8$\pm$2.4 & 
   1.3$\pm$3.3 & 
   2.6$\pm$4.9 & 
    \nodata \\ 
       J1441 & 
   1.0$\pm$1.0 & 
   1.0$\pm$1.1 & 
   3.0$\pm$2.4 & 
   1.5$\pm$3.2 & 
   1.3$\pm$3.9 & 
   0.9$\pm$4.0 & 
    \nodata \\ 
            cB58 & 
   4.3$\pm$0.9 & 
   9.3$\pm$1.0 & 
  21.6$\pm$1.6 & 
  24.9$\pm$2.4 & 
  16.0$\pm$3.3 & 
  10.9$\pm$3.6 & 
   1.1$\pm$0.3   \tablenotemark{b} \\ 
       J1527 & 
   0.7$\pm$0.8 & 
   1.9$\pm$1.2 & 
   3.0$\pm$1.6 & 
   0.3$\pm$4.0 & 
   0.5$\pm$4.0 & 
   0.8$\pm$4.4 & 
    \nodata \\ 
             Eye & 
   5.4$\pm$0.7 & 
  10.9$\pm$1.2 & 
  17.4$\pm$1.7 & 
  26.8$\pm$7.0 & 
  33.1$\pm$8.7 & 
  19.1$\pm$  10.0 & 
   1.6$\pm$0.3 \\ 
     Eyelash & 
   8.0$\pm$1.0 & 
  32.9$\pm$1.3 & 
 115.4$\pm$3.6 & 
 335.4$\pm$  19.5 & 
 438.9$\pm$9.0 & 
 353.1$\pm$  10.3 & 
  25.5$\pm$0.4   \tablenotemark{c} 
\enddata
\tablenotetext{a}{Including a factor 2.4 scaling to account for aperture effects (see \S \ref{mambo}).}
\tablenotetext{b}{\citet{baker01}}
\tablenotetext{c}{SMA measurement from \citet{swinbank10}}
\end{deluxetable*}

\subsubsection{PACS observations and data reduction}
We have obtained 70, 100 and 160$\mu$m `mini-scanmaps' of our targets using 
the  PACS instrument \citep{poglitsch10} on board the Herschel Space
Observatory \citep{pilbratt10}.
Total observing time per source was 1~hour at 70 and 100$\mu$m each, and 
2~hours at 160$\mu$m wavelength, which is observed in parallel to both of 
the shorter wavelengths. The resulting PACS maps cover an area of 
$\sim 3\arcmin\times 1.5$\arcmin\ at more
than half of the peak coverage, and have useful information over an area 
$\sim 6\arcmin\times 2.5$\arcmin\/. We processed the PACS data to maps using 
standard procedures similar to those described for the PEP project in 
\citet{lutz11},
in build 7.0.1786 of the Herschel HIPE software \citep{ott10}. The {\it Herschel} blind pointing
accuracy is $\sim 2\arcsec$ RMS \citep{pilbratt10}.   To secure the astrometry, we therefore inspected the PACS maps to 
identify far-infrared sources clearly associated with counterparts having 
accurate astrometry (typically from SDSS).  These reference positions were used  
to correct the astrometry of the PACS data to sub-arcsecond accuracy. 
 
\subsubsection{SPIRE observations and data reduction}
We used the SPIRE instrument \citep{griffin10} to simultaneously obtain
250, 350, and 500$\mu$m `small maps' of our sources, using 14 repetitions and
a total observing time of 35~minutes per source.  Maps were produced with the standard 
reduction pipeline in HIPE (version 4.0.1349).  Following the recommendation in the SPIRE Photometer Instrument Description, the maps are scaled with the appropriate flux correction factors of 1.02, 1.05, and 0.94 at 250$\mu$m, 350$\mu$m, and 500$\mu$m, respectively.  The typical calibration accuracy of SPIRE maps is 15\%.  A preliminary source extraction at 250$\mu$m was performed, and the sources with counterparts in the PACS images (shifted as described above to the appropriate coordinate zero point) were used to correct the astrometry in all SPIRE bands.  

 \subsubsection{Far-infrared flux measurements}
We developed a measurement technique that combines aspects of blind source extraction, and guided extraction using prior information on the position of the sources.  The procedure is aimed at measuring reliable fluxes for the lensed sources across the six {\it Herschel} bands, rather than produce complete catalogs in any given band.   

The first step is to perform a blind extraction with StarFinder \citep{diolaiti00} on the 100$\mu$m, 160$\mu$m and 250$\mu$m maps.  The resulting catalogs are used as prior information on the position of sources brighter than 3$\sigma$, which is used to extract accurate fluxes.  The main advantages of this PSF-fitting technique are that it accounts for most of the blending which could be an issue in these typically crowded fields at the longer wavelengths, and allows for photometry on any specific object that is reliable across the different bands \citep{magnelli09, magnelli11, lutz11}.   For the PACS images, we use the merged blind 100$\mu$m and 160$\mu$m catalogs as a prior, while the 250$\mu$m blind catalog is used as the prior for the SPIRE bands.  In cases where the lensed galaxy of interest is detected at less than 3$\sigma$, the position of this galaxy is added in the prior catalogs based on its coordinates at shorter wavelengths.  In most cases, the closest object with a 3$\sigma$ detection in the priors catalog, be it the lens or another object, is located 20-40\arcsec\ from the lensed galaxy and therefore easily separable at both 160 and 250$\mu$m.  Only in the cases of cB58, the Cosmic Eye and J1133 is the closest neighboring object closer, at a distance of 10-15\arcsec.  Even in these cases the objects are well separated in the PACS imaging, which we use as a guide for the SPIRE priors.

The {\it Herschel} fluxes obtained from the PSF-fitting are finally aperture-corrected using factors of (0.883, 0.866, 0.811, 0.835, 0.848, 0.898), derived specifically for the PSFs used at (70, 100, 160, 250, 350, 500)$\mu$m.  Unless the lens is detected in the PACS/SPIRE images with $>3\sigma$, it is assumed that its FIR emission is negligible and does not contaminate the measurement for the lensed galaxy of interest.  The final {\it Herschel} fluxes for the 17 lensed galaxies are given in Table \ref{FIRtab}.  Values $<2\sigma$ should be interpreted as upper limits, which occurs for 5 galaxies at both PACS 160$\mu$m and SPIRE 250$\mu$m.

\subsection{IRAM-MAMBO photometry}
\label{mambo}
Photometric observations in the 1.2mm continuum were obtained for part of our
sample during the pool observing sessions at the IRAM 
30m telescope in the winters 2006/2007 and 2007/2008. We used the 117 element 
version of 
the Max Planck Millimeter Bolometer (MAMBO) array \citep{kreysa98}.
On-off observations were typically obtained in blocks of 6 scans of 20 
subscans each, and repeated in later observing nights unless a detection
was already reached. The data were reduced with standard procedures 
in the MOPSIC package developed by R.~Zylka, using the default calibration 
files for the applicable pool periods. Table~\ref{FIRtab} lists the 
measured 1.2mm fluxes and their statistical uncertainties. We add the
MAMBO detection which we already obtained for cB58 \citep{baker01} and the SMA flux obtained by \citet{swinbank10} for the Eyelash.
For J0901, we obtained a clear MAMBO detection at the target position
centered on the southern bright lensed component but both the PACS maps and 
IRAM-PdB CO maps clearly indicate that the 11\arcsec\ 
MAMBO beam is missing flux.  Based on the CO and PACS 100$\mu$m maps, the ratio between the total flux of J0901 and the flux in the southern component is independently measured to be 2.4 and 2.5, respectively.  We hence scale the MAMBO flux from the southern component by a factor of 2.4 to infer the total flux before further use.

\subsection{IRAM-PdBI CO mapping}

Molecular gas mass measurements through observations of the CO(3-2) line have been performed for 10 of the 17 galaxies in the {\it Herschel} sample, and we report here on both the previously published measurements and those coming from new PdBI observations.  In sections \ref{results} and \ref{summary}, the full sample of 17 galaxies is used for the elements of the analysis that do not invoke molecular gas, and this subset of 10 galaxies for the rest of the analysis.

\subsubsection{Previous CO observations}

The CO line has been previously observed in seven galaxies from our {\it Herschel} sample.  We briefly review the specifics of the observations of each object below, while a summary of the CO measurements is given in Table \ref{COtable}. 

\paragraph{Eye} 
The Cosmic Eye was observed in the CO(3-2) line with the IRAM PdBI \citep{coppin07}. The line width is $190\pm24$ \kms\, and the total line flux integrated over the line is $0.50\pm0.07$ Jy \kms.  The CO emission appears to be spatially associated with component B$_1$ of the system, and \citet{coppin07} therefore suggest that the appropriate magnification correction factor for the CO line flux is 8 rather than the value of 30 found for the entire system \citep{dye07}.  However, \citet{riechers10} report a detection of the CO(1-0) with the VLA that is spatially consistent with the bulk of the rest-frame UV emission, leading them to conclude that the total magnification value of the system should be used.   This is the approach we adopt here. 

\paragraph{Eyelash}
Multiple CO lines of both $^{12}$CO and $^{13}$CO were observed for this object  \citep{danielson11}.  Although a $^{12}$CO(1-0) flux has been measured, for uniformity with the rest of the sample we adopt for the Eyelash the flux of $13.20\pm0.10$ Jy \kms\ measured in the (3-2) line.

\paragraph{J0901} \citet{sharon13} reports on both EVLA CO(1-0) and PdBI CO(3-2) observations.  For consistency with the rest of the dataset, we adopt the CO(3-2) flux of $19.8\pm2.0$ Jy \kms, which is obtained after primary beam correction given the large angular size of the source. 

\paragraph{cB58} We adopt for this object the CO(3-2) flux of 0.37$\pm$0.08 Jy \kms\ measured by \citet{baker04}, which itself is consistent with an upper limit previously set by \citet{frayer97} and with the VLA CO(1--0) detection by \citet{riechers10} for the value of $R_{13}$ we adopt for the bulk of our sample (see \S \ref{MH2section}).

\paragraph{\arc} The ``eight o'clock arc" (thereafter the \arc) was observed in 2007 May with the IRAM PdBI in compact configuration, using the 3mm SIS receivers to target the redshifted CO(3-2) line (Baker et al. in prep.).  Following a standard reduction process, a 0.45 mJy continuum source associated with the lens was subtracted \citep[see also][]{volino10}, and a final line flux of 0.85$\pm$0.24 Jy \kms\ is measured.

\paragraph{Clone, Horseshoe} Baker et al. (in prep.) report on IRAM PdBI compact 
configuration observations of these two sources, combining datasets taken 
in July to October 2007 and April 2009. Using standard reductions and integrating over the spatially 
extended CO emission, total line fluxes for the redshifted CO(3-2) line of 
$0.48\pm0.15$ (Clone) and $0.44\pm0.18$ (Horseshoe) are derived.

\subsubsection{New PdBI observations: J0900, J1137 and J1226}
In July-October 2011 we obtained CO(3-2) maps (rest frequency of 345.998 GHz) for three additional UV-bright lensed objects with the IRAM PdBI \citep{guilloteau92}.  The three targets, J0900, J1137 and J1226, were chosen to extend the parameter space of objects with molecular gas measurements.  Specifically, J0900 has the lowest metallicity of the sources with reliable {\it Herschel} flux measurements, and J1137 and J1226 are at the very low and very high ends of the redshift range of the sample, respectively.  

The lower redshift source, J1137, was observed in the 2mm band, while for the other two objects the CO(3-2) line was visible in the 3mm band.  Observations were carried out under average to mediocre summer conditions with five of the six 15-m antennae in operation and in compact configuration.  A total of 6-10 tracks of various duration per object were necessary to reach the required line sensitivities.  Data were recorded with the dual polarization, large bandwidth WideX correlators, providing spectral resolution of 1.95 MHz over a total bandwidth of 3.6 GHz.  

The data were calibrated using the CLIC package and maps produced with MAPPING, within the IRAM GILDAS\footnote{http://www.iram.fr/IRAMFR/GILDAS} software environment.   A standard passband calibration scheme was first applied, followed by phase and amplitude calibration.  Due to the poor observing conditions during some of the runs (e.g. high precipitable water vapour, strong winds, low elevation), particular care was taken to flag data with high phase noise.  The absolute flux calibration was done using observation of reference bright quasars, and is typically accurate to better than 20\% \citep{tacconi10}.

\begin{figure*}
\epsscale{1.2}
\plotone{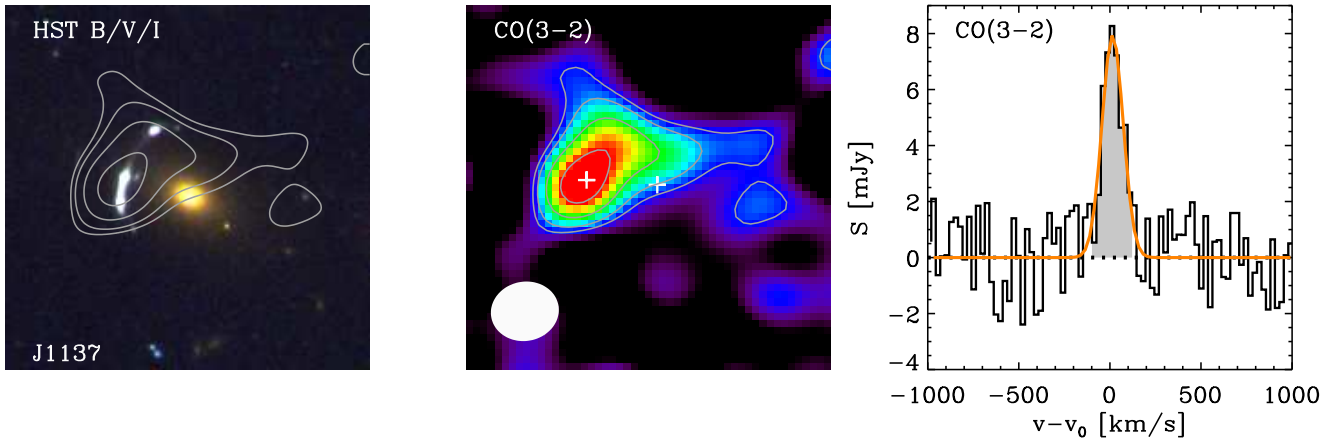}
\caption{Results of the IRAM-PdBI observations of J1137.   The velocity-integrated map is shown in the middle panel, with contours showing the 2,3,5 and 8 $\sigma$ levels.  The shape of the PdBI beam for these data (3.75\arcsec$\times$3.26\arcsec) is shown in the bottom-left corner, and the crosses show the position of the lens and of the main arc.  On the left panel, the CO(3-2) contours are overlaid on top of the three-color HST ACS image.   Both images are centered on $\alpha_{J2000}=11:37:40.1$, $\delta_{J2000}=+49:36:36.1$ and have a size of 20\arcsec$\times$20\arcsec.  Finally, the CO(3-2) line spectrum is shown in the left panel, with the best-fitting gaussian model shown.  The line has a FWHM of 137 \kms\ and a total integrated flux of $1.16\pm0.12$ Jy \kms.    \label{J1137spectrum}}
\end{figure*}

The data cubes were examined for sources at the expected spatial and spectral positions.  The CO(3-2) line is clearly detected in J1137, but not in J0900 and J1226.   Assuming 200 \kms\ line widths, the $3\sigma$ upper limit on the line flux is 0.16 Jy \kms\ for J1226 and 0.44 Jy \kms\ for J0900.  The measured CO(3-2) integrated line flux of J1137 is $1.16\pm0.12$ Jy \kms, and the line FWHM is 137 \kms\ as determined by a gaussian fit. The integrated CO(3-2) line map and spectrum of J1137 are shown in Figure \ref{J1137spectrum}.   

%===================================
\section{Derived quantities}
\label{quantities}

\subsection{Stellar masses}
\label{mstarsection}

Stellar masses found in the literature can vary significantly for the same object depending on the measurement technique, and specific assumptions made regarding star formation histories, metallicities, and stellar population ages. For example there is an order of magnitude difference for cB58 between \citet{siana08} and \citet{wuyts12},  for the Eye between \citet{richard11} and \citet{sommariva12}, and for the \arc\ between \citet{finkelstein09} and \citet{dessauges12} or \citet{richard11}.  Since homogeneity is paramount for the analysis we conduct here, we derive new stellar masses consistently for all the lensed galaxies in our sample. 

This is done with {\it Spitzer}/IRAC imaging, which is available for all the lensed galaxies with the exception of J1441.  The calibrated images at 3.6 and 4.5 $\mu$m were retrieved from the {\it Spitzer} archive, and fluxes measured using a custom-made pipeline.  Since most of the lensed galaxies appear as resolved arcs in the IRAC images, and since they are also often situated in the wings of bright sources (generally, the lensing galaxy), standard photometric tools are not adequate.  The details of the extraction technique are given in Appendix \ref{iracphoto}, and the measured 3.6 and 4.5 $\mu$m fluxes (un-corrected for lensing) are given in Table \ref{masstab}.

\begin{deluxetable*}{lccccc}
\tabletypesize{\small}
\tablewidth{0pt}
\tablecaption{IRAC photometry and derived stellar masses\label{masstab}}
\tablehead{
\colhead{Name} & \colhead{$S_{3.6}$} & \colhead{$S_{4.5}$} & \colhead{$\log M_{\ast}/M_{\odot}$} & \colhead{$\log M_{\ast}/M_{\odot}$}  & \colhead{reference}  \\
 & \colhead{$\mu$Jy} & \colhead{$\mu$Jy} & \colhead{this work} & \colhead{literature} & }
\startdata
        \arc  & 165.4$\pm$  6.4 & 200.3$\pm$  8.3 & 10.88$\pm$ 0.21 & 11.20$\pm$ 0.40 &       \citet{finkelstein09} \\
       J0712 & \nodata & \nodata & \nodata & 10.23$\pm$ 0.48 &           \citet{richard11} \\
       J0744 &  21.7$\pm$  1.3 &  23.0$\pm$  1.4 &  9.70$\pm$ 0.11 &  9.76$\pm$ 0.08 &           \citet{richard11} \\
       J0900 &  44.2$\pm$  5.1 &  41.1$\pm$  7.1 & 10.45$\pm$ 0.16 & 10.28$\pm$ 0.20 &              \citet{bian10} \\
       J0901 & 477.6$\pm$ 20.2 & 556.3$\pm$ 26.4 & 11.49$\pm$ 0.14 &           \nodata & \nodata \\
       J1133 & 112.2$\pm$ 26.0 & 130.5$\pm$ 16.7 & 10.17$\pm$ 0.15 &           \nodata & \nodata \\
       J1137 & 197.0$\pm$ 22.5 & 192.6$\pm$ 14.0 & 10.21$\pm$ 0.13 &           \nodata & \nodata \\
          Horseshoe & 158.2$\pm$ 11.0 & 123.1$\pm$ 16.0 & 10.30$\pm$ 0.13 & 10.56$\pm$ 0.19 &         \citet{sommariva12} \\
       J1149 &  27.0$\pm$  2.0 &  25.2$\pm$  1.6 &  9.26$\pm$ 0.13 &           \nodata & \nodata \\
              Clone & 230.8$\pm$  8.2 & 218.7$\pm$  9.9 & 10.39$\pm$ 0.04 &           \nodata & \nodata \\
       J1226 &  52.8$\pm$  6.0 &  47.2$\pm$  4.2 &  9.72$\pm$ 0.15 &      $9.54\pm0.07 $ & \citet{wuyts12} \\
       J1343 &  69.6$\pm$  7.7 &      \nodata &  9.76$\pm$ 0.10 &  9.40$\pm$ 0.20 &            \citet{wuyts12b} \\
       J1441 & \nodata & \nodata & \nodata & \nodata & \nodata \\
               cB58 &  79.0$\pm$  4.3 &  85.2$\pm$  5.2 & 10.03$\pm$ 0.18 &  9.69$\pm$ 0.14 &             \citet{wuyts12} \\
       J1527 &  36.3$\pm$  2.6 &  38.2$\pm$  4.1 & 10.02$\pm$ 0.15 &        $9.82\pm0.04 $ & \citet{wuyts12} \\
                Eye & 204.7$\pm$  6.7 & 185.2$\pm$  9.1 & 10.51$\pm$ 0.09 & 10.53$\pm$ 0.08 &           \citet{richard11} \\
     Eyelash & 132.8$\pm$  8.3 & 206.8$\pm$ 12.3 & 10.34$\pm$ 0.10 & 10.28$\pm$ 0.29 &          \citet{swinbank10} 
\enddata
\tablecomments{Alternative published values of stellar masses (in $\log M_{\odot}$ units) are $10.02\pm0.36$ for the \arc\  \citep{richard11}, $9.55\pm0.14$ for the Cosmic Eye \citep{sommariva12}, and $8.94\pm0.15$ for cB58 \citep{siana08}.}
\end{deluxetable*}

To compute stellar masses from these IRAC fluxes that will be consistent with the masses of the comparison sample, we use the catalog of \citet{wuyts11} for the GOODS fields as a calibration set.  For each of our lensed galaxies, we extract from the calibration set all galaxies within $\Delta z = \pm0.2$ that are star forming based on their location in the SFR-\mstar\ plane.  This subset of galaxies is used to determine the relation between observed 3.6 and 4.5 $\mu$m fluxes and stellar mass.  The scatter in these empirical relations varies from 0.13 dex for the lensed galaxies at $z\sim1.5$ to 0.19 dex at the highest redshifts.  These uncertainties are comparable to the typical variations in stellar masses derived though SED modeling under varying assumptions \citep[e.g.][]{forster04,shapley05,maraston10}.   Stellar masses for the lensed galaxies are then obtained by taking the observed IRAC fluxes (Table \ref{masstab}), correcting them for the lensing magnification, and then applying the empirical calibration determined for each object.  The masses derived from the 3.6 and 4.5 $\mu$m images are consistent within the errors, and we adopt as our stellar masses the mean between the two values.  These values are summarized in Table \ref{masstab} with the errors quoted obtained by propagating the uncertainties on the parameters of the fit to the calibration data set, the measurement errors on the IRAC fluxes, and the uncertainty on the magnification factor.   The measurement and calibration uncertainties account for $\sim10-30\%$ of the error budget, the rest coming from the uncertainty on the magnification.

\begin{figure}
\epsscale{1.0}
\plotone{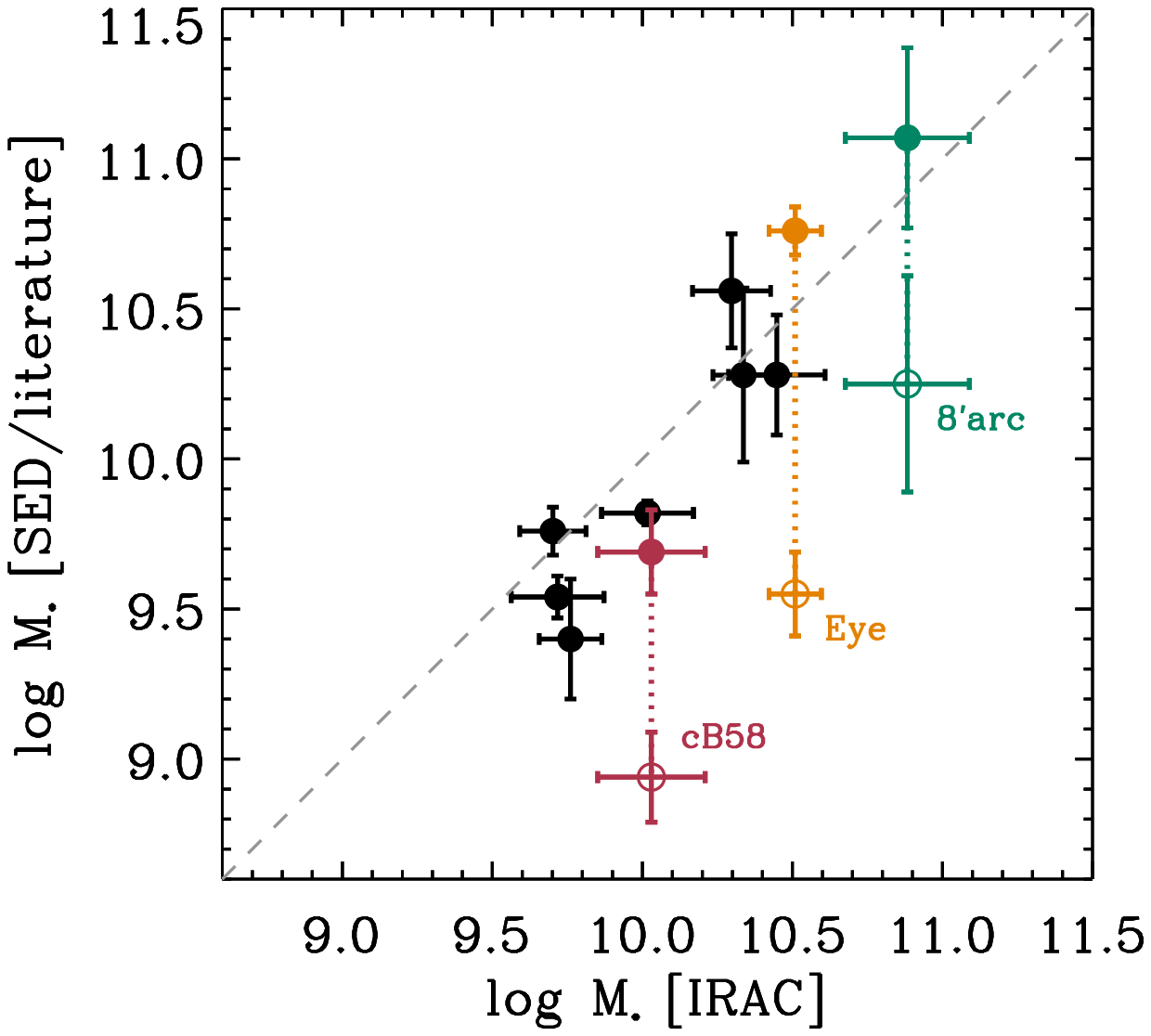}
\caption{Comparison between the stellar masses derived from the IRAC photometry and the values recovered from the literature and converted to a Chabrier IMF when necessary (see Table \ref{masstab}).   Alternative published values for the \arc, the Eye and cB58 are shown as open symbols. \label{masses}}
\end{figure}

When available, we compare in Figure \ref{masses} previously published stellar masses (typically from SED-fitting) to our IRAC-derived masses.  For cB58, the Cosmic Eye and the \arc, two previously published values differing by $\sim$one order of magnitude are shown.  The IRAC-derived masses are always consistent with the higher of the two values.   Adopting these higher estimates for these three galaxies, the scatter between the IRAC- and SED-derived stellar masses is 0.23 dex.  This scatter and the outlier points are caused by the uncertainty on the calibration of our IRAC-based stellar masses, and by different assumptions about star formation histories in the SED modeling. 

\subsection{Dust masses, dust temperatures and IR luminosities}
\label{FIRproducts}

\begin{figure*}
\epsscale{1.1}
\plotone{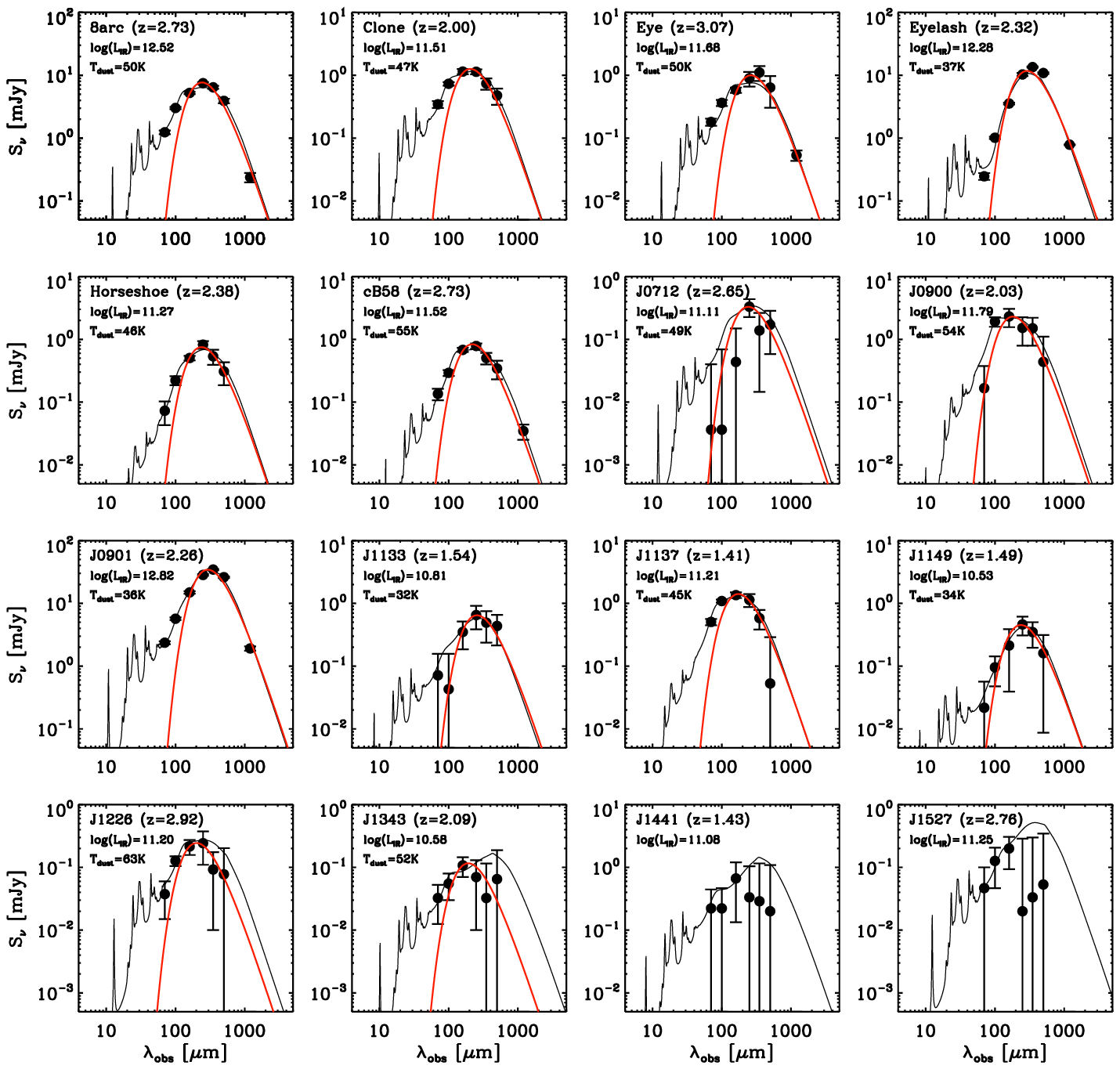}
\caption{Far-infrared spectral energy distributions.  The fluxes have been corrected for the magnification.  The red line is the best-fitting modified black body function, and the black line is from the \citet{draineli07} models.  \label{SED}}
\end{figure*}

The far-infrared SEDs from the PACS, SPIRE and MAMBO photometry are shown in Figure \ref{SED}.  They represent some of the highest fidelity {\it individual} Herschel SEDs of star-forming galaxies at $z>1.5$.  These SEDs are used to derive dust temperatures, dust masses, and total infrared luminosities.  The procedure is identical to the one applied on the comparison samples.  We summarize the key elements here, with the full details given in \citet{magnellisaintonge}. 

Dust masses are calculated using the models of \citet{draineli07} (DL07).  A grid of models is created to sample the expected values of PAH abundances, radiation field intensities and dust fractions in the diffuse ISM.  At each grid point, the model SED is compared with the Herschel photometry, with the dust mass given by the normalization of the SED minimizing the $\chi^2$.  For each galaxy, the final dust mass assigned is the mean value over the grid points where $\chi^2<\chi^2_{min}+1$.  To derive dust temperatures, a single modified black-body model with dust emissivity $\beta=1.5$ is then fit to all the SED points with $\lambda_{rest}>50 \mu$m.  Dust temperatures and masses, along with their measured uncertainties, are given in Table \ref{IRproducts}. 

While infrared luminosities can be derived from the DL07 modeling at the same time as the dust masses, we adopt here the values that are derived from the 160$\mu$m fluxes (although for completeness we also give in Tab. \ref{IRproducts} the values of $L_{IR}$ obtained by integrating in the wavelength range 8-1000$\mu$m the best fitting DL07 model SED).  \citet{nordon12} have demonstrated that for our $z\sim 2$ galaxies this method is robust, as the uncertainties on the 160$\mu$m-to-$L_{IR}$ conversion factors provided by the \citet{ce01} template library are $\lesssim 0.1$ dex.  Even in the cases where the lensed galaxies are not detected in the SPIRE bands, making the fitting of DL07 model templates difficult, there is always a $\gtrsim3\sigma$ detection at 160$\mu$m. For the galaxies with good detections in all PACS and SPIRE bands, there is an excellent agreement between the infrared luminosities derived by the DL07 modeling and by the extrapolation from 160$\mu$m flux  \citep[see also][]{elbaz10}.

\begin{deluxetable*}{lcccccc}
\tablewidth{0pt}
\tablecaption{Dust temperatures, dust masses and star formation rates \label{IRproducts}}
\tablehead{
\colhead{Name} & \colhead{$T_{dust}$} & \colhead{$M_{dust}$} & \colhead{$L_{IR,DL07}$} & \colhead{SFR$_{IR}$} &
\colhead{SFR$_{tot}$} & \colhead{SFR$_{UV}$/SFR$_{IR}$} \\
 & \colhead{K} & \colhead{$10^7 M_{\odot}$} & \colhead{$10^{11} L_{\odot}$} & \colhead{$M_{\odot} {\rm yr}^{-1}$} & \colhead{$M_{\odot} {\rm yr}^{-1}$} & }
\startdata
        \arc & 50$^{+ 1}_{- 1}$& 31.00$^{+  1.52}_{-  1.56}$&    33.44$^{+  0.00}_{-  0.21}$&    342.6$\pm$  13.9 &  354.1$\pm$  14.4 &    0.033$\pm$0.010 \\
       J0712 & 49$^{+20}_{-18}$&  1.91$^{+  2.96}_{-  0.69}$&     1.29$^{+  0.38}_{-  0.37}$&      9.8$\pm$  23.7 &   12.7$\pm$  23.8 &    0.293$\pm$0.713 \\
     J0744 & \nodata  &  \nodata  &     1.20$^{+  0.00}_{-  0.56}$&     40.5$\pm$ 159.2 &   41.8$\pm$ 159.2 &    0.032$\pm$0.127 \\
       J0900 & 54$^{+15}_{-10}$&  4.89$^{+  5.05}_{-  0.00}$&     6.22$^{+  0.52}_{-  0.00}$&     68.2$\pm$  22.1 &  129.9$\pm$  26.2 &    0.904$\pm$0.358 \\
       J0901 & 36$^{+ 1}_{- 1}$&281.67$^{+  0.00}_{-  0.00}$&    67.56$^{+  0.00}_{-  0.00}$&    584.9$\pm$  26.5 &  608.3$\pm$  27.0 &    0.040$\pm$0.009 \\
       J1133 & 32$^{+ 9}_{- 6}$&  3.82$^{+  9.99}_{-  2.77}$&     0.65$^{+  0.19}_{-  0.23}$&      6.3$\pm$3.0 &   14.1$\pm$3.2 &    1.234$\pm$0.606 \\
       J1137 & 45$^{+ 6}_{- 3}$&  2.00$^{+  0.40}_{-  0.22}$&     1.65$^{+  0.14}_{-  0.01}$&     15.6$\pm$1.6 &   17.5$\pm$1.7 &    0.121$\pm$0.027 \\
          Horseshoe & 46$^{+ 1}_{- 2}$&  2.99$^{+  0.89}_{-  0.27}$&     1.89$^{+  0.20}_{-  0.02}$&     25.4$\pm$2.2 &   37.3$\pm$3.1 &    0.467$\pm$0.097 \\
       J1149 & 34$^{+12}_{-10}$&  2.86$^{+  6.87}_{-  1.66}$&     0.34$^{+  0.09}_{-  0.10}$&      3.4$\pm$2.8 &    5.1$\pm$2.8 &    0.500$\pm$0.435 \\
              Clone & 47$^{+ 4}_{- 1}$&  3.58$^{+  0.70}_{-  0.34}$&     3.25$^{+  0.02}_{-  0.15}$&     33.8$\pm$3.3 &   44.4$\pm$3.3 &    0.315$\pm$0.036 \\
       J1226 & 63$^{+ 6}_{-13}$&  1.42$^{+  1.10}_{-  0.28}$&     1.61$^{+  0.00}_{-  0.26}$&     21.0$\pm$5.6 &   29.5$\pm$5.9 &    0.409$\pm$0.138 \\
     J1343 & \nodata  &  \nodata  &     0.38$^{+  0.07}_{-  0.13}$&      5.5$\pm$1.9 &    8.9$\pm$2.0 &    0.624$\pm$0.257 \\
     J1441 & \nodata  &  \nodata  &     1.20$^{+  0.27}_{-  0.51}$&     19.1$\pm$  15.3 &   27.0$\pm$  15.5 &    0.417$\pm$0.362 \\
             cB58 & 55$^{+ 1}_{- 1}$&  2.74$^{+  0.48}_{-  0.18}$&     3.34$^{+  0.19}_{-  0.10}$&     52.9$\pm$3.9 &   61.5$\pm$4.6 &    0.162$\pm$0.046 \\
     J1527 & \nodata  &  \nodata  &     1.78$^{+  0.64}_{-  0.68}$&     26.0$\pm$  13.9 &   49.5$\pm$  14.8 &    0.900$\pm$0.517 \\
              Eye & 50$^{+ 4}_{- 4}$&  4.30$^{+  0.05}_{-  0.00}$&     4.88$^{+  0.00}_{-  0.07}$&     68.0$\pm$6.6 &   80.2$\pm$6.8 &    0.180$\pm$0.027 \\
     Eyelash & 37$^{+ 1}_{- 1}$& 77.99$^{+  0.00}_{-  0.00}$&    19.31$^{+  0.00}_{-  0.00}$&    155.6$\pm$4.9 &  155.9$\pm$4.9 &    0.002$\pm$0.000 
\enddata
\tablecomments{SFR$_{IR}=L_{IR,160\mu m}\times 10^{-10}$, with $L_{IR,160\mu m}$ the infrared luminosity derived from the PACS 160$\mu$m flux.}
\end{deluxetable*}

\subsection{Star formation rates}
\label{sfrsection}

Dust-obscured SFRs can be obtained simply from the infrared luminosities as ${\rm SFR}_{IR}=10^{-10} L_{IR}$, with $L_{IR}$ in units of solar luminosity and SFR$_{IR}$ in \msun yr$^{-1}$, assuming a Chabrier IMF.  At high redshifts and at high total SFR (SFR$_{tot}$), SFR$_{IR}$ is the dominant contribution \citep[e.g.][]{pannella09,reddy10,wuyts11,whitaker12,nordon13,heinis13} and it is commonly assumed that  ${\rm SFR}_{tot} \sim {\rm SFR}_{IR}$, neglecting the un-obscured component (SFR$_{UV}$).  Since the lensed galaxies in the sample were mostly selected for being bright blue arcs in optical images (i.e. bright in rest frame UV), and since due to their large magnification factors their total intrinsic SFRs are modest, it is important in this case to estimate the contribution of SFR$_{UV}$ to SFR$_{tot}$.  Following \citet{kennicutt98}, we measure this for a Chabrier IMF as ${\rm SFR}_{UV}=8.2\times10^{-29} L_{\nu,1600}$.  The rest-frame 1600\AA\ luminosity (in erg/s/Hz) is taken to be: 
\begin{equation}
L_{\nu,1600}=\frac{4\pi D_L^2}{\mu (1+z)} 10^{(48.6+m_{1600})/(-2.5)}, 
\label{eq_sfruv}
\end{equation}
where $D_L$ is the luminosity distance in cm, $\mu$ is the magnification factor from Table \ref{Sampletab},  and $m_{1600}$ the apparent AB-magnitude at a rest wavelength 1600\AA.  The most accurate approach to obtain $m_{1600}$ would be to interpolate between all available bands as was done for example by \citet{nordon13}, but its main drawback for the lensed galaxy sample is in the lack of homogenous, multi-wavelength photometry.  While some galaxies have published photometry in several HST bands, some have been observed only with a single HST filter, while some others only have ground based photometry available.  Given the nature of the available data and the fact that the SEDs of galaxies are typically relatively flat at these wavelengths, we directly adopt for $m_{1600}$ the observed magnitude in the available optical band the closest to rest-frame 1600\AA.  Given the typical UV slopes $\beta$ in the range from -2.0 to -1.0 (see Fig. \ref{irxbeta}), the conversion factor from AB magnitude to SFR$_{UV}$ has only a very weak wavelength dependence, and therefore this approximation of adopting the closest band leads to systematic uncertainties of no more than 10\%.

We therefore adopt for the lensed galaxies a total SFR given by ${\rm SFR}_{UV}+{\rm SFR}_{IR}$.  For the GOODS-N comparison sample, we retrieve the ACS photometry from the multi-wavelength catalog of \citet{berta11} \footnote{available at http://www.mpe.mpg.de/ir/Research/PEP/index.php}  and compute SFR$_{UV}$ from Eq. \ref{eq_sfruv} using either B- or V-band magnitudes, depending on the redshift of each galaxy.  For the SMG sample, we assume that ${\rm SFR}_{tot}={\rm SFR}_{IR}$, based on their high IR luminosities and location in the SFR-\mstar\ plane, indicating a very high attenuation of the UV light and therefore a negligible contribution of  SFR$_{UV}$ \citep{buat05,chapman05,wuyts11sfr,nordon13,casey13}. 

\subsection{Metallicities}
\label{metalsection}
In order to obtain a homogenous set of metallicities for all galaxies in our Herschel/IRAM sample, we have compiled from the literature [NII] and H$\alpha$ line fluxes (Table \ref{metaltab}).  Such fluxes were available for 11 of the 17 lensed galaxies.  Using the observed [NII]$/H{\alpha}$ ratio, we compute the nebular abundance using the calibration from \citet{denicolo02}.  This is the calibration that produces the least amount of scatter in the mass-metallicity relation \citep{kewley08} and that was used by \citet{genzel12} to derive a prescription for the metallicity-dependence of the CO-to-H$_2$ conversion factor.  

In the absence of a measured [NII]$/H{\alpha}$ ratio, metallicities measured from the strong line indicator $R_{23}$ are adopted for the Eye, J1226 (see Appendix \ref{J1226metal}) and J1441.  These individual measurements are converted to the \citet{denicolo02} scale using the appropriate relation given in \citet{kewley08}.  For the remaining three galaxies (Eyelash, J1133 and J1137), we use the mass-metallicity (MZ) relation as given in \citet{genzel12} to infer a metallicity.  For the special case of J0901, we adopt the metallicity value derived from the MZ relation even in the presence of a measured [NII]/H$\alpha$ ratio as this value is likely affected by the presence of an AGN (details in \S \ref{J0901agn}). For galaxies in the comparison sample, the latter method is used as explained in \citet{magnellisaintonge}.   

All the metallicities, as well as the H$\alpha$ luminosities and [NII]/H$\alpha$ line ratios when available, are summarized in Table \ref{metaltab}.

\begin{deluxetable*}{lccccc}
\tabletypesize{\small}
\tablewidth{0pt}
\tablecaption{[NII] and H$\alpha$ line fluxes and derived gas-phase metallicites\label{metaltab}}
\tablehead{
\colhead{Name} & \colhead{H${\alpha}$} & \colhead{[NII]/H$\alpha$} & \colhead{12$+\log$O/H} & \colhead{method}  & \colhead{reference}  \\
 & \colhead{$10^{-17}$ erg s$^{-1}$ cm$^{-2}$} &  &  &  & \\ }
\startdata
        \arc &  74.5$\pm$  0.9 &  0.153$\pm$ 0.005 &  8.52$\pm$ 0.10 &                N2 & \citet{richard11} \\     
       J0712 &  10.7$\pm$  0.3 &  0.210$\pm$ 0.019 &  8.63$\pm$ 0.11 &                N2 & \citet{richard11} \\
       J0744 &  15.4$\pm$  0.4 &  0.299$\pm$ 0.017 &  8.74$\pm$ 0.10 &                N2 & \citet{richard11} \\
       J0900 &  84.7$\pm$  3.2 &  0.054$\pm$ 0.011 &  8.20$\pm$ 0.13 &                   N2 & \citet{bian10} \\
       J0901 &  72.0$\pm$  1.0 &  0.889$\pm$ 0.019\tablenotemark{a} &  8.91$\pm$ 0.15 &               MZ & \citet{hainline09} \\
       J1133 &   \nodata & \nodata &  8.55$\pm$ 0.20 &                        MZ & this work \\
       J1137 &   \nodata & \nodata &  8.61$\pm$ 0.20 &                        MZ & this work \\
     Horseshoe &  43.4$\pm$  0.9 &  0.092$\pm$ 0.016 &  8.36$\pm$ 0.13 &               N2 & \citet{hainline09} \\
       J1149 &   \nodata &  0.112$\pm$ 0.050 &  8.43$\pm$ 0.22 &                   N2 & \citet{yuan11} \\
             Clone & 202.0$\pm$  5.0 &  0.193$\pm$ 0.011 &  8.60$\pm$ 0.10 &               N2 & \citet{hainline09} \\    
       J1226 &   \nodata & \nodata &  8.27$\pm$ 0.19 &                 R23 & \citet{wuyts12} \\
       J1343 & 155.0$\pm$ 10.0 &  0.148$\pm$ 0.028 &  8.52$\pm$ 0.13 &                 N2 & \citet{wuyts12b} \\
       J1441 &   \nodata & \nodata &  8.14$\pm$ 0.40 &               R23 & \citet{pettini10} \\
               cB58 & 125.6$\pm$  3.7 &  0.091$\pm$ 0.021 &  8.36$\pm$ 0.14 &                N2 & \citet{teplitz00} \\
       J1527 &   \nodata & $<$ 0.224$\pm$ 0.075 & $<$ 8.65$\pm$ 0.18 &                     N2 & \citet{wuyts12} \\
       Eye &   \nodata & \nodata &  8.64$\pm$ 0.14 &               R23 & \citet{richard11} \\
     Eyelash &   \nodata & \nodata &  8.61$\pm$ 0.20 &                        MZ &this work 
\enddata
\tablenotetext{a}{See Appendix \ref{J0901agn} for a discussion about the presence of an AGN in this galaxy. }
\tablecomments{The metallicities in this table are either from the N2 indicator using the values of [NII]/H$\alpha$ from Column (3), the $R_{23}$ indicator, or derived from the mass-metallicity relation (MZ). In all cases, the metallicities are converted to the scale of the \citet{denicolo02} calibration.}
\end{deluxetable*}

\subsection{Molecular gas masses}
\label{MH2section}

\begin{figure*}
\epsscale{0.9}
\plotone{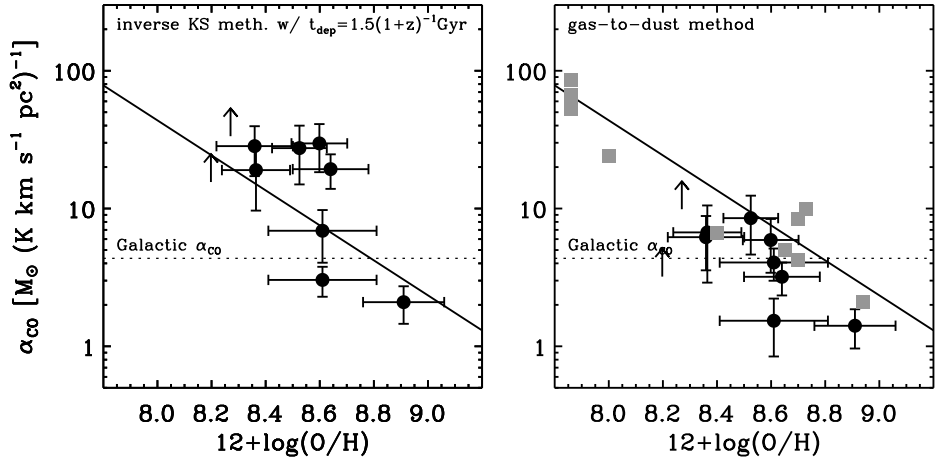}
\caption{Comparison between the conversion factor \xco\ derived either using the ``inverse KS relation" method (left) or using the gas-to-dust ratio method (right) as a function of metallicity.  Only the lensed galaxies with a direct measurement of metallicity are represented (filled circles).   The error bars are obtained by propagating the measurement errors on $L_{CO}$, $M_{dust}$ and SFR, but do not include the systematic uncertainty in the calibration of the two techniques. The $z=0$ data points from \citet{leroy11} are shown as filled squares for the gas-to-dust method.  Finally, the empirical relation for the metallicity-dependence of \xco\ from \citet{genzel12} is shown as a solid line, and the value of 4.35 \msun\ (K km/s pc$^2)^{-1}$ derived from Milky Way clouds and typically assumed for solar-metallicity star forming galaxies is highlighted with the horizontal dotted line.  \label{xco}}
\end{figure*}

Starting from $I_{CO(3-2)}$, the integrated CO(3-2) line fluxes in Jy \kms\ presented in Table \ref{COtable}, we calculate the CO luminosity of the lensed galaxies, $L^{\prime}_{CO}$, in units of (K \kms\ pc$^2$) following \citet{solomon97}:
\begin{equation}
L^{\prime}_{CO}=3.25\times10^7 \frac{I_{CO(3-2)} R_{13} }{\mu} \frac{D_L^2}{\nu_{obs}^2 (1+z)^3}, 
\label{LCOeq}
\end{equation}
where $\mu$ is the magnification factor as given in Table \ref{Sampletab}, $D_L$ is the luminosity distance in Mpc and $\nu_{obs}$ the observed frequency in GHz.  The factor $R_{13}\equiv I_{CO(1-0)}/I_{CO(3-2)}$ is the excitation correction to extrapolate the CO(1-0) line flux from our CO(3-2) observations.  As in \citet{tacconi13}, we adopt a value of $R_{13}=2$ based on recent studies of the CO spectral line energy distribution \citep[e.g.][]{weiss07,dannerbauer09,harris10,ivison11,bauermeister13,bothwell13}.  These studies have targeted both normal high-$z$ star-forming galaxies and SMGs, with similar results pointing to a characteristic value of $R_{13}\sim2$ that may be due to low excitation or to the typical filling factors of the two lines.  We therefore apply this value of $R_{13}$ uniformly across our sample, and assume a conservative 20\% uncertainty on this value compared to the typical errors presented in these individual studies.

Total molecular gas masses are inferred from $L^{\prime}_{CO}$ using the CO-to-H$_2$ conversion factor, \xco\ ($M_{H2}=\alpha_{CO} L^{\prime}_{CO}$).  The specific value of \xco\ to be applied for each galaxy must be determined with care, as most have sub-solar metallicities \citep[$(12+\log{\rm O/H})_{\odot}=8.69\pm0.05$;][]{asplund09}.  Under such low metallicity conditions, both observations and models suggest that the value of \xco\ increases, as the CO molecule becomes a poor proxy for H$_2$ \citep[e.g.][]{israel97,dame01,rosolowsky03,blitz07,leroy11, glover11,shetty11,feldmann12,genzel12}.  Here, different methods to estimate \xco\ are investigated. 

The first is the ``inverse Kennicutt-Schmidt relation" method.  Under the assumption that a tight relation exists between the surface density of molecular gas and the SFR surface density, the value of \xco\ can be estimated knowing $\Sigma_{SFR}$ and the CO luminosity.  Using a compilation of local and high redshift galaxies, \citet{genzel12} calibrated a relation between \xco\ and metallicity using this approach.   Since most recent studies suggest that the KS relation is near-linear \citep[e.g.][]{leroy08,blanc09,genzel10,daddi10KS,bigiel11,rahman12,saintonge12,shetty13,feldmann13}, the problem is further simplified and \xco\ can be estimated as:
\begin{equation}
\alpha_{CO,KS}=\frac{{\rm SFR}~t_{dep}(H_2)}{L^{\prime}_{CO}},
\label{xcoks}
\end{equation}
where $L^{\prime}_{CO}$ is the CO(1-0) line luminosity in (K \kms\ pc$^2$), and \xco\ the CO-to-H$_2$ conversion factor in \msun~(K \kms\ pc$^2)^{-1}$. In applying eq. \ref{xcoks}, we adopt a redshift-dependent depletion time, \tdep$=1.5(1+z)^{-1}$, as suggested by \citet{tacconi13} and further supported here to $z=2-3$ in \S \ref{evolution}.  In Figure \ref{xco}, left panel, the value of $\alpha_{CO,KS}$ is plotted against metallicity, for the lensed galaxies. The values of \xco\ scatter around the relation derived by \citet{genzel12} using this method and a sample of both local and high redshift galaxies, 
\begin{equation}
\log \alpha_{CO}=-1.27 [12+\log{\rm(O/H)}]+11.8.
\label{XCOG12}
\end{equation}

The second approach, the ``gas-to-dust ratio method", relies on a measurement of the dust mass and a motivated choice of a gas-to-dust ratio.  This method has been used successfully in the local Universe \citep[e.g.][]{israel97,gratier10a,leroy11,bolatto11,sandstrom13}, and shown to be applicable also at high redshifts \citep{magdis11,magdis12b, magnellisaintonge}.  In the case of high redshift galaxies, the conversion factor can be estimated simply as:
\begin{equation}
\alpha_{CO,dust}=\frac{\delta_{GDR}(Z) M_{dust}}{L^{\prime}_{CO}},
\label{xcodust}
\end{equation}
where $M_{dust}$ is in solar masses, and we adopt a metallicity-dependent gas-to-dust ratio, $\delta_{GDR}(Z)$, from \citet{leroy11}.  All the assumptions required to apply equation \ref{xcodust} to high redshift galaxies are extensively described in \S 5.1 of \citet{magnellisaintonge}.  In particular, it needs to be assumed that the CO lines and the FIR continuum are emitted from the same physical regions of the galaxies given that {\it Herschel} does not resolve them, and that at high redshift $M_{H2} \gg M_{HI}$ and therefore that the atomic component of the ISM can be neglected in Eq. \ref{xcodust} (see \S \ref{missinggas} for a justification of this assumption). A significant additional uncertainty  lies in the assumption that the $\delta_{GDR}(Z)$ relation of \citet{leroy11}, calibrated on a handful of very nearby galaxies, applies directly at $z\sim1-3$.   Furthermore, other studies of nearby galaxies indicate that the scatter in the $\delta_{GDR}-Z$ relation may be larger than suggested by this specific $z=0$ sample, especially at low metallicities \citep{draine07,galametz11}.

In Figure \ref{xco}, right panel, $\alpha_{CO,dust}$ measured from eq. \ref{xcodust} is shown as a function of metallicity.  The $z>2$ galaxies follow the inverse relation between \xco\ and metallicity seen in the $z=0$ data and in the empirical relation of \citet{genzel12}, although with a small systematic offset to lower values of \xco\ at fixed metallicity, the implications of which are discussed in \S \ref{gdrsection}.

Although affected by different sets of uncertainties and assumptions, the two methods of estimating \xco\ produce consistent results, which in the mean and within their errors reproduce the metallicity-dependence of \xco\  recently calibrated at $z>1$ by \citet{genzel12}, and previously also observed locally \citep[e.g.][]{wilson95,israel97,boselli02,bolatto11,leroy11}. Throughout the rest of this paper, we therefore adopt as a consensus between $\alpha_{CO,KS}$ and $\alpha_{CO,dust}$ the value of \xco\ obtained from the prescription of \citet{genzel12}, as given in eq. \ref{XCOG12}, using our best estimates of metallicities given in Table \ref{metaltab}.  These values of \xco, as well as the CO luminosities and derived molecular gas masses are presented in Table \ref{COtable}.

\begin{deluxetable*}{lccccccccc}
\tabletypesize{\small}
\tablewidth{0pt}
\tablecaption{Lensed galaxies CO line fluxes and molecular gas masses \label{COtable}}
\tablehead{
\colhead{Name} & \colhead{$I_{CO(3-2)}$\tablenotemark{a}} & \colhead{Reference} & \colhead{$L^{\prime}_{CO}$\tablenotemark{b}} & \colhead{$\alpha_{CO,KS}$} &
\colhead{$\alpha_{CO,dust}$} & \colhead{$\alpha_{CO,Z}$} & \colhead{$\log M_{H2}/M_{\odot}$\tablenotemark{c}} & \colhead{$f_{gas}$\tablenotemark{d}}  & \colhead{$t_{dep}$}\\ 
 & \colhead{[Jy km/s]} & & \colhead{[$10^9$(K km/s pc$^2$)]} & & & & & & \colhead{[Myr]}}
\startdata
        \arc &   0.85$\pm$0.24         & Baker in prep.  &   5.18$\pm$  2.35 &  27.46 &   8.52 &   9.41 &  10.69 &   0.39  & 137\\
                  J0900 & \nodata & this work          & $<$  4.13 & $>$ 15.58 & $>$  3.21 &  24.50 & $<$ 11.00 & \nodata & $<$ 777\\
       J0901 &  19.80$\pm$2.00         & \citet{sharon13}    & 133.91$\pm$ 40.22 &   2.09 &   1.41 &   3.05 &  11.61 &   0.57  & 671\\
        J1137 &   1.16$\pm$0.12         & this work    &   1.57$\pm$  0.45 &   6.90 &   1.53 &   7.33 &  10.06 &   0.42 &  661 \\      
   Horseshoe &   0.44$\pm$0.18         & Baker in prep.          &   0.87$\pm$  0.42 &  19.02 &   6.71 &  15.05 &  10.12 &   0.40 & 351 \\       
       Clone &   0.48$\pm$0.15         & Baker in prep. &   0.75$\pm$  0.28 &  29.70 &   5.91 &   7.58 &   9.75 &   0.19 & 127 \\
        J1226 & \nodata & this work          & $<$  0.34 & $>$ 33.51 & $>$  9.89 &  19.82 & $<$  9.82 & \nodata & $<$ 226 \\
       cB58 &   0.37$\pm$0.08         & \citet{baker04}     &   0.87$\pm$  0.34 &  28.38 &   6.20 &  15.27 &  10.12 &   0.55  & 216 \\
         Eye &   0.50$\pm$0.07         & \citet{coppin07}    &   1.53$\pm$  0.41 &  19.32 &   3.20 &   6.72 &  10.01 &   0.24 &  128 \\
     Eyelash &  13.20$\pm$0.10         & \citet{danielson11} &  23.17$\pm$  5.64 &   3.03 &   4.06 &   7.33 &  11.23 &   0.89 & 1089
 \enddata
\tablenotetext{a}{Observed CO(3-2) integrated line fluxes, uncorrected for lensing.}
\tablenotetext{b}{CO line luminosity, including the magnification correction and an excitation correction to bring this measurement on the CO(1-0) scale (see eq. \ref{LCOeq}). }
\tablenotetext{c}{Measured from $L^{\prime}_{CO}$ and $\alpha_{CO,Z}$, the metallicity-dependent conversion factor of \citet{genzel12}. These values of $M_{H2}$ do include the contribution of helium and are therefore total molecular gas masses.}
\tablenotetext{d}{Molecular gas mass fraction, as defined in eq. \ref{fgaseq}.}
\end{deluxetable*}

%===================================
\section{Results}
\label{results}

Having derived accurate and homogeneous measures of \mstar, SFR, $T_{dust}$, $M_{dust}$, $M_{H2}$ and $12+\log{\rm O/H}$ across the lensed galaxies and comparison samples, we now investigate the gas and dust properties of galaxies as a function of their position in the SFR-\mstar-$z$ parameter space.

\subsection{High dust temperatures in $z>$2 lensed galaxies}
\label{hightdust}

\begin{figure*}
\epsscale{0.9}
\plotone{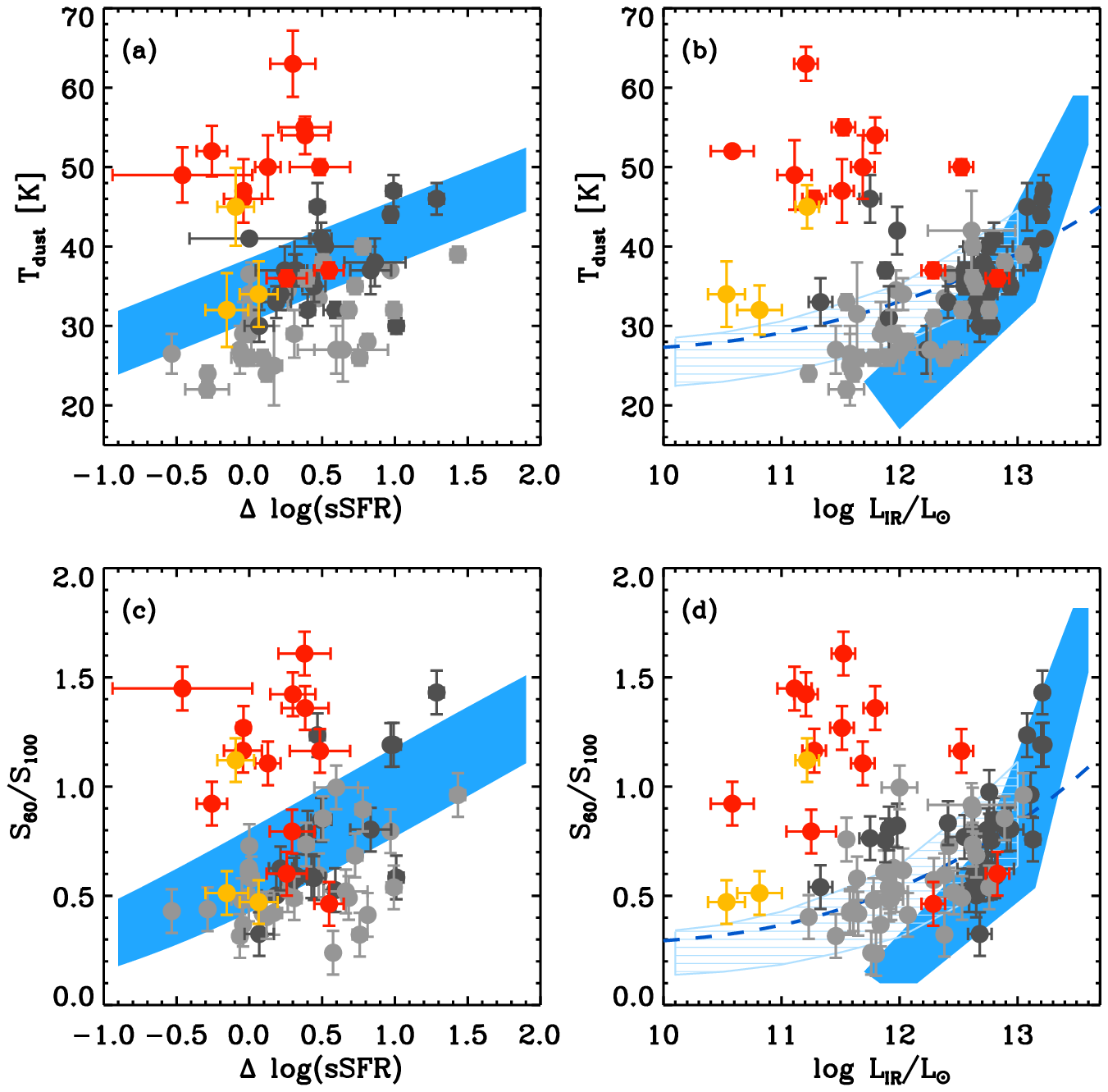}
\caption{Dust temperature as a function of (a) offset from the star formation main sequence, as defined using the GOODS and EGS catalogs of \citet{wuyts11} as a reference, and (b) infrared luminosity.  The infrared color ($S_{60}/S_{100}$), a proxy for dust temperature, is also shown as a function of (c) $\Delta \log(sSFR)$ and (d) $L_{IR}$.  The lensed galaxies are shown in colored symbols (red: $z>2$, orange: $z<2$), and the GOODS-N and SMG reference sample (see \S \ref{sample}) is in gray (dark:  $z>2$, light: $z<2$).   The $T_{dust}-\Delta$(MS) relations of Magnelli et al. (2013) between $z\sim1$ and $z\sim2$ is shown as the shaded region on the left-hand panels (note that this relation is derived using a sample of galaxies more massive than our typical lensed galaxies).  On the right-hand panels, we show as a comparison the local relation of \citet{chapman03} (filled dashed region), the locus of high-redshift SMGs \citep{chapman05} (solid filled region), and the relation of \citet{roseboom12} (dashed line).  To convert the different reference relations from $T_{dust}$ to $S_{60}/S_{100}$ (or vice-versa), we have assumed a simple modified black body with $\beta=1.5$.   \label{tdustfig}}
\end{figure*}

In Figure \ref{tdustfig}, we plot the dust temperature against the offset from the star formation main sequence and the total infrared luminosity, both for the lensed and comparison galaxy samples.  While we recover with the comparison sample the known trend between \tdust\ and these two quantities \citep[][]{dale01,chapman03,hwang10,magnellisaintonge,symeonidis13}, the $z>2$ lensed galaxies occupy a significantly different region of the plot.  Dust temperatures in these galaxies are very high (\tdust$\sim50$K), even though they are located on the main sequence and have modest infrared luminosities ($L_{IR}\sim10^{11}-10^{12}L_{\odot}$) given their redshifts and stellar masses.   Plotting dust temperatures against specific star formation rate rather than main sequence offset produces qualitatively equivalent results. 

There is evidence that the tightness of the $T_{dust}-L_{IR}$ relation observed in classical samples of SMGs \citep[e.g.][see also Fig. \ref{tdustfig}b]{chapman05} is due to selection biases \citep{casey09,magnelli10,magnelli12}.  In particular, the sub-millimeter selection technique favors colder objects, especially at low $L_{IR}$.  When also considering samples of dusty high-redshift star-forming galaxies selected through other techniques \citep[e.g. the optically-faint radio galaxies of][]{chapman04}, dust temperatures can be significantly higher at fixed $L_{IR}$.  The lensed galaxies shown in Fig. \ref{tdustfig}b extend these results to galaxies with even lower IR luminosities and higher dust temperatures.

To confirm the high \tdust\ among the lensed galaxies, we also show in Figure \ref{tdustfig}c-d similar relations using the infrared color, defined as the ratio between rest-frame 60$\mu$m and 100$\mu$m fluxes ($S_{60}/S_{100}$, where $S_{60}$ and $S_{100}$ are determined by linear interpolation of the Herschel photometry).  This quantity, commonly used in IRAS-based studies, is a proxy for \tdust\ but is independent of any model assumptions.  The same behavior is observed for the $z>2$ lensed galaxies, with these objects having unusually high $S_{60}/S_{100}$ ratios given their modest infrared luminosities. 

Among the $z>2$ lensed galaxies, there are however two exceptions, J0901 and the Eyelash.  With \tdust$\sim36$K and $S_{60}/S_{100}\sim0.5$, these two galaxies follow the general trend between \tdust\ and MS offset (or $L_{IR}$) traced by the various comparison samples \citep{magnelli12,roseboom12}.  These two galaxies also differ from the rest of the $z>2$ lensed galaxy sample in other regards.  For example, they have the largest dust masses (\mdust$\sim10^9$\msun\ as compared to $\sim1-3\times10^7$ \msun), and although no direct metallicity measurement is available for the Eyelash, J0901 has the highest metallicity of all the galaxies in the sample (but see Appendix \ref{J0901agn}).   The high dust temperatures in the rest of the $z>2$ lensed galaxies therefore seem to be linked to their low dust contents compared to their SFRs.  As we could not find evidence in \ref{samplebias} that the $z>2$ lensed galaxies form a biased subsample of the underlying galaxy population,  this observations suggests that $\sim10^{10}$\msun\ galaxies at $z=2-3$ have higher dust temperatures and lower dust masses than similar galaxies at lower redshifts.   Using deep Herschel PACS and SPIRE blind fields and a stacking technique, the mean $T_{dust}$ of normal star-forming galaxies with $M_{\ast}\sim10^{10}$\msun\  at $z>2$ is just barely measurable. Such studies indeed suggest a rise in temperature and intensity of the radiation field on the main sequence in the mass/redshift regime that we have now probed directly with the lensed galaxies \citep[][]{reddy12,magdis12a,magnelli13}.

\subsection{Gas-to-dust ratio}
\label{gdrsection}

\begin{figure}
\epsscale{1.2}
\plotone{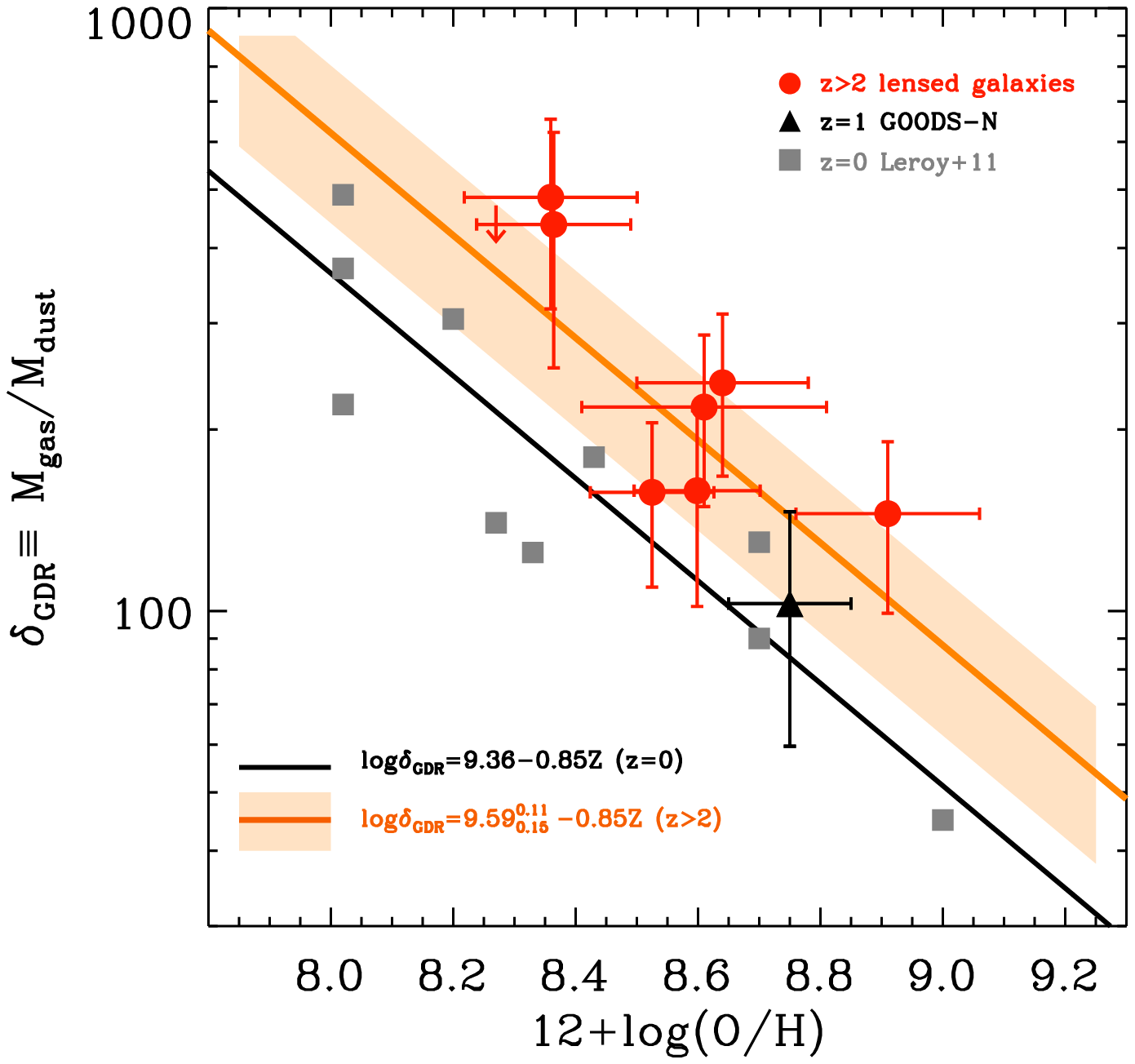}
\caption{Gas-to-dust ratio as a function of gas-phase metallicity for the $z>2$ lensed galaxies (red filled circles) and the $z\sim1.5$ reference sample in GOODS-N (black filled square; the mean value in the sample is given since no direct metallicities are available only estimates from the mass-metallicity relation).   Only galaxies within 0.5 dex of the star formation main sequence are considered.   For all high-redshift galaxies, $M_{dust}$ is computed from the Herschel photometry using the DL07 models, and $M_{gas}$ is the molecular gas mass computed from $L^{\prime}_{CO}$ and a metallicity-dependent conversion factor \xco\ \citep{genzel12}.   As a comparison, the open squares and the dashed line are from the study of \citet{leroy11} of local galaxies.  Fitting the $z>2$ galaxies while keeping the slope fixed (solid orange line and shaded region) shows a mean increase of $\delta_{GDR}$ by a factor of 1.7.  \label{gdr}}
\end{figure}

In the local Universe, the typical gas-to-dust ratio, $\delta_{GDR}$, for star-forming galaxies with solar metallicities is of order 100 \citep{draine07}.  This ratio has been shown to increase in low metallicity environments \citep[e.g.][]{hunt05,engelbracht08,leroy11}, as predicted by dust formation models \citep[e.g.][]{edmunds01}.  The value of $\delta_{GDR}$ is also expected to vary in high density environments, such as the nuclear regions of starbursts, but observations suggest only a mild decrease of no more than a factor of 2 \citep{wilson08,clements10,santini10}. 

In the previous section, we inferred that the high dust temperatures of the $z>2$ lensed galaxies were due to their low dust masses and metallicities.  To understand whether the dust content of these galaxies is abnormally low for their other properties we plot in Figure \ref{gdr} the measured value of $\delta_{GDR}$ as a function of metallicity for the galaxies within 0.5 dex of the star formation main sequence. The gas-to-dust ratio is measured as $\delta_{GDR}=M_{gas}/M_{dust}$, where $M_{gas}$ is the molecular gas mass from Table \ref{COtable} calculated from the CO luminosity using the metallicity-dependent conversion factor of \citet{genzel12}, and $M_{dust}$ is derived from the Herschel photometry using the DL07 models (see \S \ref{FIRproducts}).   Figure \ref{gdr} reveals that the high redshift galaxies have a gas-to-dust ratio that scales inversely with metallicity like in the local Universe.  However, fitting the $\delta_{GDR}-Z$ relation of the $z>2$ galaxies while keeping the slope fixed to the $z=0$ reference relation reveals an increase in the mean $\delta_{GDR}$ at fixed metallicity by a factor of 1.7.    

Since measuring CO in normal star-forming galaxies at high redshifts is challenging, there has been significant interest recently in using dust masses as a proxy for the total molecular gas contents \citep[e.g.][]{magdis11,magdis12a,scoville12}.  The method consists in measuring $M_{dust}$ using far-infrared and/or sub-mm photometry, and then applying the estimated gas-to-dust ratio ($\delta_{GDR}(Z)$, see \S \ref{MH2section}) to arrive at the gas mass.   The systematic offset between the $z>2$ normal star-forming galaxies and the reference relation of $\delta_{GDR}(Z)$ therefore needs to be taken into consideration.  There are at least three possible explanations:

 \paragraph{Are dust properties different at $z>2$?}  As detailed in \S \ref{FIRproducts}, dust masses for the lensed galaxies are computed using the model of DL07.  In this model, dust is considered to be a combination of carbonaceous and amorphous silicate grains with a specific size distribution set to reproduce the Milky Way extinction curve.  The model also assumes that a large fraction of the dust is found in the diffuse ISM, the rest in photodissociation regions, with the two components exposed to different radiation fields, the parameters of which are set to reproduce Milky Way conditions or left as free parameters.   As argued by \citet{bolatto13}, not only are dust grain properties poorly understood even in the local universe, the lack of a clear understanding of how dust is exactly produced and destroyed leads to significant uncertainties in applying the DL07 model to the denser, hotter interstellar medium of high redshift galaxies.  

\paragraph{Is \xco\ for the lensed galaxies smaller than expected?} If we instead assume that the dust masses are accurately measured in the lensed galaxies, the offset in $\delta_{GDR}(Z)$ compared to $z=0$ could be explained by a conversion factor \xco\ smaller on average by a factor of $\sim1.7$.   The lensed galaxies are located on the main-sequence, which is interpreted as being the locus of galaxies where star formation takes place in virialized GMCs rather than in a centrally-concentrated, dense starburst mode.  Under such conditions, \xco\ should scale as $\rho^{0.5}/T$  \citep{tacconi08,bolatto13}.  Assuming that gas and dust are thermalized, the large measured values of $T_{dust}$ may indicate a reduction of \xco, unless the gas density $\rho$ also increases in proportion.  The two $z>2$ lensed galaxies with normal temperatures of $\sim35$K (J0901 and the Eyelash, see Fig. \ref{tdustfig}) also have gas-to-dust ratios that are elevated compared to the $z=0$ relation, arguing that the ISM in the rest of the sample is not only warm but also denser such that \xco\ is not affected by the temperature variations. 

\paragraph{Is the gas-to-dust ratio really higher at $z>2$?} As a last step, we can assume that none of these concerns apply and that both $M_{gas}$ and $M_{dust}$ are accurately measured in the lensed galaxies, leaving us with the conclusion that the gas-to-dust ratio is higher at fixed metallicity than in the local universe.   This could happen for example if a smaller fraction of the metals are locked up in dust grains under the specific conditions prevailing in the ISM of the high redshift galaxies, or if the far-infrared emission used to compute $M_{dust}$ and the rest-frame optical line emission used to compute the metallicity are not emitted from the same physical regions.    

However, irrespectively of which one of these possible explanations is valid, we can conclude that when measuring dust and gas masses though standard techniques, an offset of 0.23 dex is obtained between the measured gas-to-dust ratio and the standard $z=0$ prescription of \citet{leroy11}.   Care must therefore be taken in applying the gas-to-dust ratio method to estimate molecular gas masses of high redshift galaxies, as small but systematic differences with CO-based measurements may otherwise occur.

\subsection{Star formation efficiencies and gas fractions in $z>$2 normal star-forming galaxies}
\label{evolution}

\begin{figure*}
\epsscale{0.9}
\plotone{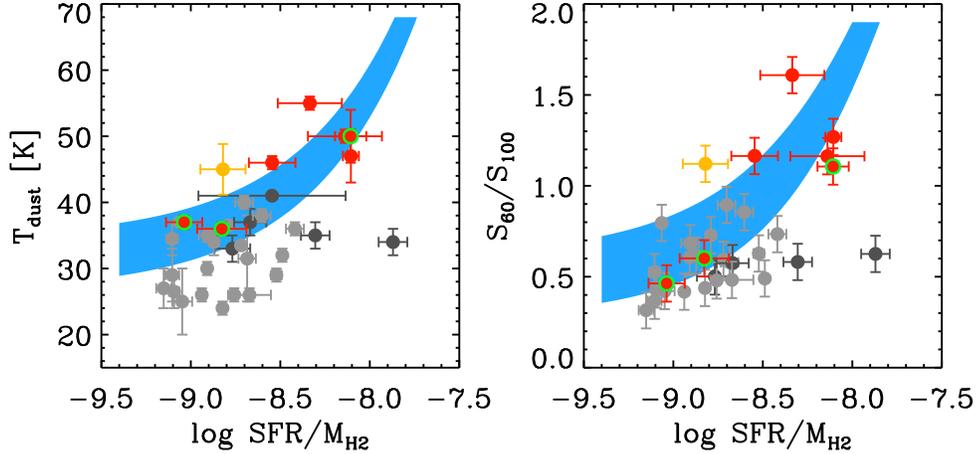}
\caption{Dust temperature (left) and $S_{60}/S_{100}$ ratio (right) as a function of star formation efficiency ($\equiv SFR/M_{H2}$).  The lensed galaxies are shown in colored symbols (red: $z>2$, orange: $z<2$), and the reference sample in gray (dark:  $z>2$, light: $z<2$). The $z>2$ lensed galaxies with $\sim$solar metallicities ($12+\log$O/H$>8.6$) are marked with open green circles.   The blue band is the $T_{dust}-\Delta \log(sSFR)$ relation of Fig. \ref{tdustfig} converted to a  $T_{dust}-$SFE relation using the relation of \citet{magdis12a} between main sequence offset and star formation efficiency, SFE$\propto \Delta \log(sSFR)^{1.34\pm0.13}$ \citep[see also][]{saintonge12,sargent13}.} \label{tdustsfe}
\end{figure*}

In \S \ref{hightdust} above, we suggested that the high $T_{dust}$ of the lensed galaxies is caused by their low dust contents.  To further illustrate this, we repeat the analysis done in Figure \ref{tdustfig}, but this time normalizing $L_{IR}$ by the dust or the gas mass.  We first show in Figure \ref{tdustsfe} how both $T_{dust}$ and the $S_{60}/S_{100}$ ratio depend on the star formation efficiency (SFE$\equiv SFR/M_{H2}$).  A correlation between these quantities is expected as $T_{dust}$ increases as a function of $\Delta \log(sSFR)$ and $L_{IR}$ (Fig. \ref{tdustfig}), and there is also a positive correlation between SFE and $\Delta \log(sSFR)$ \citep{saintonge12,magdis12a,sargent13}.   Figure \ref{tdustsfe} reveals that when the gas mass in each galaxy is taken into account, most of the offset between the $z>2$ lensed galaxies and the comparison samples disappears, although the $z>2$ normal star-forming galaxies still appear to have dust temperatures higher by $\sim5-10$K at fixed SFE compared to the reference sample (gray symbols).  The offset is strongest for those galaxies with the lowest metallicities.  While metallicity may account for part of the remaining offset in $T_{dust}$ at fixed SFE, there could also be a contribution from differential lensing (see \S \ref{difflens}), which if present at all, could manifest itself in an overestimation of the dust temperatures.

\begin{figure*}
\epsscale{0.9}
\plotone{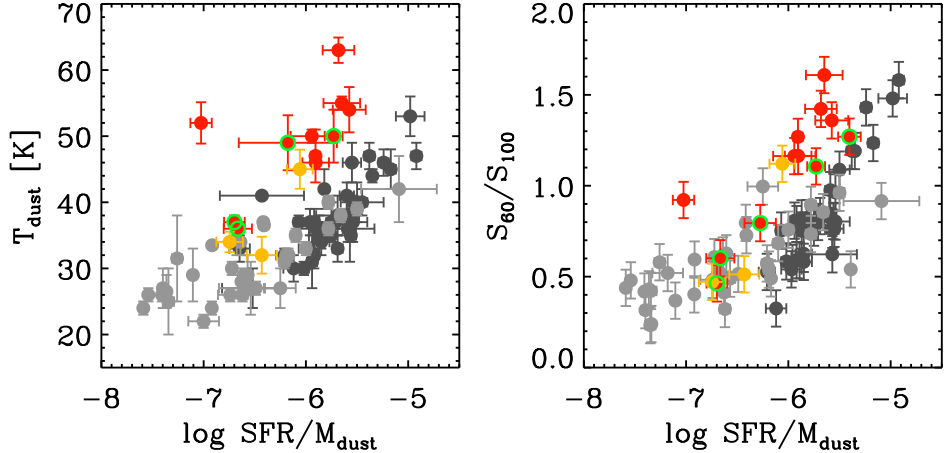}
\caption{Same as Figure \ref{tdustsfe}, both with the SFR on the $x-$axis normalized by $M_{dust}$ rather than $M_{H2}$.   The location of the subset of the most  metal-rich lensed galaxies in the figure, closer the locus of the reference samples, suggests that the remaining offset in $T_{dust}$ and $S_{60}/S_{100}$ ratio is a consequence of the low metallicities (see e.g. Fig. \ref{gdr}).}  \label{tdustmdust}
\end{figure*}

In Figure \ref{tdustmdust} we then show a similar relation, but using the dust rather than the gas mass as the normalization factor, allowing us to include a larger number of galaxies for which dust masses are measured, but gas masses are not available.   In this case again, some of the lensed galaxies appear to be systematically offset to higher $T_{dust}$ at fixed value of SFR$/M_{dust}$.  The $z>2$ lensed galaxies having typically low metallicities,  each dust grain will be exposed to more UV photons for a fixed SFR per unit gas mass, and therefore will have a hotter temperature.  There is evidence for this in the fact that the galaxies with $\sim$solar metallicities follow more closely the relation traced by the reference sample, in particular when using the $S_{60}/S_{100}$ ratio (see Fig. \ref{tdustmdust}).  The combination of Figures \ref{gdr}-\ref{tdustmdust} thus argues that in these galaxies, the balance between $M_{dust}$, $M_{gas}$ and $Z$ follows expected relations (modulo the possible change of $\delta_{GDR}$ with redshift), and that the high dust temperatures are the result of the high SFRs per unit gas and dust mass, i.e. the high star formation efficiencies. 

This assertion that the $z>2$ lensed galaxies have high star formation efficiencies (or put differently, short molecular gas depletion timescales, \tdep) needs to be investigated in more detail.  In the local Universe, normal star-forming galaxies have \tdep$\sim1.5$ Gyr \citep{leroy08,bigiel11,COLDGASS2,saintonge12}.  In the PHIBSS survey, \citet{tacconi13} show that at $z=1-1.5$ the depletion time  for main-sequence galaxies is reduced to $\sim$700 Myr (with 0.24 dex scatter in the $\log$\tdep\ distribution, and assuming a Galactic conversion factor $\alpha_{CO}$).  Based on these observations, \citet{tacconi13} infer a redshift-dependence of the form \tdep$=1.5(1+z)^{-1}$. This redshift evolution is only slightly slower than the dependence of $(1+z)^{-1.5}$ expected if \tdep\ is proportional to the dynamical timescale \citep{dave12}. 

\begin{figure*}
\epsscale{1.0}
\plotone{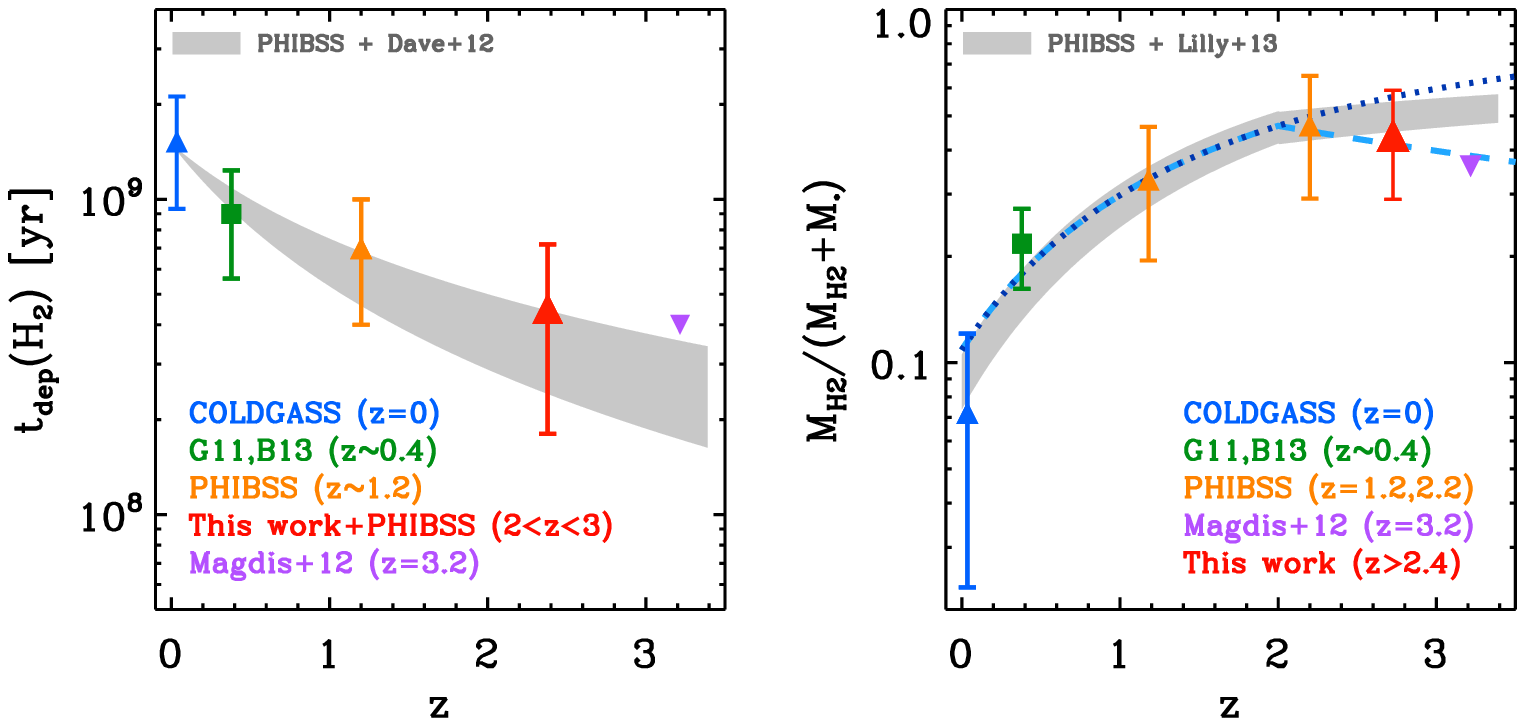}
\caption{Redshift evolution of the mean gas depletion time (left) and gas mass fraction (right) for main-sequence galaxies from our lensed and reference samples for which a CO-based measurement of $M_{H2}$ is available.  Error bars show the standard deviation within each redshift bin.  The different colors represent the following datasets:  {\it blue}: representative $z=0$ sample from COLD GASS, {\it green}: galaxies from \citet{geach11} and \citet{bauermeister13} at $z\sim0.4$, {\it orange}: the incompleteness-corrected mean values from \citet{tacconi13}.  The red points show the contribution from this study and include all the galaxies in the specified redshift intervals from the lensed and comparison samples as well as from PHIBSS, and corrects for sample incompleteness.  On the left plot, the gray shaded region shows the expected trend between between \tdep\ and $z$ described by eq. \ref{tdepeq}, for $\alpha=[-1.0,-1.5]$.  On the right panel, the gray shaded region is the expected redshift dependence of $f_{gas}$ derived from equations \ref{fgaseq}-\ref{ssfreq}, assuming that $\alpha=-1.0$ \citep{tacconi13} and sSFR follows eq. \ref{ssfreq} \citep{lilly13}.   Alternative relations for $f_{gas}(z)$ are obtained by assuming that sSFR$\propto (1+z)^{2.8}$ at all redshifts (dotted dark blue line), or else reaches a plateau at $z=2$ (dashed light blue line).
\label{fgas}}
\end{figure*}

The lensed and reference galaxies at $z>2$ can be used to track the behavior of the depletion time at even higher redshifts.  In Figure \ref{fgas}, left panel, the redshift evolution of \tdep\ is shown.  The gray band shows the expected trend based on the empirical $(1+z)^{-1}$ dependence (upper envelope), and the analytical expectation of a $(1+z)^{-1.5}$ dependence (lower envelope). The mean values in four different redshift intervals are shown.  For the highest redshift interval ($2<z<3$), we combine galaxies from the lensed and comparison samples as well as from PHIBSS.   While the mean depletion time is $\sim$700 Myr at $z\sim1.2$, it decreases further to $\sim450$ Myr at $z\sim2.2$, consistently with the predicted redshift evolution.  We caution however that the sample at $z>2$ is by no means complete or representative of all normal star forming galaxies, but in terms of stellar masses and main sequence offset it is at least comparable to the $z\sim1-1.5$ PHIBSS sample and the $z=0$ reference sample from COLD GASS.  This measurement at $z>2$ suggests that the redshift-dependence of the molecular gas depletion time for main sequence galaxies, as predicted analytically \citep[e.g.][]{dave11,dave12} and recently observed up to $z\sim1.5$ \citep{tacconi13}, extends to $z\sim3$.

Using the lensed galaxies and the various comparison samples, we can also trace the redshift evolution of the molecular gas mass fraction,  
\begin{equation}
f_{gas}=\frac{M_{H2}}{M_{H2}+M_{\ast}}.
\label{fgaseq}
\end{equation}
In what follows, we restrict the samples only to galaxies found within 0.5 dex of the star formation main sequence at the appropriate redshift, i.e. we study the gas contents of normal, star-forming discs as a function of redshift.  Several studies have now reported a rapid increase of \fgas\ with redshift \citep{tacconi10,daddi10,geach11,magdis12a}.  The most robust analysis so far was performed by \citet{tacconi13}.  In that work, the PHIBSS data at $z\sim1$ and $z\sim2$ were corrected for incompleteness and compared to a matched local control sample extracted from the COLD GASS catalog, revealing an increase of \fgas\ from 8\% at $z=0$ to 33\% at $z\sim1$ and 47\% at $z\sim2$.  These three secure measurements are reproduced in Figure \ref{fgas} (right panel).  There are very few galaxies at $z\sim0.5$ with published CO measurements, but in Fig. \ref{fgas} the few systems found in \citet{geach11} and \citet{bauermeister13} are compiled (CO measurements for several galaxies with $0.6<z<1.0$ are published also in \citet{combes12}, but we do not include them here as they are above-main sequence objects).  

As the PHIBSS sample extends to $z=2.4$, we combine all the main-sequence galaxies in our lensed and comparison samples above that redshift to derive a mean gas fraction of $40\pm15$\% at $<z>=2.8$.   We then apply the methodology of \citet{tacconi13} to correct for sample incompleteness.  As the sample of lensed galaxies with $z>2.4$ is richer in on- and below-main sequence galaxies, accounting for this bias rises the mean gas fraction to 44\%.   Our observations therefore suggest that the trend for increasing gas fraction with redshift does not extend beyond $z\sim2$, and may even be reversing. 

Can this flattening of the relation between gas fractions and redshift at $z>2$ be expected under the equilibrium model?   The definition of the gas fraction (eq. \ref{fgaseq}) can be re-expressed as:
\begin{equation}
f_{gas}=\frac{1}{1+ (t_{dep}~{\rm sSFR})^{-1}},
\label{fgaseq}
\end{equation}
 and the best predictions available for the redshift evolution of $t_{dep}$ and sSFR used to compute the expected behavior of $f_{gas}(z)$.  As explained in \S \ref{evolution}, it is estimated that
\begin{equation}
t_{dep}(z)=1.5 (1+z)^{\alpha}~~ [{\rm Gyr}], 
\label{tdepeq}
\end{equation}
with $\alpha$ measured to be -1.0 by \citet{tacconi13} and predicted to be -1.5 in the analytic model of \citet{dave12}.  The relation is normalized to the typical depletion time of 1.5 Gyr observed in local galaxies \citep{leroy08,bigiel11,COLDGASS2,saintonge12}.   Based on studies of the slope and redshift evolution of the star formation main sequence, the typical sSFR (in Gyr$^{-1}$) of a star-forming galaxy of mass $M_{\ast}$ at redshift $z$ is
\begin{equation}
{\rm sSFR}(M_{\ast},z)=\Bigg \{ \begin{array}{ll}
0.07 \left (  \frac{M_{\ast}}{10^{10.5}M_{\odot}} \right)^{-0.1} (1+z)^{3}  & \mbox{if $z<2$}\\
0.30 \left (  \frac{M_{\ast}}{10^{10.5}M_{\odot}} \right)^{-0.1} (1+z)^{5/3}  & \mbox{if $z>2$}.
\end{array}
\label{ssfreq}
\end{equation}
The above equation is presentd by \citet{lilly13} based on results from a number of recent high-redshift imaging surveys \citep{noeske07,elbaz07,daddi07,pannella09,stark12}.  The expected redshift evolution of the gas fraction for galaxies of a given stellar mass can then be obtained by combining equations \ref{fgaseq}-\ref{ssfreq}.  For galaxies in the mass range $10^{10}-5\times10^{11}$ and for $\alpha=-1.0$ in eq. \ref{tdepeq}, the expected trend is shown in Figure \ref{fgas} as the light gray band.  At $z>2$, \fgas\ flattens out because of the shallower evolution of sSFR with redshift mostly canceling out  the $(1+z)^{-1}$ term from the \tdep\ relation (eq. \ref{tdepeq}).  This model predicts a very modest evolution of the mean gas fraction from 47\% at $z=2.2$ to 49\% at $z=2.8$, consistently with our measurement. 

A different behavior for $f_{gas}(z)$ can be obtained by modifying the redshift-dependence of either $t_{dep}$ or sSFR (eq. \ref{fgaseq}).  For example, if a value of $\alpha=-1.5$ is used to set the evolution of $t_{dep}$, the mean gas fraction expected at any redshift is lower than in the first model.  However, since there is no observational evidence for a redshift evolution of $t_{dep}$ different from what has been assumed so far, we turn instead our attention to sSFR.   

Equation \ref{ssfreq} assumes that at fixed stellar mass, the specific star formation rate increases as $(1+z)^3$ out to $z=2$, and that this evolution then slows down to $(1+z)^{5/3}$.  If instead we took sSFR to keep increasing steadily with redshift, for example as sSFR$\propto$\mstar$^{-0.3}(1+z)^{2.8}$ \citep{tacconi13}, the predicted mean gas fraction at $z=2.8$ would become 57\%, in tension with our measured value of $40\pm15\%$.  

Alternatively, it has been reported that the characteristic sSFR of main-sequence galaxies reaches a plateau at $z\sim2$ \citep{stark09,gonzalez10,rodighiero10,weinmann11,reddy12}.  If we assume that sSFR increases up to $z=2$ according to eq. \ref{ssfreq} and then remains constant at $z>2$, the mean gas fraction is predicted to drop to 41\% at $z=2.8$.    While this behavior is consistent with the observations presented here, the existence of such a plateau in sSFR at $z\sim2-7$ has recently been challenged. After accounting for contamination by strong line emission, different authors suggest an increase of the mean sSFR by a factor of 2-5 between $z=2$ and $z=7$ \citep[e.g.][]{bouwens12,stark12,gonzalez12}.  This decrease in \fgas\ at $z>2$ should therefore be considered as a lower limit, with the relation based on eq. \ref{ssfreq} providing the most realistic prediction of the redshift evolution of the gas fraction, as supported by our observations. 

\subsubsection{Missing cold gas at $z>2$? \label{missinggas}}
Another factor that could contribute to the redshift evolution of the gas fraction is the relative contribution of atomic hydrogen to the cold gas budget of the galaxies (M$_{HI}+$M$_{H2}$) at the different epochs.  For star forming galaxies with $10.5<\log M_{\ast}/M_{\odot}<11.5$ at $z=1-2$ such as those in the PHIBSS sample, it is generally assumed that $M_{H2}>>M_{HI}$, at least within the parts of the discs where star formation is actively taking place. This assumption is based on the observed high surface densities, above the characteristic threshold for the atomic-to-molecular conversion.  On the other hand, for nearby galaxies with $\log M_{\ast}/M_{\odot}>10.0$, it is observed that $M_{HI}\sim3M_{H2}$, albeit with large galaxy-galaxy variations \citep{COLDGASS1}.  An important fraction of this atomic gas is located in the outer regions of galaxies, outside of the actively star-forming disks, but even within the central regions a significant fraction of the cold gas is in atomic form as the HI-to-H$_2$ transition typically occurs at a radius of $\sim0.4r_{25}$ \citep{bigiel08,leroy08}.  We can therefore estimate that on average within $r_{25}$, $M_{HI}\sim M_{H2}$, raising the cold gas mass fraction from 8\% when only the molecular phase was considered to $\sim15\%$.  We cannot directly quantify the fraction of atomic gas in our high redshift lensed galaxies, but it could be significant given the low stellar masses.  Not only do we know locally that the HI fractions increase as stellar masses decrease \citep{GASS1,GASS6,cortese11,huang12}, it can also be expected that the gas surface densities are lower in the $z>2$,  $\log M_{\ast}/M_{\odot}\sim10$ lensed galaxies than in others at the high-mass end of the star-formation main sequence.  Therefore, an additional interpretation for the low mean value of $f_{H2}$ at $<z>=2.8$ is that a higher fraction of the cold gas mass is in neutral form than in the sample at $<z>=1.2$ and 2.2.  However, were we missing a significant cold gas component, we would have measured lower than expected gas-to-dust ratios, as any dust should be mixed with both the molecular and atomic phases of the ISM.  Figure \ref{gdr} shows the opposite behavior, suggesting that the lower gas fraction measured at $z=2.8$ is not the result of neglecting the atomic component.

%===================================
\section{Summary of observational results and conclusions}
\label{summary}

The main observational results presented in this paper can be summarized as:

\begin{enumerate}

\item We measure a mean molecular gas depletion time of 450 Myr at $<z>=2.5$.  This decrease of $t_{dep}$ by a factor of 5 since $z=0$ and 1.5 since $z=1$ is  consistent with the expected scaling of $t_{dep}=1.5(1+z)^\alpha$ ($\alpha=-1.0,-1.5$) if the depletion time is linked to the dynamical timescale.  Our results validate up to $z\sim3$ and down to \mstar$\sim10^{10}$\msun\ the calibration of this relation established by \citet{tacconi13} out to $z=1.5$ using data from the PHIBSS and COLD GASS surveys and a metallicity-dependent conversion factor.  

\item The mean gas fraction measured at $<z>=2.8$ is $40\pm15\%$ (44\% after accounting for sample incompleteness), suggesting that the trend of increasing gas fraction with redshift ($<f_{gas}>=[0.08, 0.33, 0.47]$ at $z=[0, 1, 2]$) does not extend beyond $z\sim2$. This observation is consistent with recent studies suggesting that the evolution of the mean sSFR of main-sequence galaxies slows down or even reaches a plateau beyond $z=2$. 

\item The lensed galaxies at $z>2$ exhibit high dust temperatures, with values of $\sim50$K such as found typically only in galaxies with extreme IR luminosities.  The high values of $T_{dust}$ are a consequence of the fact that the lensed galaxies have low metallicities and short gas depletion times, as expected for their high redshifts and low stellar masses.  

\item Using CO line luminosities, dust masses, and direct metallicity measurements, the conversion factor \xco\ for the lensed galaxies is estimated using the ``inverse Kennicutt-Schmidt" and the gas-to-dust ratio methods, and in both cases agrees with a scaling of \xco$\propto Z^{-1.3}$ as parametrized by \citet{genzel12}.  

\item The gas-to-dust ratios in the $z>2$ lensed galaxies exhibit the same metallicity dependence as observed in the local Universe, but with a systematic offset (Fig. \ref{gdr}).  At fixed metallicity, we observe $\delta_{GDR}$ to be larger by a factor of 1.7 at $z>2$, suggesting that applying the local calibration of the $\delta_{GDR}$-metallicity relation to infer the molecular gas mass of high redshift galaxies may lead to systematic differences with CO-based estimates.
\end{enumerate}

Most of these results could be explained by assuming that our lensed galaxies sample, mostly selected in SDSS imaging based on high luminosities at rest-UV wavelengths, is heavily biased toward dust-poor objects.  This could serve to explain for example the high dust temperatures, the low dust and gas masses, and the short depletion times.  However, we have conducted extensive tests (see \S \ref{samplebias}), and could not find evidence of any such bias.  Instead, the sample appears to be representative of the overall population of main sequence galaxies at $<z>=2.5$ with $M_{\odot}\sim10^{10}$\msun.  While it should be kept in mind that sample biases may be present, in the absence of evidence to the contrary we have proceeded with the analysis under the assumption that our results apply to the bulk of the high redshift galaxy population.   

The combined Herschel and IRAM observations presented in this paper then suggest that main sequence galaxies with modest stellar masses ($9.5<\log M_{\ast}/M_{\odot}<10.5$) at $<z>=2.5$ have high star formation efficiencies and a molecular gas mass fraction no larger than measured at $z=1-2$, consistently with a simple model where the redshift evolution of the characteristic sSFR of main-sequence galaxies can be explained by a slowly varying gas depletion time and the measured gas fractions.    The short depletion times and the possible redshift evolution of the gas-to-dust ratio imply that these high redshift galaxies have less dust at fixed SFR,  producing the high $T_{dust}$ values we measure as each dust grain is exposed to more radiation (Fig. \ref{tdustfig}).    

Before concluding, we wish to point out that the equilibrium model on which this analysis relies requires that the gas depletion timescale is shorter than the accretion timescale.  This balance is reached at a redshift $z_{eq}$.  At $z>z_{eq}$, the star formation process cannot keep up with the accretion of new gas, and this is thus the epoch where the gas reservoirs of the galaxies are filling up with gas fractions expected to be high.  The exact value of $z_{eq}$ is however still debated.  For example, \citet{dave12} found that $z_{eq}\sim2$ in the absence of outflows, but that even modest outflows raised the threshold to $z_{eq}\sim7$, the latter value being in agreement with the analytical work of \citet{bouche10}.  On the other hand, \citet{krumholz12} suggest that $z_{eq}\sim2$, and \citet{papovich11} estimate that $z_{eq}\sim4$, using a sample of Lyman break galaxies and an indirect measurement of gas fractions.  

In this study, we have used direct measurements of CO line fluxes and the power of gravitational lensing to push to higher redshifts the study of the redshift evolution of gas fractions.  After correcting for sample incompleteness, \citet{tacconi13} measure a mean gas fraction of 47\% at $z\sim2.2$.  In this study, we directly measure a mean gas fraction of 40\% at $<z>=2.8$, which is then corrected upward to 44\% after accounting for sample incompleteness.  This observation suggests that $f_{gas}$ does not increase significantly between $z=2$ and 3.  A similar conclusion was reached by \citet{magdis12b} based on CO observations of two $z\sim3$ LBGs.  Since gas fractions are expected to be high during the gas accretion phase, these results may indicate that we have not yet reached observationally the regime where galaxies are out of equilibrium, and therefore that $z_{eq}>3$.   The improved determination of stellar mass and star formation rates in large samples of galaxies beyond $z=2$, as well as the direct measurement of gas masses in normal star-forming galaxies at $z>3$ with ALMA and NOEMA will be essential to refine this picture.

%===================================
\acknowledgements
We thank C. Feruglio for assistance with the reduction of the IRAM data and J. Richard for sharing the magnification factor for J1133 ahead of publication. We also thank T. Jones and C. Sharon for useful input, as well as the anonymous referee for constructive comments.

PACS has been developed by a consortium of institutes led by MPE (Germany) and including UVIE (Austria); KU Leuven, CSL, IMEC (Belgium); CEA, LAM (France); MPIA (Germany); INAF-IFSI/OAA/OAP/OAT, LENS, SISSA (Italy); IAC (Spain). This development has been supported by the funding agencies BMVIT (Austria), ESA-PRODEX (Belgium), CEA/CNES (France), DLR (Germany), ASI/INAF (Italy), and CICYT/MCYT (Spain).  SPIRE has been developed by a consortium of institutes led by Cardiff University (UK) and including Univ. Lethbridge (Canada); NAOC (China); CEA, LAM (France); IFSI, Univ. Padua (Italy); IAC (Spain); Stockholm Observatory (Sweden); Imperial College London, RAL, UCL-MSSL, UKATC, Univ. Sussex (UK); and Caltech, JPL, NHSC, Univ. Colorado (USA). This development has been supported by national funding agencies: CSA (Canada); NAOC (China); CEA, CNES, CNRS (France); ASI (Italy); MCINN (Spain); SNSB (Sweden); STFC (UK); and NASA (USA).

%===================================
\bibliographystyle{apj}

\begin{thebibliography}{196}
\expandafter\ifx\csname natexlab\endcsname\relax\def\natexlab#1{#1}\fi

\bibitem[{{Allam} {et~al.}(2007){Allam}, {Tucker}, {Lin}, {Diehl}, {Annis},
  {Buckley-Geer}, \& {Frieman}}]{allam07}
{Allam}, S.~S., {Tucker}, D.~L., {Lin}, H., {Diehl}, H.~T., {Annis}, J.,
  {Buckley-Geer}, E.~J., \& {Frieman}, J.~A. 2007, \apjl, 662, L51

\bibitem[{{Asplund} {et~al.}(2009){Asplund}, {Grevesse}, {Sauval}, \&
  {Scott}}]{asplund09}
{Asplund}, M., {Grevesse}, N., {Sauval}, A.~J., \& {Scott}, P. 2009, \araa, 47,
  481

\bibitem[{{Baker} {et~al.}(2001){Baker}, {Lutz}, {Genzel}, {Tacconi}, \&
  {Lehnert}}]{baker01}
{Baker}, A.~J., {Lutz}, D., {Genzel}, R., {Tacconi}, L.~J., \& {Lehnert}, M.~D.
  2001, \aap, 372, L37

\bibitem[{{Baker} {et~al.}(2004){Baker}, {Tacconi}, {Genzel}, {Lehnert}, \&
  {Lutz}}]{baker04}
{Baker}, A.~J., {Tacconi}, L.~J., {Genzel}, R., {Lehnert}, M.~D., \& {Lutz}, D.
  2004, \apj, 604, 125

\bibitem[{{Bauermeister} {et~al.}(2013){Bauermeister}, {Blitz}, {Bolatto},
  {Bureau}, {Teuben}, {Wong}, \& {Wright}}]{bauermeister13}
{Bauermeister}, A., {Blitz}, L., {Bolatto}, A., {Bureau}, M., {Teuben}, P.,
  {Wong}, T., \& {Wright}, M. 2013, \apj, 763, 64

\bibitem[{{Belokurov} {et~al.}(2007){Belokurov}, {Evans}, {Moiseev}, {King},
  {Hewett}, {Pettini}, {Wyrzykowski}, {McMahon}, {Smith}, {Gilmore}, {Sanchez},
  {Udalski}, {Koposov}, {Zucker}, \& {Walcher}}]{belokurov07}
{Belokurov}, V., {Evans}, N.~W., {Moiseev}, A., {King}, L.~J., {Hewett}, P.~C.,
  {Pettini}, M., {Wyrzykowski}, L., {McMahon}, R.~G., {Smith}, M.~C.,
  {Gilmore}, G., {Sanchez}, S.~F., {Udalski}, A., {Koposov}, S., {Zucker},
  D.~B., \& {Walcher}, C.~J. 2007, \apjl, 671, L9

\bibitem[{{Berta} {et~al.}(2011){Berta}, {Magnelli}, {Nordon}, {Lutz}, {Wuyts},
  {Altieri}, {Andreani}, {Aussel}, \& et~al.}]{berta11}
{Berta}, S., {Magnelli}, B., {Nordon}, R., {Lutz}, D., {Wuyts}, S., {Altieri},
  B., {Andreani}, P., {Aussel}, H., \& et~al. 2011, \aap, 532, A49

\bibitem[{{Bertin} \& {Arnouts}(1996)}]{bertin96}
{Bertin}, E., \& {Arnouts}, S. 1996, \aaps, 117, 393

\bibitem[{{Bian} {et~al.}(2010){Bian}, {Fan}, {Bechtold}, {McGreer}, {Just},
  {Sand}, {Green}, {Thompson}, {Peng}, {Seifert}, {Ageorges}, {Juette},
  {Knierim}, \& {Buschkamp}}]{bian10}
{Bian}, F., {Fan}, X., {Bechtold}, J., {McGreer}, I.~D., {Just}, D.~W., {Sand},
  D.~J., {Green}, R.~F., {Thompson}, D., {Peng}, C.~Y., {Seifert}, W.,
  {Ageorges}, N., {Juette}, M., {Knierim}, V., \& {Buschkamp}, P. 2010, \apj,
  725, 1877

\bibitem[{{Bigiel} {et~al.}(2008){Bigiel}, {Leroy}, {Walter}, {Brinks}, {de
  Blok}, {Madore}, \& {Thornley}}]{bigiel08}
{Bigiel}, F., {Leroy}, A., {Walter}, F., {Brinks}, E., {de Blok}, W.~J.~G.,
  {Madore}, B., \& {Thornley}, M.~D. 2008, \aj, 136, 2846

\bibitem[{{Bigiel} {et~al.}(2011){Bigiel}, {Leroy}, {Walter}, {Brinks}, {de
  Blok}, {Kramer}, {Rix}, {Schruba}, {Schuster}, {Usero}, \&
  {Wiesemeyer}}]{bigiel11}
{Bigiel}, F., {Leroy}, A.~K., {Walter}, F., {Brinks}, E., {de Blok}, W.~J.~G.,
  {Kramer}, C., {Rix}, H.~W., {Schruba}, A., {Schuster}, K.-F., {Usero}, A., \&
  {Wiesemeyer}, H.~W. 2011, \apjl, 730, L13+

\bibitem[{{Blanc} {et~al.}(2009){Blanc}, {Heiderman}, {Gebhardt}, {Evans}, \&
  {Adams}}]{blanc09}
{Blanc}, G.~A., {Heiderman}, A., {Gebhardt}, K., {Evans}, N.~J., \& {Adams}, J.
  2009, \apj, 704, 842

\bibitem[{{Blandford} \& {Narayan}(1992)}]{blandford92}
{Blandford}, R.~D., \& {Narayan}, R. 1992, \araa, 30, 311

\bibitem[{{Blitz} {et~al.}(2007){Blitz}, {Fukui}, {Kawamura}, {Leroy},
  {Mizuno}, \& {Rosolowsky}}]{blitz07}
{Blitz}, L., {Fukui}, Y., {Kawamura}, A., {Leroy}, A., {Mizuno}, N., \&
  {Rosolowsky}, E. 2007, Protostars and Planets V, 81

\bibitem[{{Bolatto} {et~al.}(2011){Bolatto}, {Leroy}, {Jameson}, {Ostriker},
  x~{Gordon}, {Lawton}, {Stanimirovi{\'c}}, {Israel}, {Madden}, {Hony},
  {Sandstrom}, {Bot}, {Rubio}, {Winkler}, {Roman-Duval}, {van Loon},
  {Oliveira}, \& {Indebetouw}}]{bolatto11}
{Bolatto}, A.~D., {Leroy}, A.~K., {Jameson}, K., {Ostriker}, E., x~{Gordon},
  K., {Lawton}, B., {Stanimirovi{\'c}}, S., {Israel}, F.~P., {Madden}, S.~C.,
  {Hony}, S., {Sandstrom}, K.~M., {Bot}, C., {Rubio}, M., {Winkler}, P.~F.,
  {Roman-Duval}, J., {van Loon}, J.~T., {Oliveira}, J.~M., \& {Indebetouw}, R.
  2011, \apj, 741, 12

\bibitem[{{Bolatto} {et~al.}(2013){Bolatto}, {Wolfire}, \& {Leroy}}]{bolatto13}
{Bolatto}, A.~D., {Wolfire}, M., \& {Leroy}, A.~K. 2013, ArXiv e-prints

\bibitem[{{Boselli} {et~al.}(2002){Boselli}, {Lequeux}, \&
  {Gavazzi}}]{boselli02}
{Boselli}, A., {Lequeux}, J., \& {Gavazzi}, G. 2002, \aap, 384, 33

\bibitem[{{Bothwell} {et~al.}(2013){Bothwell}, {Smail}, {Chapman}, {Genzel},
  {Ivison}, {Tacconi}, {Alaghband-Zadeh}, {Bertoldi}, {Blain}, {Casey}, {Cox},
  {Greve}, {Lutz}, {Neri}, {Omont}, \& {Swinbank}}]{bothwell13}
{Bothwell}, M.~S., {Smail}, I., {Chapman}, S.~C., {Genzel}, R., {Ivison},
  R.~J., {Tacconi}, L.~J., {Alaghband-Zadeh}, S., {Bertoldi}, F., {Blain},
  A.~W., {Casey}, C.~M., {Cox}, P., {Greve}, T.~R., {Lutz}, D., {Neri}, R.,
  {Omont}, A., \& {Swinbank}, A.~M. 2013, \mnras, 563

\bibitem[{{Bouch{\'e}} {et~al.}(2010){Bouch{\'e}}, {Dekel}, {Genzel}, {Genel},
  {Cresci}, {F{\"o}rster Schreiber}, {Shapiro}, {Davies}, \&
  {Tacconi}}]{bouche10}
{Bouch{\'e}}, N., {Dekel}, A., {Genzel}, R., {Genel}, S., {Cresci}, G.,
  {F{\"o}rster Schreiber}, N.~M., {Shapiro}, K.~L., {Davies}, R.~I., \&
  {Tacconi}, L. 2010, \apj, 718, 1001

\bibitem[{{Bouwens} {et~al.}(2012){Bouwens}, {Illingworth}, {Oesch}, {Franx},
  {Labb{\'e}}, {Trenti}, {van Dokkum}, {Carollo}, {Gonz{\'a}lez}, {Smit}, \&
  {Magee}}]{bouwens12}
{Bouwens}, R.~J., {Illingworth}, G.~D., {Oesch}, P.~A., {Franx}, M.,
  {Labb{\'e}}, I., {Trenti}, M., {van Dokkum}, P., {Carollo}, C.~M.,
  {Gonz{\'a}lez}, V., {Smit}, R., \& {Magee}, D. 2012, \apj, 754, 83

\bibitem[{{Buat} {et~al.}(2005){Buat}, {Iglesias-P{\'a}ramo}, {Seibert},
  {Burgarella}, {Charlot}, {Martin}, {Xu}, {Heckman}, \& et~al.}]{buat05}
{Buat}, V., {Iglesias-P{\'a}ramo}, J., {Seibert}, M., {Burgarella}, D.,
  {Charlot}, S., {Martin}, D.~C., {Xu}, C.~K., {Heckman}, T.~M., \& et~al.
  2005, \apjl, 619, L51

\bibitem[{{Calzetti} {et~al.}(2000){Calzetti}, {Armus}, {Bohlin}, {Kinney},
  {Koornneef}, \& {Storchi-Bergmann}}]{calzetti00}
{Calzetti}, D., {Armus}, L., {Bohlin}, R.~C., {Kinney}, A.~L., {Koornneef}, J.,
  \& {Storchi-Bergmann}, T. 2000, \apj, 533, 682

\bibitem[{{Casey} {et~al.}(2009){Casey}, {Chapman}, {Beswick}, {Biggs},
  {Blain}, {Hainline}, {Ivison}, {Muxlow}, \& {Smail}}]{casey09}
{Casey}, C.~M., {Chapman}, S.~C., {Beswick}, R.~J., {Biggs}, A.~D., {Blain},
  A.~W., {Hainline}, L.~J., {Ivison}, R.~J., {Muxlow}, T.~W.~B., \& {Smail}, I.
  2009, \mnras, 399, 121

\bibitem[{{Casey} {et~al.}(2013){Casey}, {Chen}, {Cowie}, {Barger}, {Capak},
  {Ilbert}, {Koss}, {Lee}, {Le Floc'h}, {Sanders}, \& {Williams}}]{casey13}
{Casey}, C.~M., {Chen}, C.-C., {Cowie}, L., {Barger}, A., {Capak}, P.,
  {Ilbert}, O., {Koss}, M., {Lee}, N., {Le Floc'h}, E., {Sanders}, D.~B., \&
  {Williams}, J.~P. 2013, arXiv:1302.2619

\bibitem[{{Catinella} {et~al.}(2012){Catinella}, {Schiminovich}, {Kauffmann},
  {Fabello}, {Hummels}, {Lemonias}, {Moran}, {Wu}, {Cooper}, \& {Wang}}]{GASS6}
{Catinella}, B., {Schiminovich}, D., {Kauffmann}, G., {Fabello}, S., {Hummels},
  C., {Lemonias}, J., {Moran}, S.~M., {Wu}, R., {Cooper}, A., \& {Wang}, J.
  2012, \aap, 544, A65

\bibitem[{{Catinella} {et~al.}(2010){Catinella}, {Schiminovich}, {Kauffmann},
  {Fabello}, {Wang}, \& et~al.}]{GASS1}
{Catinella}, B., {Schiminovich}, D., {Kauffmann}, G., {Fabello}, S., {Wang},
  J., \& et~al. 2010, \mnras, 403, 683

\bibitem[{{Chabrier}(2003)}]{chabrier03}
{Chabrier}, G. 2003, \pasp, 115, 763

\bibitem[{{Chapman} {et~al.}(2005){Chapman}, {Blain}, {Smail}, \&
  {Ivison}}]{chapman05}
{Chapman}, S.~C., {Blain}, A.~W., {Smail}, I., \& {Ivison}, R.~J. 2005, \apj,
  622, 772

\bibitem[{{Chapman} {et~al.}(2003){Chapman}, {Helou}, {Lewis}, \&
  {Dale}}]{chapman03}
{Chapman}, S.~C., {Helou}, G., {Lewis}, G.~F., \& {Dale}, D.~A. 2003, \apj,
  588, 186

\bibitem[{{Chapman} {et~al.}(2004){Chapman}, {Smail}, {Blain}, \&
  {Ivison}}]{chapman04}
{Chapman}, S.~C., {Smail}, I., {Blain}, A.~W., \& {Ivison}, R.~J. 2004, \apj,
  614, 671

\bibitem[{{Chary} \& {Elbaz}(2001)}]{ce01}
{Chary}, R., \& {Elbaz}, D. 2001, \apj, 556, 562

\bibitem[{{Clements} {et~al.}(2010){Clements}, {Dunne}, \&
  {Eales}}]{clements10}
{Clements}, D.~L., {Dunne}, L., \& {Eales}, S. 2010, \mnras, 403, 274

\bibitem[{{Combes} {et~al.}(2012){Combes}, {Garcia-Burillo}, {Braine},
  {Schinnerer}, {Walter}, \& {Colina}}]{combes12}
{Combes}, F., {Garcia-Burillo}, S., {Braine}, J., {Schinnerer}, E., {Walter},
  F., \& {Colina}, L. 2012, ArXiv e-prints

\bibitem[{{Coppin} {et~al.}(2007){Coppin}, {Swinbank}, {Neri}, {Cox}, {Smail},
  {Ellis}, {Geach}, {Siana}, {Teplitz}, {Dye}, {Kneib}, {Edge}, \&
  {Richard}}]{coppin07}
{Coppin}, K.~E.~K., {Swinbank}, A.~M., {Neri}, R., {Cox}, P., {Smail}, I.,
  {Ellis}, R.~S., {Geach}, J.~E., {Siana}, B., {Teplitz}, H., {Dye}, S.,
  {Kneib}, J.-P., {Edge}, A.~C., \& {Richard}, J. 2007, \apj, 665, 936

\bibitem[{{Cortese} {et~al.}(2011){Cortese}, {Catinella}, {Boissier},
  {Boselli}, \& {Heinis}}]{cortese11}
{Cortese}, L., {Catinella}, B., {Boissier}, S., {Boselli}, A., \& {Heinis}, S.
  2011, \mnras, 415, 1797

\bibitem[{{Daddi} {et~al.}(2010{\natexlab{a}}){Daddi}, {Bournaud}, {Walter},
  {Dannerbauer}, {Carilli}, {Dickinson}, {Elbaz}, {Morrison}, {Riechers},
  {Onodera}, {Salmi}, {Krips}, \& {Stern}}]{daddi10}
{Daddi}, E., {Bournaud}, F., {Walter}, F., {Dannerbauer}, H., {Carilli}, C.~L.,
  {Dickinson}, M., {Elbaz}, D., {Morrison}, G.~E., {Riechers}, D., {Onodera},
  M., {Salmi}, F., {Krips}, M., \& {Stern}, D. 2010{\natexlab{a}}, \apj, 713,
  686

\bibitem[{{Daddi} {et~al.}(2007){Daddi}, {Dickinson}, {Morrison}, {Chary},
  {Cimatti}, {Elbaz}, {Frayer}, {Renzini}, {Pope}, {Alexander}, {Bauer},
  {Giavalisco}, {Huynh}, {Kurk}, \& {Mignoli}}]{daddi07}
{Daddi}, E., {Dickinson}, M., {Morrison}, G., {Chary}, R., {Cimatti}, A.,
  {Elbaz}, D., {Frayer}, D., {Renzini}, A., {Pope}, A., {Alexander}, D.~M.,
  {Bauer}, F.~E., {Giavalisco}, M., {Huynh}, M., {Kurk}, J., \& {Mignoli}, M.
  2007, \apj, 670, 156

\bibitem[{{Daddi} {et~al.}(2010{\natexlab{b}}){Daddi}, {Elbaz}, {Walter},
  {Bournaud}, {Salmi}, {Carilli}, {Dannerbauer}, {Dickinson}, {Monaco}, \&
  {Riechers}}]{daddi10KS}
{Daddi}, E., {Elbaz}, D., {Walter}, F., {Bournaud}, F., {Salmi}, F., {Carilli},
  C., {Dannerbauer}, H., {Dickinson}, M., {Monaco}, P., \& {Riechers}, D.
  2010{\natexlab{b}}, \apjl, 714, L118

\bibitem[{{Dale} {et~al.}(2001){Dale}, {Helou}, {Contursi}, {Silbermann}, \&
  {Kolhatkar}}]{dale01}
{Dale}, D.~A., {Helou}, G., {Contursi}, A., {Silbermann}, N.~A., \&
  {Kolhatkar}, S. 2001, \apj, 549, 215

\bibitem[{{Dame} {et~al.}(2001){Dame}, {Hartmann}, \& {Thaddeus}}]{dame01}
{Dame}, T.~M., {Hartmann}, D., \& {Thaddeus}, P. 2001, \apj, 547, 792

\bibitem[{{Danielson} {et~al.}(2011){Danielson}, {Swinbank}, {Smail}, {Cox},
  {Edge}, {Weiss}, {Harris}, {Baker}, {De Breuck}, {Geach}, {Ivison}, {Krips},
  {Lundgren}, {Longmore}, {Neri}, \& {Flaquer}}]{danielson11}
{Danielson}, A.~L.~R., {Swinbank}, A.~M., {Smail}, I., {Cox}, P., {Edge},
  A.~C., {Weiss}, A., {Harris}, A.~I., {Baker}, A.~J., {De Breuck}, C.,
  {Geach}, J.~E., {Ivison}, R.~J., {Krips}, M., {Lundgren}, A., {Longmore}, S.,
  {Neri}, R., \& {Flaquer}, B.~O. 2011, \mnras, 410, 1687

\bibitem[{{Dannerbauer} {et~al.}(2009){Dannerbauer}, {Daddi}, {Riechers},
  {Walter}, {Carilli}, {Dickinson}, {Elbaz}, \& {Morrison}}]{dannerbauer09}
{Dannerbauer}, H., {Daddi}, E., {Riechers}, D.~A., {Walter}, F., {Carilli},
  C.~L., {Dickinson}, M., {Elbaz}, D., \& {Morrison}, G.~E. 2009, \apjl, 698,
  L178

\bibitem[{{Dav{\'e}} {et~al.}(2012){Dav{\'e}}, {Finlator}, \&
  {Oppenheimer}}]{dave12}
{Dav{\'e}}, R., {Finlator}, K., \& {Oppenheimer}, B.~D. 2012, \mnras, 421, 98

\bibitem[{{Dav{\'e}} {et~al.}(2011){Dav{\'e}}, {Oppenheimer}, \&
  {Finlator}}]{dave11}
{Dav{\'e}}, R., {Oppenheimer}, B.~D., \& {Finlator}, K. 2011, \mnras, 415, 11

\bibitem[{{Denicol{\'o}} {et~al.}(2002){Denicol{\'o}}, {Terlevich}, \&
  {Terlevich}}]{denicolo02}
{Denicol{\'o}}, G., {Terlevich}, R., \& {Terlevich}, E. 2002, \mnras, 330, 69

\bibitem[{{Dessauges-Zavadsky} {et~al.}(2011){Dessauges-Zavadsky},
  {Christensen}, {D'Odorico}, {Schaerer}, \& {Richard}}]{dessauges12}
{Dessauges-Zavadsky}, M., {Christensen}, L., {D'Odorico}, S., {Schaerer}, D.,
  \& {Richard}, J. 2011, \aap, 533, A15

\bibitem[{{Diehl} {et~al.}(2009){Diehl}, {Allam}, {Annis}, {Buckley-Geer},
  {Frieman}, {Kubik}, {Kubo}, {Lin}, {Tucker}, \& {West}}]{diehl09}
{Diehl}, H.~T., {Allam}, S.~S., {Annis}, J., {Buckley-Geer}, E.~J., {Frieman},
  J.~A., {Kubik}, D., {Kubo}, J.~M., {Lin}, H., {Tucker}, D., \& {West}, A.
  2009, \apj, 707, 686

\bibitem[{{Diolaiti} {et~al.}(2000){Diolaiti}, {Bendinelli}, {Bonaccini},
  {Close}, {Currie}, \& {Parmeggiani}}]{diolaiti00}
{Diolaiti}, E., {Bendinelli}, O., {Bonaccini}, D., {Close}, L., {Currie}, D.,
  \& {Parmeggiani}, G. 2000, \aaps, 147, 335

\bibitem[{{Downes} {et~al.}(1995){Downes}, {Solomon}, \& {Radford}}]{downes95}
{Downes}, D., {Solomon}, P.~M., \& {Radford}, S.~J.~E. 1995, \apjl, 453, L65

\bibitem[{{Draine} {et~al.}(2007){Draine}, {Dale}, {Bendo}, {Gordon}, {Smith},
  \& et~al.}]{draine07}
{Draine}, B.~T., {Dale}, D.~A., {Bendo}, G., {Gordon}, K.~D., {Smith},
  J.~D.~T., \& et~al. 2007, \apj, 663, 866

\bibitem[{{Draine} \& {Li}(2007)}]{draineli07}
{Draine}, B.~T., \& {Li}, A. 2007, \apj, 657, 810

\bibitem[{{Dye} {et~al.}(2007){Dye}, {Smail}, {Swinbank}, {Ebeling}, \&
  {Edge}}]{dye07}
{Dye}, S., {Smail}, I., {Swinbank}, A.~M., {Ebeling}, H., \& {Edge}, A.~C.
  2007, \mnras, 379, 308

\bibitem[{{Edmunds}(2001)}]{edmunds01}
{Edmunds}, M.~G. 2001, \mnras, 328, 223

\bibitem[{{Elbaz} {et~al.}(2007){Elbaz}, {Daddi}, {Le Borgne}, {Dickinson},
  {Alexander}, {Chary}, {Starck}, {Brandt}, {Kitzbichler}, {MacDonald},
  {Nonino}, {Popesso}, {Stern}, \& {Vanzella}}]{elbaz07}
{Elbaz}, D., {Daddi}, E., {Le Borgne}, D., {Dickinson}, M., {Alexander}, D.~M.,
  {Chary}, R., {Starck}, J., {Brandt}, W.~N., {Kitzbichler}, M., {MacDonald},
  E., {Nonino}, M., {Popesso}, P., {Stern}, D., \& {Vanzella}, E. 2007, \aap,
  468, 33

\bibitem[{{Elbaz} {et~al.}(2011){Elbaz}, {Dickinson}, {Hwang},
  {D{\'{\i}}az-Santos}, {Magdis}, {Magnelli}, {Le Borgne}, {Galliano},
  {Pannella}, {Chanial}, \& et~al.}]{elbaz11}
{Elbaz}, D., {Dickinson}, M., {Hwang}, H.~S., {D{\'{\i}}az-Santos}, T.,
  {Magdis}, G., {Magnelli}, B., {Le Borgne}, D., {Galliano}, F., {Pannella},
  M., {Chanial}, P., \& et~al. 2011, \aap, 533, A119+

\bibitem[{{Elbaz} {et~al.}(2010){Elbaz}, {Hwang}, {Magnelli}, {Daddi},
  {Aussel}, {Altieri}, {Amblard}, {Andreani}, \& et~al.}]{elbaz10}
{Elbaz}, D., {Hwang}, H.~S., {Magnelli}, B., {Daddi}, E., {Aussel}, H.,
  {Altieri}, B., {Amblard}, A., {Andreani}, P., \& et~al. 2010, \aap, 518, L29+

\bibitem[{{Engelbracht} {et~al.}(2008){Engelbracht}, {Rieke}, {Gordon},
  {Smith}, {Werner}, {Moustakas}, {Willmer}, \& {Vanzi}}]{engelbracht08}
{Engelbracht}, C.~W., {Rieke}, G.~H., {Gordon}, K.~D., {Smith}, J.-D.~T.,
  {Werner}, M.~W., {Moustakas}, J., {Willmer}, C.~N.~A., \& {Vanzi}, L. 2008,
  \apj, 678, 804

\bibitem[{{Epinat} {et~al.}(2012){Epinat}, {Tasca}, {Amram}, {Contini}, {Le
  F{\`e}vre}, {Queyrel}, {Vergani}, {Garilli}, {Kissler-Patig}, {Moultaka},
  {Paioro}, {Tresse}, {Bournaud}, {L{\'o}pez-Sanjuan}, \& {Perret}}]{epinat12}
{Epinat}, B., {Tasca}, L., {Amram}, P., {Contini}, T., {Le F{\`e}vre}, O.,
  {Queyrel}, J., {Vergani}, D., {Garilli}, B., {Kissler-Patig}, M., {Moultaka},
  J., {Paioro}, L., {Tresse}, L., {Bournaud}, F., {L{\'o}pez-Sanjuan}, C., \&
  {Perret}, V. 2012, \aap, 539, A92

\bibitem[{{Fadely} {et~al.}(2010){Fadely}, {Allam}, {Baker}, {Lin}, {Lutz},
  {Shapley}, {Shin}, {Allyn Smith}, {Strauss}, \& {Tucker}}]{fadely10}
{Fadely}, R., {Allam}, S.~S., {Baker}, A.~J., {Lin}, H., {Lutz}, D., {Shapley},
  A.~E., {Shin}, M.-S., {Allyn Smith}, J., {Strauss}, M.~A., \& {Tucker}, D.~L.
  2010, \apj, 723, 729

\bibitem[{{Feldmann}(2013)}]{feldmann13}
{Feldmann}, R. 2013, \mnras, 433, 1910

\bibitem[{{Feldmann} {et~al.}(2012){Feldmann}, {Gnedin}, \&
  {Kravtsov}}]{feldmann12}
{Feldmann}, R., {Gnedin}, N.~Y., \& {Kravtsov}, A.~V. 2012, \apj, 747, 124

\bibitem[{{Finkelstein} {et~al.}(2009){Finkelstein}, {Papovich}, {Rudnick},
  {Egami}, {Le Floc'h}, {Rieke}, {Rigby}, \& {Willmer}}]{finkelstein09}
{Finkelstein}, S.~L., {Papovich}, C., {Rudnick}, G., {Egami}, E., {Le Floc'h},
  E., {Rieke}, M.~J., {Rigby}, J.~R., \& {Willmer}, C.~N.~A. 2009, \apj, 700,
  376

\bibitem[{{F{\"o}rster Schreiber} {et~al.}(2009){F{\"o}rster Schreiber},
  {Genzel}, {Bouch{\'e}}, {Cresci}, {Davies}, {Buschkamp}, {Shapiro},
  {Tacconi}, {Hicks}, {Genel}, {Shapley}, {Erb}, {Steidel}, {Lutz},
  {Eisenhauer}, {Gillessen}, {Sternberg}, {Renzini}, {Cimatti}, {Daddi},
  {Kurk}, {Lilly}, {Kong}, {Lehnert}, {Nesvadba}, {Verma}, {McCracken},
  {Arimoto}, {Mignoli}, \& {Onodera}}]{forster09}
{F{\"o}rster Schreiber}, N.~M., {Genzel}, R., {Bouch{\'e}}, N., {Cresci}, G.,
  {Davies}, R., {Buschkamp}, P., {Shapiro}, K., {Tacconi}, L.~J., {Hicks},
  E.~K.~S., {Genel}, S., {Shapley}, A.~E., {Erb}, D.~K., {Steidel}, C.~C.,
  {Lutz}, D., {Eisenhauer}, F., {Gillessen}, S., {Sternberg}, A., {Renzini},
  A., {Cimatti}, A., {Daddi}, E., {Kurk}, J., {Lilly}, S., {Kong}, X.,
  {Lehnert}, M.~D., {Nesvadba}, N., {Verma}, A., {McCracken}, H., {Arimoto},
  N., {Mignoli}, M., \& {Onodera}, M. 2009, \apj, 706, 1364

\bibitem[{{F{\"o}rster Schreiber} {et~al.}(2006){F{\"o}rster Schreiber},
  {Genzel}, {Lehnert}, {Bouch{\'e}}, {Verma}, {Erb}, {Shapley}, {Steidel},
  {Davies}, {Lutz}, {Nesvadba}, {Tacconi}, {Eisenhauer}, {Abuter}, {Gilbert},
  {Gillessen}, \& {Sternberg}}]{forster06}
{F{\"o}rster Schreiber}, N.~M., {Genzel}, R., {Lehnert}, M.~D., {Bouch{\'e}},
  N., {Verma}, A., {Erb}, D.~K., {Shapley}, A.~E., {Steidel}, C.~C., {Davies},
  R., {Lutz}, D., {Nesvadba}, N., {Tacconi}, L.~J., {Eisenhauer}, F., {Abuter},
  R., {Gilbert}, A., {Gillessen}, S., \& {Sternberg}, A. 2006, \apj, 645, 1062

\bibitem[{{F{\"o}rster Schreiber} {et~al.}(2004){F{\"o}rster Schreiber}, {van
  Dokkum}, {Franx}, {Labb{\'e}}, {Rudnick}, {Daddi}, {Illingworth}, {Kriek},
  {Moorwood}, {Rix}, {R{\"o}ttgering}, {Trujillo}, {van der Werf}, {van
  Starkenburg}, \& {Wuyts}}]{forster04}
{F{\"o}rster Schreiber}, N.~M., {van Dokkum}, P.~G., {Franx}, M., {Labb{\'e}},
  I., {Rudnick}, G., {Daddi}, E., {Illingworth}, G.~D., {Kriek}, M.,
  {Moorwood}, A.~F.~M., {Rix}, H.-W., {R{\"o}ttgering}, H., {Trujillo}, I.,
  {van der Werf}, P., {van Starkenburg}, L., \& {Wuyts}, S. 2004, \apj, 616, 40

\bibitem[{{Frayer} {et~al.}(1997){Frayer}, {Papadopoulos}, {Bechtold},
  {Seaquist}, {Yee}, \& {Scoville}}]{frayer97}
{Frayer}, D.~T., {Papadopoulos}, P.~P., {Bechtold}, J., {Seaquist}, E.~R.,
  {Yee}, H.~K.~C., \& {Scoville}, N.~Z. 1997, \aj, 113, 562

\bibitem[{{Fu} {et~al.}(2012){Fu}, {Jullo}, {Cooray}, {Bussmann}, {Ivison},
  {P{\'e}rez-Fournon}, {Djorgovski}, {Scoville}, \& et~al.}]{fu12}
{Fu}, H., {Jullo}, E., {Cooray}, A., {Bussmann}, R.~S., {Ivison}, R.~J.,
  {P{\'e}rez-Fournon}, I., {Djorgovski}, S.~G., {Scoville}, N., \& et~al. 2012,
  \apj, 753, 134

\bibitem[{{Galametz} {et~al.}(2011){Galametz}, {Madden}, {Galliano}, {Hony},
  {Bendo}, \& {Sauvage}}]{galametz11}
{Galametz}, M., {Madden}, S.~C., {Galliano}, F., {Hony}, S., {Bendo}, G.~J., \&
  {Sauvage}, M. 2011, \aap, 532, A56

\bibitem[{{Geach} {et~al.}(2011){Geach}, {Smail}, {Moran}, {MacArthur},
  {Lagos}, \& {Edge}}]{geach11}
{Geach}, J.~E., {Smail}, I., {Moran}, S.~M., {MacArthur}, L.~A., {Lagos},
  C.~d.~P., \& {Edge}, A.~C. 2011, \apjl, 730, L19

\bibitem[{{Genel} {et~al.}(2008){Genel}, {Genzel}, {Bouch{\'e}}, {Sternberg},
  {Naab}, {Schreiber}, {Shapiro}, {Tacconi}, {Lutz}, {Cresci}, {Buschkamp},
  {Davies}, \& {Hicks}}]{genel08}
{Genel}, S., {Genzel}, R., {Bouch{\'e}}, N., {Sternberg}, A., {Naab}, T.,
  {Schreiber}, N.~M.~F., {Shapiro}, K.~L., {Tacconi}, L.~J., {Lutz}, D.,
  {Cresci}, G., {Buschkamp}, P., {Davies}, R.~I., \& {Hicks}, E.~K.~S. 2008,
  \apj, 688, 789

\bibitem[{{Genzel} {et~al.}(2008){Genzel}, {Burkert}, {Bouch{\'e}}, {Cresci},
  {F{\"o}rster Schreiber}, {Shapley}, {Shapiro}, {Tacconi}, {Buschkamp},
  {Cimatti}, {Daddi}, {Davies}, {Eisenhauer}, {Erb}, {Genel}, {Gerhard},
  {Hicks}, {Lutz}, {Naab}, {Ott}, {Rabien}, {Renzini}, {Steidel}, {Sternberg},
  \& {Lilly}}]{genzel08}
{Genzel}, R., {Burkert}, A., {Bouch{\'e}}, N., {Cresci}, G., {F{\"o}rster
  Schreiber}, N.~M., {Shapley}, A., {Shapiro}, K., {Tacconi}, L.~J.,
  {Buschkamp}, P., {Cimatti}, A., {Daddi}, E., {Davies}, R., {Eisenhauer}, F.,
  {Erb}, D.~K., {Genel}, S., {Gerhard}, O., {Hicks}, E., {Lutz}, D., {Naab},
  T., {Ott}, T., {Rabien}, S., {Renzini}, A., {Steidel}, C.~C., {Sternberg},
  A., \& {Lilly}, S.~J. 2008, \apj, 687, 59

\bibitem[{{Genzel} {et~al.}(2012){Genzel}, {Tacconi}, {Combes}, {Bolatto},
  {Neri}, {Sternberg}, {Cooper}, {Bouch{\'e}}, \& et~al.}]{genzel12}
{Genzel}, R., {Tacconi}, L.~J., {Combes}, F., {Bolatto}, A., {Neri}, R.,
  {Sternberg}, A., {Cooper}, M.~C., {Bouch{\'e}}, N., \& et~al. 2012, \apj,
  746, 69

\bibitem[{{Genzel} {et~al.}(2006){Genzel}, {Tacconi}, {Eisenhauer},
  {F{\"o}rster Schreiber}, {Cimatti}, {Daddi}, {Bouch{\'e}}, {Davies},
  {Lehnert}, {Lutz}, {Nesvadba}, {Verma}, {Abuter}, {Shapiro}, {Sternberg},
  {Renzini}, {Kong}, {Arimoto}, \& {Mignoli}}]{genzel06}
{Genzel}, R., {Tacconi}, L.~J., {Eisenhauer}, F., {F{\"o}rster Schreiber},
  N.~M., {Cimatti}, A., {Daddi}, E., {Bouch{\'e}}, N., {Davies}, R., {Lehnert},
  M.~D., {Lutz}, D., {Nesvadba}, N., {Verma}, A., {Abuter}, R., {Shapiro}, K.,
  {Sternberg}, A., {Renzini}, A., {Kong}, X., {Arimoto}, N., \& {Mignoli}, M.
  2006, \nat, 442, 786

\bibitem[{{Genzel} {et~al.}(2010){Genzel}, {Tacconi}, {Gracia-Carpio},
  {Sternberg}, {Cooper}, \& et~al.}]{genzel10}
{Genzel}, R., {Tacconi}, L.~J., {Gracia-Carpio}, J., {Sternberg}, A., {Cooper},
  M.~C., \& et~al. 2010, \mnras, 407, 2091

\bibitem[{{Glover} \& {Mac Low}(2011)}]{glover11}
{Glover}, S.~C.~O., \& {Mac Low}, M.-M. 2011, \mnras, 412, 337

\bibitem[{{Gnerucci} {et~al.}(2011){Gnerucci}, {Marconi}, {Cresci}, {Maiolino},
  {Mannucci}, {Calura}, {Cimatti}, {Cocchia}, {Grazian}, {Matteucci}, {Nagao},
  {Pozzetti}, \& {Troncoso}}]{gnerucci11}
{Gnerucci}, A., {Marconi}, A., {Cresci}, G., {Maiolino}, R., {Mannucci}, F.,
  {Calura}, F., {Cimatti}, A., {Cocchia}, F., {Grazian}, A., {Matteucci}, F.,
  {Nagao}, T., {Pozzetti}, L., \& {Troncoso}, P. 2011, \aap, 528, A88

\bibitem[{{Gonzalez} {et~al.}(2012){Gonzalez}, {Bouwens}, {llingworth},
  {Labbe}, {Oesch}, {Franx}, \& {Magee}}]{gonzalez12}
{Gonzalez}, V., {Bouwens}, R., {llingworth}, G., {Labbe}, I., {Oesch}, P.,
  {Franx}, M., \& {Magee}, D. 2012, arXiv:1208.4362

\bibitem[{{Gonz{\'a}lez} {et~al.}(2010){Gonz{\'a}lez}, {Labb{\'e}}, {Bouwens},
  {Illingworth}, {Franx}, {Kriek}, \& {Brammer}}]{gonzalez10}
{Gonz{\'a}lez}, V., {Labb{\'e}}, I., {Bouwens}, R.~J., {Illingworth}, G.,
  {Franx}, M., {Kriek}, M., \& {Brammer}, G.~B. 2010, \apj, 713, 115

\bibitem[{{Gratier} {et~al.}(2010){Gratier}, {Braine}, {Rodriguez-Fernandez},
  {Israel}, {Schuster}, {Brouillet}, \& {Gardan}}]{gratier10a}
{Gratier}, P., {Braine}, J., {Rodriguez-Fernandez}, N.~J., {Israel}, F.~P.,
  {Schuster}, K.~F., {Brouillet}, N., \& {Gardan}, E. 2010, \aap, 512, A68

\bibitem[{{Griffin} {et~al.}(2010){Griffin}, {Abergel}, {Abreu}, {Ade},
  {Andr{\'e}}, {Augueres}, {Babbedge}, {Bae}, \& et~al.}]{griffin10}
{Griffin}, M.~J., {Abergel}, A., {Abreu}, A., {Ade}, P.~A.~R., {Andr{\'e}}, P.,
  {Augueres}, J.-L., {Babbedge}, T., {Bae}, Y., \& et~al. 2010, \aap, 518, L3

\bibitem[{{Guilloteau} {et~al.}(1992){Guilloteau}, {Delannoy}, {Downes},
  {Greve}, {Guelin}, {Lucas}, {Morris}, {Radford}, {Wink}, {Cernicharo},
  {Forveille}, {Garcia-Burillo}, {Neri}, {Blondel}, {Perrigourad}, {Plathner},
  \& {Torres}}]{guilloteau92}
{Guilloteau}, S., {Delannoy}, J., {Downes}, D., {Greve}, A., {Guelin}, M.,
  {Lucas}, R., {Morris}, D., {Radford}, S.~J.~E., {Wink}, J., {Cernicharo}, J.,
  {Forveille}, T., {Garcia-Burillo}, S., {Neri}, R., {Blondel}, J.,
  {Perrigourad}, A., {Plathner}, D., \& {Torres}, M. 1992, \aap, 262, 624

\bibitem[{{Hainline} {et~al.}(2009){Hainline}, {Shapley}, {Kornei}, {Pettini},
  {Buckley-Geer}, {Allam}, \& {Tucker}}]{hainline09}
{Hainline}, K.~N., {Shapley}, A.~E., {Kornei}, K.~A., {Pettini}, M.,
  {Buckley-Geer}, E., {Allam}, S.~S., \& {Tucker}, D.~L. 2009, \apj, 701, 52

\bibitem[{{Harris} {et~al.}(2010){Harris}, {Baker}, {Zonak}, {Sharon},
  {Genzel}, {Rauch}, {Watts}, \& {Creager}}]{harris10}
{Harris}, A.~I., {Baker}, A.~J., {Zonak}, S.~G., {Sharon}, C.~E., {Genzel}, R.,
  {Rauch}, K., {Watts}, G., \& {Creager}, R. 2010, \apj, 723, 1139

\bibitem[{{Heinis} {et~al.}(2013){Heinis}, {Buat}, {B{\'e}thermin}, {Aussel},
  {Bock}, {Boselli}, {Burgarella}, {Conley}, \& et~al.}]{heinis13}
{Heinis}, S., {Buat}, V., {B{\'e}thermin}, M., {Aussel}, H., {Bock}, J.,
  {Boselli}, A., {Burgarella}, D., {Conley}, A., \& et~al. 2013, \mnras, 429,
  1113

\bibitem[{{Hezaveh} {et~al.}(2012){Hezaveh}, {Marrone}, \&
  {Holder}}]{hezaveh12}
{Hezaveh}, Y.~D., {Marrone}, D.~P., \& {Holder}, G.~P. 2012, \apj, 761, 20

\bibitem[{{Huang} {et~al.}(2012){Huang}, {Haynes}, {Giovanelli}, \&
  {Brinchmann}}]{huang12}
{Huang}, S., {Haynes}, M.~P., {Giovanelli}, R., \& {Brinchmann}, J. 2012, \apj,
  756, 113

\bibitem[{{Hunt} {et~al.}(2005){Hunt}, {Bianchi}, \& {Maiolino}}]{hunt05}
{Hunt}, L., {Bianchi}, S., \& {Maiolino}, R. 2005, \aap, 434, 849

\bibitem[{{Hwang} {et~al.}(2010){Hwang}, {Elbaz}, {Magdis}, {Daddi},
  {Symeonidis}, {Altieri}, {Amblard}, {Andreani}, \& et~al.}]{hwang10}
{Hwang}, H.~S., {Elbaz}, D., {Magdis}, G., {Daddi}, E., {Symeonidis}, M.,
  {Altieri}, B., {Amblard}, A., {Andreani}, P., \& et~al. 2010, \mnras, 409, 75

\bibitem[{{Israel}(1997)}]{israel97}
{Israel}, F.~P. 1997, \aap, 328, 471

\bibitem[{{Ivison} {et~al.}(2011){Ivison}, {Papadopoulos}, {Smail}, {Greve},
  {Thomson}, {Xilouris}, \& {Chapman}}]{ivison11}
{Ivison}, R.~J., {Papadopoulos}, P.~P., {Smail}, I., {Greve}, T.~R., {Thomson},
  A.~P., {Xilouris}, E.~M., \& {Chapman}, S.~C. 2011, \mnras, 412, 1913

\bibitem[{{Jones} {et~al.}(2010{\natexlab{a}}){Jones}, {Ellis}, {Jullo}, \&
  {Richard}}]{jones10a}
{Jones}, T., {Ellis}, R., {Jullo}, E., \& {Richard}, J. 2010{\natexlab{a}},
  \apjl, 725, L176

\bibitem[{{Jones} {et~al.}(2013){Jones}, {Ellis}, {Richard}, \&
  {Jullo}}]{jones13}
{Jones}, T., {Ellis}, R.~S., {Richard}, J., \& {Jullo}, E. 2013, \apj, 765, 48

\bibitem[{{Jones} {et~al.}(2010{\natexlab{b}}){Jones}, {Swinbank}, {Ellis},
  {Richard}, \& {Stark}}]{jones10b}
{Jones}, T.~A., {Swinbank}, A.~M., {Ellis}, R.~S., {Richard}, J., \& {Stark},
  D.~P. 2010{\natexlab{b}}, \mnras, 404, 1247

\bibitem[{{Kaviraj} {et~al.}(2013){Kaviraj}, {Cohen}, {Windhorst}, {Silk},
  {O'Connell}, {Dopita}, {Dekel}, {Hathi}, {Straughn}, \&
  {Rutkowski}}]{kaviraj13}
{Kaviraj}, S., {Cohen}, S., {Windhorst}, R.~A., {Silk}, J., {O'Connell}, R.~W.,
  {Dopita}, M.~A., {Dekel}, A., {Hathi}, N.~P., {Straughn}, A., \& {Rutkowski},
  M. 2013, \mnras, 429, L40

\bibitem[{{Kennicutt}(1998)}]{kennicutt98}
{Kennicutt}, Jr., R.~C. 1998, \araa, 36, 189

\bibitem[{{Kewley} {et~al.}(2001){Kewley}, {Dopita}, {Sutherland}, {Heisler},
  \& {Trevena}}]{kewley01}
{Kewley}, L.~J., {Dopita}, M.~A., {Sutherland}, R.~S., {Heisler}, C.~A., \&
  {Trevena}, J. 2001, \apj, 556, 121

\bibitem[{{Kewley} \& {Ellison}(2008)}]{kewley08}
{Kewley}, L.~J., \& {Ellison}, S.~L. 2008, \apj, 681, 1183

\bibitem[{{Kochanek} {et~al.}(2001){Kochanek}, {Keeton}, \&
  {McLeod}}]{kochanek01}
{Kochanek}, C.~S., {Keeton}, C.~R., \& {McLeod}, B.~A. 2001, \apj, 547, 50

\bibitem[{{Koester} {et~al.}(2010){Koester}, {Gladders}, {Hennawi}, {Sharon},
  {Wuyts}, {Rigby}, {Bayliss}, \& {Dahle}}]{koester10}
{Koester}, B.~P., {Gladders}, M.~D., {Hennawi}, J.~F., {Sharon}, K., {Wuyts},
  E., {Rigby}, J.~R., {Bayliss}, M.~B., \& {Dahle}, H. 2010, \apjl, 723, L73

\bibitem[{{Kormann} {et~al.}(1994){Kormann}, {Schneider}, \&
  {Bartelmann}}]{kormann94}
{Kormann}, R., {Schneider}, P., \& {Bartelmann}, M. 1994, \aap, 284, 285

\bibitem[{{Kreysa} {et~al.}(1998){Kreysa}, {Gemuend}, {Gromke}, {Haslam},
  {Reichertz}, {Haller}, {Beeman}, {Hansen}, {Sievers}, \& {Zylka}}]{kreysa98}
{Kreysa}, E., {Gemuend}, H.-P., {Gromke}, J., {Haslam}, C.~G., {Reichertz}, L.,
  {Haller}, E.~E., {Beeman}, J.~W., {Hansen}, V., {Sievers}, A., \& {Zylka}, R.
  1998, in Society of Photo-Optical Instrumentation Engineers (SPIE) Conference
  Series, Vol. 3357, Society of Photo-Optical Instrumentation Engineers (SPIE)
  Conference Series, ed. T.~G. {Phillips}, 319--325

\bibitem[{{Krumholz} \& {Dekel}(2012)}]{krumholz12}
{Krumholz}, M.~R., \& {Dekel}, A. 2012, \apj, 753, 16

\bibitem[{{Kubo} {et~al.}(2009){Kubo}, {Allam}, {Annis}, {Buckley-Geer},
  {Diehl}, {Kubik}, {Lin}, \& {Tucker}}]{kubo09}
{Kubo}, J.~M., {Allam}, S.~S., {Annis}, J., {Buckley-Geer}, E.~J., {Diehl},
  H.~T., {Kubik}, D., {Lin}, H., \& {Tucker}, D. 2009, \apjl, 696, L61

\bibitem[{{Kuzio de Naray} {et~al.}(2004){Kuzio de Naray}, {McGaugh}, \& {de
  Blok}}]{denaray04}
{Kuzio de Naray}, R., {McGaugh}, S.~S., \& {de Blok}, W.~J.~G. 2004, \mnras,
  355, 887

\bibitem[{{Leroy} {et~al.}(2011){Leroy}, {Bolatto}, {Gordon}, {Sandstrom},
  {Gratier}, {Rosolowsky}, {Engelbracht}, {Mizuno}, {Corbelli}, {Fukui}, \&
  {Kawamura}}]{leroy11}
{Leroy}, A.~K., {Bolatto}, A., {Gordon}, K., {Sandstrom}, K., {Gratier}, P.,
  {Rosolowsky}, E., {Engelbracht}, C.~W., {Mizuno}, N., {Corbelli}, E.,
  {Fukui}, Y., \& {Kawamura}, A. 2011, \apj, 737, 12

\bibitem[{{Leroy} {et~al.}(2008){Leroy}, {Walter}, {Brinks}, {Bigiel}, {de
  Blok}, {Madore}, \& {Thornley}}]{leroy08}
{Leroy}, A.~K., {Walter}, F., {Brinks}, E., {Bigiel}, F., {de Blok}, W.~J.~G.,
  {Madore}, B., \& {Thornley}, M.~D. 2008, \aj, 136, 2782

\bibitem[{{Levesque} {et~al.}(2010){Levesque}, {Kewley}, \&
  {Larson}}]{levesque10}
{Levesque}, E.~M., {Kewley}, L.~J., \& {Larson}, K.~L. 2010, \aj, 139, 712

\bibitem[{{Lilly} {et~al.}(2013){Lilly}, {Carollo}, {Pipino}, {Renzini}, \&
  {Peng}}]{lilly13}
{Lilly}, S.~J., {Carollo}, C.~M., {Pipino}, A., {Renzini}, A., \& {Peng}, Y.
  2013, ArXiv e-prints

\bibitem[{{Lin} {et~al.}(2009){Lin}, {Buckley-Geer}, {Allam}, {Tucker},
  {Diehl}, {Kubik}, {Kubo}, {Annis}, {Frieman}, {Oguri}, \& {Inada}}]{lin09}
{Lin}, H., {Buckley-Geer}, E., {Allam}, S.~S., {Tucker}, D.~L., {Diehl}, H.~T.,
  {Kubik}, D., {Kubo}, J.~M., {Annis}, J., {Frieman}, J.~A., {Oguri}, M., \&
  {Inada}, N. 2009, \apj, 699, 1242

\bibitem[{{Livermore} {et~al.}(2012){Livermore}, {Jones}, {Richard}, {Bower},
  {Ellis}, {Swinbank}, {Rigby}, {Smail}, {Arribas}, {Rodriguez Zaurin},
  {Colina}, {Ebeling}, \& {Crain}}]{livermore12}
{Livermore}, R.~C., {Jones}, T., {Richard}, J., {Bower}, R.~G., {Ellis}, R.~S.,
  {Swinbank}, A.~M., {Rigby}, J.~R., {Smail}, I., {Arribas}, S., {Rodriguez
  Zaurin}, J., {Colina}, L., {Ebeling}, H., \& {Crain}, R.~A. 2012, \mnras,
  427, 688

\bibitem[{{Lutz} {et~al.}(2011){Lutz}, {Poglitsch}, {Altieri}, {Andreani},
  {Aussel}, {Berta}, {Bongiovanni}, {Brisbin}, \& et~al.}]{lutz11}
{Lutz}, D., {Poglitsch}, A., {Altieri}, B., {Andreani}, P., {Aussel}, H.,
  {Berta}, S., {Bongiovanni}, A., {Brisbin}, D., \& et~al. 2011, \aap, 532,
  A90+

\bibitem[{{Magdis} {et~al.}(2012{\natexlab{a}}){Magdis}, {Daddi},
  {B{\'e}thermin}, {Sargent}, {Elbaz}, {Pannella}, {Dickinson}, {Dannerbauer},
  {da Cunha}, {Walter}, {Rigopoulou}, {Charmandaris}, {Hwang}, \&
  {Kartaltepe}}]{magdis12a}
{Magdis}, G.~E., {Daddi}, E., {B{\'e}thermin}, M., {Sargent}, M., {Elbaz}, D.,
  {Pannella}, M., {Dickinson}, M., {Dannerbauer}, H., {da Cunha}, E., {Walter},
  F., {Rigopoulou}, D., {Charmandaris}, V., {Hwang}, H.~S., \& {Kartaltepe}, J.
  2012{\natexlab{a}}, \apj, 760, 6

\bibitem[{{Magdis} {et~al.}(2011){Magdis}, {Daddi}, {Elbaz}, {Sargent},
  {Dickinson}, {Dannerbauer}, {Aussel}, {Walter}, {Hwang}, {Charmandaris},
  {Hodge}, {Riechers}, {Rigopoulou}, {Carilli}, {Pannella}, {Mullaney},
  {Leiton}, \& {Scott}}]{magdis11}
{Magdis}, G.~E., {Daddi}, E., {Elbaz}, D., {Sargent}, M., {Dickinson}, M.,
  {Dannerbauer}, H., {Aussel}, H., {Walter}, F., {Hwang}, H.~S.,
  {Charmandaris}, V., {Hodge}, J., {Riechers}, D., {Rigopoulou}, D., {Carilli},
  C., {Pannella}, M., {Mullaney}, J., {Leiton}, R., \& {Scott}, D. 2011, \apjl,
  740, L15

\bibitem[{{Magdis} {et~al.}(2012{\natexlab{b}}){Magdis}, {Daddi}, {Sargent},
  {Elbaz}, {Gobat}, {Dannerbauer}, {Feruglio}, {Tan}, {Rigopoulou},
  {Charmandaris}, {Dickinson}, {Reddy}, \& {Aussel}}]{magdis12b}
{Magdis}, G.~E., {Daddi}, E., {Sargent}, M., {Elbaz}, D., {Gobat}, R.,
  {Dannerbauer}, H., {Feruglio}, C., {Tan}, Q., {Rigopoulou}, D.,
  {Charmandaris}, V., {Dickinson}, M., {Reddy}, N., \& {Aussel}, H.
  2012{\natexlab{b}}, \apjl, 758, L9

\bibitem[{{Magnelli} {et~al.}(2009){Magnelli}, {Elbaz}, {Chary}, {Dickinson},
  {Le Borgne}, {Frayer}, \& {Willmer}}]{magnelli09}
{Magnelli}, B., {Elbaz}, D., {Chary}, R.~R., {Dickinson}, M., {Le Borgne}, D.,
  {Frayer}, D.~T., \& {Willmer}, C.~N.~A. 2009, \aap, 496, 57

\bibitem[{{Magnelli} {et~al.}(2011){Magnelli}, {Elbaz}, {Chary}, {Dickinson},
  {Le Borgne}, {Frayer}, \& {Willmer}}]{magnelli11}
---. 2011, \aap, 528, A35

\bibitem[{{Magnelli} {et~al.}(2010){Magnelli}, {Lutz}, {Berta}, {Altieri},
  {Andreani}, {Aussel}, {Casta{\~n}eda}, {Cava}, \& et~al.}]{magnelli10}
{Magnelli}, B., {Lutz}, D., {Berta}, S., {Altieri}, B., {Andreani}, P.,
  {Aussel}, H., {Casta{\~n}eda}, H., {Cava}, A., \& et~al. 2010, \aap, 518,
  L28+

\bibitem[{{Magnelli} {et~al.}(2012{\natexlab{a}}){Magnelli}, {Lutz}, {Santini},
  {Saintonge}, {Berta}, {Albrecht}, {Altieri}, \& {Andreani}}]{magnelli12}
{Magnelli}, B., {Lutz}, D., {Santini}, P., {Saintonge}, A., {Berta}, S.,
  {Albrecht}, M., {Altieri}, B., \& {Andreani}, P. e.~a. 2012{\natexlab{a}},
  \aap, 539, A155

\bibitem[{{Magnelli} {et~al.}(2013){Magnelli}, {Lutz}, {Santini}, {Saintonge},
  {Berta}, {Albrecht}, {Altieri}, \& {Andreani}}]{magnelli13}
---. 2013, \aap\ submitted

\bibitem[{{Magnelli} {et~al.}(2012{\natexlab{b}}){Magnelli}, {Saintonge},
  {Lutz}, {Tacconi}, {Berta}, {Bournaud}, {Charmandaris}, {Dannerbauer},
  {Elbaz}, {F{\"o}rster-Schreiber}, {Graci{\'a}-Carpio}, {Ivison}, {Maiolino},
  {Nordon}, {Popesso}, {Rodighiero}, {Santini}, \& {Wuyts}}]{magnellisaintonge}
{Magnelli}, B., {Saintonge}, A., {Lutz}, D., {Tacconi}, L.~J., {Berta}, S.,
  {Bournaud}, F., {Charmandaris}, V., {Dannerbauer}, H., {Elbaz}, D.,
  {F{\"o}rster-Schreiber}, N.~M., {Graci{\'a}-Carpio}, J., {Ivison}, R.,
  {Maiolino}, R., {Nordon}, R., {Popesso}, P., {Rodighiero}, G., {Santini}, P.,
  \& {Wuyts}, S. 2012{\natexlab{b}}, \aap, 548, A22

\bibitem[{{Maraston} {et~al.}(2010){Maraston}, {Pforr}, {Renzini}, {Daddi},
  {Dickinson}, {Cimatti}, \& {Tonini}}]{maraston10}
{Maraston}, C., {Pforr}, J., {Renzini}, A., {Daddi}, E., {Dickinson}, M.,
  {Cimatti}, A., \& {Tonini}, C. 2010, \mnras, 407, 830

\bibitem[Marshall et al.(2007)]{marshall07} Marshall, P.~J., Treu, 
T., Melbourne, J., et al.\ 2007, \apj, 671, 1196 

\bibitem[{{McGaugh}(1991)}]{mcgaugh91}
{McGaugh}, S.~S. 1991, \apj, 380, 140

\bibitem[{{Meurer} {et~al.}(1999){Meurer}, {Heckman}, \& {Calzetti}}]{meurer99}
{Meurer}, G.~R., {Heckman}, T.~M., \& {Calzetti}, D. 1999, \apj, 521, 64

\bibitem[{{Newman} {et~al.}(2013){Newman}, {Genzel}, {F{\"o}rster Schreiber},
  {Shapiro Griffin}, {Mancini}, {Lilly}, {Renzini}, {Bouch{\'e}}, {Burkert},
  {Buschkamp}, {Carollo}, {Cresci}, {Davies}, {Eisenhauer}, {Genel}, {Hicks},
  {Kurk}, {Lutz}, {Naab}, {Peng}, {Sternberg}, {Tacconi}, {Wuyts}, {Zamorani},
  \& {Vergani}}]{newman13}
{Newman}, S.~F., {Genzel}, R., {F{\"o}rster Schreiber}, N.~M., {Shapiro
  Griffin}, K., {Mancini}, C., {Lilly}, S.~J., {Renzini}, A., {Bouch{\'e}}, N.,
  {Burkert}, A., {Buschkamp}, P., {Carollo}, C.~M., {Cresci}, G., {Davies}, R.,
  {Eisenhauer}, F., {Genel}, S., {Hicks}, E.~K.~S., {Kurk}, J., {Lutz}, D.,
  {Naab}, T., {Peng}, Y., {Sternberg}, A., {Tacconi}, L.~J., {Wuyts}, S.,
  {Zamorani}, G., \& {Vergani}, D. 2013, \apj, 767, 104

\bibitem[{{Noeske} {et~al.}(2007){Noeske}, {Weiner}, {Faber}, {Papovich},
  {Koo}, {Somerville}, {Bundy}, {Conselice}, \& et~al.}]{noeske07}
{Noeske}, K.~G., {Weiner}, B.~J., {Faber}, S.~M., {Papovich}, C., {Koo}, D.~C.,
  {Somerville}, R.~S., {Bundy}, K., {Conselice}, C.~J., \& et~al. 2007, \apjl,
  660, L43

\bibitem[{{Nordon} {et~al.}(2012){Nordon}, {Lutz}, {Genzel}, {Berta}, {Wuyts},
  \& et~al.}]{nordon12}
{Nordon}, R., {Lutz}, D., {Genzel}, R., {Berta}, S., {Wuyts}, S., \& et~al.
  2012, \apj, 745, 182

\bibitem[{{Nordon} {et~al.}(2013){Nordon}, {Lutz}, {Saintonge}, {Berta},
  {Wuyts}, {F{\"o}rster Schreiber}, {Genzel}, {Magnelli}, {Poglitsch},
  {Popesso}, {Rosario}, {Sturm}, \& {Tacconi}}]{nordon13}
{Nordon}, R., {Lutz}, D., {Saintonge}, A., {Berta}, S., {Wuyts}, S.,
  {F{\"o}rster Schreiber}, N.~M., {Genzel}, R., {Magnelli}, B., {Poglitsch},
  A., {Popesso}, P., {Rosario}, D., {Sturm}, E., \& {Tacconi}, L.~J. 2013,
  \apj, 762, 125

\bibitem[{{Nordon} {et~al.}(2010){Nordon}, {Lutz}, {Shao}, {Magnelli}, {Berta},
  {Altieri}, {Andreani}, {Aussel}, \& et~al.}]{nordon10}
{Nordon}, R., {Lutz}, D., {Shao}, L., {Magnelli}, B., {Berta}, S., {Altieri},
  B., {Andreani}, P., {Aussel}, H., \& et~al. 2010, \aap, 518, L24+

\bibitem[{{Ott}(2010)}]{ott10}
{Ott}, S. 2010, in Astronomical Society of the Pacific Conference Series, Vol.
  434, Astronomical Data Analysis Software and Systems XIX, ed. Y.~{Mizumoto},
  K.-I. {Morita}, \& M.~{Ohishi}, 139

\bibitem[{{Pannella} {et~al.}(2009){Pannella}, {Carilli}, {Daddi}, {McCracken},
  {Owen}, {Renzini}, {Strazzullo}, {Civano}, {Koekemoer}, {Schinnerer},
  {Scoville}, {Smol{\v c}i{\'c}}, {Taniguchi}, {Aussel}, {Kneib}, {Ilbert},
  {Mellier}, {Salvato}, {Thompson}, \& {Willott}}]{pannella09}
{Pannella}, M., {Carilli}, C.~L., {Daddi}, E., {McCracken}, H.~J., {Owen},
  F.~N., {Renzini}, A., {Strazzullo}, V., {Civano}, F., {Koekemoer}, A.~M.,
  {Schinnerer}, E., {Scoville}, N., {Smol{\v c}i{\'c}}, V., {Taniguchi}, Y.,
  {Aussel}, H., {Kneib}, J.~P., {Ilbert}, O., {Mellier}, Y., {Salvato}, M.,
  {Thompson}, D., \& {Willott}, C.~J. 2009, \apjl, 698, L116

\bibitem[{{Papovich} {et~al.}(2011){Papovich}, {Finkelstein}, {Ferguson},
  {Lotz}, \& {Giavalisco}}]{papovich11}
{Papovich}, C., {Finkelstein}, S.~L., {Ferguson}, H.~C., {Lotz}, J.~M., \&
  {Giavalisco}, M. 2011, \mnras, 412, 1123

\bibitem[{{Peng} {et~al.}(2002){Peng}, {Ho}, {Impey}, \& {Rix}}]{peng02}
{Peng}, C.~Y., {Ho}, L.~C., {Impey}, C.~D., \& {Rix}, H.-W. 2002, \aj, 124, 266

\bibitem[{{Peng} {et~al.}(2010){Peng}, {Ho}, {Impey}, \& {Rix}}]{pengc10}
---. 2010, \aj, 139, 2097

\bibitem[{{Peng} {et~al.}(2006){Peng}, {Impey}, {Rix}, {Kochanek}, {Keeton},
  {Falco}, {Leh{\'a}r}, \& {McLeod}}]{peng06}
{Peng}, C.~Y., {Impey}, C.~D., {Rix}, H.-W., {Kochanek}, C.~S., {Keeton},
  C.~R., {Falco}, E.~E., {Leh{\'a}r}, J., \& {McLeod}, B.~A. 2006, \apj, 649,
  616

\bibitem[{{Pettini} {et~al.}(2010){Pettini}, {Christensen}, {D'Odorico},
  {Belokurov}, {Evans}, {Hewett}, {Koposov}, {Mason}, \& {Vernet}}]{pettini10}
{Pettini}, M., {Christensen}, L., {D'Odorico}, S., {Belokurov}, V., {Evans},
  N.~W., {Hewett}, P.~C., {Koposov}, S., {Mason}, E., \& {Vernet}, J. 2010,
  \mnras, 402, 2335

\bibitem[{{Pilbratt} {et~al.}(2010){Pilbratt}, {Riedinger}, {Passvogel},
  {Crone}, {Doyle}, {Gageur}, {Heras}, {Jewell}, {Metcalfe}, {Ott}, \&
  {Schmidt}}]{pilbratt10}
{Pilbratt}, G.~L., {Riedinger}, J.~R., {Passvogel}, T., {Crone}, G., {Doyle},
  D., {Gageur}, U., {Heras}, A.~M., {Jewell}, C., {Metcalfe}, L., {Ott}, S., \&
  {Schmidt}, M. 2010, \aap, 518, L1

\bibitem[{{Poglitsch} {et~al.}(2010){Poglitsch}, {Waelkens}, {Geis},
  {Feuchtgruber}, {Vandenbussche}, {Rodriguez}, {Krause}, {Renotte}, \&
  et~al.}]{poglitsch10}
{Poglitsch}, A., {Waelkens}, C., {Geis}, N., {Feuchtgruber}, H.,
  {Vandenbussche}, B., {Rodriguez}, L., {Krause}, O., {Renotte}, E., \& et~al.
  2010, \aap, 518, L2

\bibitem[{{Polletta} {et~al.}(2007){Polletta}, {Tajer}, {Maraschi},
  {Trinchieri}, {Lonsdale}, {Chiappetti}, {Andreon}, {Pierre}, \&
  et~al.}]{polletta07}
{Polletta}, M., {Tajer}, M., {Maraschi}, L., {Trinchieri}, G., {Lonsdale},
  C.~J., {Chiappetti}, L., {Andreon}, S., {Pierre}, M., \& et~al. 2007, \apj,
  663, 81

\bibitem[{{Rahman} {et~al.}(2012){Rahman}, {Bolatto}, {Xue}, {Wong}, {Leroy},
  {Walter}, {Bigiel}, {Rosolowsky}, {Fisher}, {Vogel}, {Blitz}, {West}, \&
  {Ott}}]{rahman12}
{Rahman}, N., {Bolatto}, A.~D., {Xue}, R., {Wong}, T., {Leroy}, A.~K.,
  {Walter}, F., {Bigiel}, F., {Rosolowsky}, E., {Fisher}, D.~B., {Vogel},
  S.~N., {Blitz}, L., {West}, A.~A., \& {Ott}, J. 2012, \apj, 745, 183

\bibitem[{{Reddy} {et~al.}(2010){Reddy}, {Erb}, {Pettini}, {Steidel}, \&
  {Shapley}}]{reddy10}
{Reddy}, N.~A., {Erb}, D.~K., {Pettini}, M., {Steidel}, C.~C., \& {Shapley},
  A.~E. 2010, \apj, 712, 1070

\bibitem[{{Reddy} {et~al.}(2012){Reddy}, {Pettini}, {Steidel}, {Shapley},
  {Erb}, \& {Law}}]{reddy12}
{Reddy}, N.~A., {Pettini}, M., {Steidel}, C.~C., {Shapley}, A.~E., {Erb},
  D.~K., \& {Law}, D.~R. 2012, \apj, 754, 25

\bibitem[{{Richard} {et~al.}(2011){Richard}, {Jones}, {Ellis}, {Stark},
  {Livermore}, \& {Swinbank}}]{richard11}
{Richard}, J., {Jones}, T., {Ellis}, R., {Stark}, D.~P., {Livermore}, R., \&
  {Swinbank}, M. 2011, \mnras, 413, 643

\bibitem[{{Riechers} {et~al.}(2010){Riechers}, {Carilli}, {Walter}, \&
  {Momjian}}]{riechers10}
{Riechers}, D.~A., {Carilli}, C.~L., {Walter}, F., \& {Momjian}, E. 2010,
  \apjl, 724, L153

\bibitem[{{Robaina} {et~al.}(2009){Robaina}, {Bell}, {Skelton}, {McIntosh},
  {Somerville}, {Zheng}, {Rix}, {Bacon}, \& et~al.}]{robaina09}
{Robaina}, A.~R., {Bell}, E.~F., {Skelton}, R.~E., {McIntosh}, D.~H.,
  {Somerville}, R.~S., {Zheng}, X., {Rix}, H.-W., {Bacon}, D., \& et~al. 2009,
  \apj, 704, 324

\bibitem[{{Rodighiero} {et~al.}(2010){Rodighiero}, {Cimatti}, {Gruppioni},
  {Popesso}, {Andreani}, \& et~al.}]{rodighiero10}
{Rodighiero}, G., {Cimatti}, A., {Gruppioni}, C., {Popesso}, P., {Andreani},
  P., \& et~al. 2010, \aap, 518, L25+

\bibitem[{{Rodighiero} {et~al.}(2011){Rodighiero}, {Daddi}, {Baronchelli},
  {Cimatti}, {Renzini}, {Aussel}, {Popesso}, {Lutz}, \& et~al.}]{rodighiero11}
{Rodighiero}, G., {Daddi}, E., {Baronchelli}, I., {Cimatti}, A., {Renzini}, A.,
  {Aussel}, H., {Popesso}, P., {Lutz}, D., \& et~al. 2011, \apjl, 739, L40+

\bibitem[{{Rosario} {et~al.}(2012){Rosario}, {Santini}, {Lutz}, {Shao},
  {Maiolino}, \& et~al.}]{rosario12}
{Rosario}, D.~J., {Santini}, P., {Lutz}, D., {Shao}, L., {Maiolino}, R., \&
  et~al. 2012, \aap, 545, A45

\bibitem[{{Roseboom} {et~al.}(2012){Roseboom}, {Ivison}, {Greve}, {Amblard},
  {Arumugam}, \& et~al.}]{roseboom12}
{Roseboom}, I.~G., {Ivison}, R.~J., {Greve}, T.~R., {Amblard}, A., {Arumugam},
  V., \& et~al. 2012, \mnras, 419, 2758

\bibitem[{{Rosolowsky} {et~al.}(2003){Rosolowsky}, {Engargiola}, {Plambeck}, \&
  {Blitz}}]{rosolowsky03}
{Rosolowsky}, E., {Engargiola}, G., {Plambeck}, R., \& {Blitz}, L. 2003, \apj,
  599, 258

\bibitem[{{Saintonge} {et~al.}(2011{\natexlab{a}}){Saintonge}, {Kauffmann},
  {Kramer}, {Tacconi}, {Buchbender}, {Catinella}, {Fabello},
  {Graci{\'a}-Carpio}, \& et~al.}]{COLDGASS1}
{Saintonge}, A., {Kauffmann}, G., {Kramer}, C., {Tacconi}, L.~J., {Buchbender},
  C., {Catinella}, B., {Fabello}, S., {Graci{\'a}-Carpio}, J., \& et~al.
  2011{\natexlab{a}}, \mnras, 415, 32

\bibitem[{{Saintonge} {et~al.}(2011{\natexlab{b}}){Saintonge}, {Kauffmann},
  {Wang}, {Kramer}, {Tacconi}, {Buchbender}, {Catinella}, {Graci{\'a}-Carpio},
  \& et~al.}]{COLDGASS2}
{Saintonge}, A., {Kauffmann}, G., {Wang}, J., {Kramer}, C., {Tacconi}, L.~J.,
  {Buchbender}, C., {Catinella}, B., {Graci{\'a}-Carpio}, J., \& et~al.
  2011{\natexlab{b}}, \mnras, 415, 61

\bibitem[{{Saintonge} {et~al.}(2012){Saintonge}, {Tacconi}, {Fabello}, {Wang},
  {Catinella}, \& et~al.}]{saintonge12}
{Saintonge}, A., {Tacconi}, L.~J., {Fabello}, S., {Wang}, J., {Catinella}, B.,
  \& et~al. 2012, \apj, 758, 73

\bibitem[{{Salim} {et~al.}(2007){Salim}, {Rich}, {Charlot}, {Brinchmann},
  {Johnson}, \& et~al.}]{salim07}
{Salim}, S., {Rich}, R.~M., {Charlot}, S., {Brinchmann}, J., {Johnson}, B.~D.,
  \& et~al. 2007, \apjs, 173, 267

\bibitem[{{Sand} {et~al.}(2004){Sand}, {Treu}, {Smith}, \& {Ellis}}]{sand04}
{Sand}, D.~J., {Treu}, T., {Smith}, G.~P., \& {Ellis}, R.~S. 2004, \apj, 604,
  88

\bibitem[{{Sanders} \& {Mirabel}(1996)}]{sanders96}
{Sanders}, D.~B., \& {Mirabel}, I.~F. 1996, \araa, 34, 749

\bibitem[{{Sandstrom} {et~al.}(2012){Sandstrom}, {Leroy}, {Walter}, {Bolatto},
  {Croxall}, \& et~al.}]{sandstrom13}
{Sandstrom}, K.~M., {Leroy}, A.~K., {Walter}, F., {Bolatto}, A.~D., {Croxall},
  K.~V., \& et~al. 2012, arXiv:1212.120

\bibitem[{{Santini} {et~al.}(2010){Santini}, {Maiolino}, {Magnelli}, {Silva},
  {Grazian}, \& et~al.}]{santini10}
{Santini}, P., {Maiolino}, R., {Magnelli}, B., {Silva}, L., {Grazian}, A., \&
  et~al. 2010, \aap, 518, L154

\bibitem[{{Sargent} {et~al.}(2013){Sargent}, {Daddi}, {B{\'e}thermin},
  {Aussel}, {Magdis}, {Hwang}, {Juneau}, {Elbaz}, \& {da Cunha}}]{sargent13}
{Sargent}, M.~T., {Daddi}, E., {B{\'e}thermin}, M., {Aussel}, H., {Magdis}, G.,
  {Hwang}, H.~S., {Juneau}, S., {Elbaz}, D., \& {da Cunha}, E. 2013,
  ArXiv:1303.439

\bibitem[{{Scoville}(2012)}]{scoville12}
{Scoville}, N.~Z. 2012, arXiv:1210.6990

\bibitem[{{Seitz} {et~al.}(1998){Seitz}, {Saglia}, {Bender}, {Hopp}, {Belloni},
  \& {Ziegler}}]{seitz98}
{Seitz}, S., {Saglia}, R.~P., {Bender}, R., {Hopp}, U., {Belloni}, P., \&
  {Ziegler}, B. 1998, \mnras, 298, 945

\bibitem[{{Serjeant}(2012)}]{serjeant12}
{Serjeant}, S. 2012, \mnras, 424, 2429

\bibitem[{{Sersic}(1968)}]{sersic68}
{Sersic}, J.~L. 1968, {Atlas de galaxias australes}

\bibitem[{{Shapiro} {et~al.}(2008){Shapiro}, {Genzel}, {F{\"o}rster Schreiber},
  {Tacconi}, {Bouch{\'e}}, {Cresci}, {Davies}, {Eisenhauer}, {Johansson},
  {Krajnovi{\'c}}, {Lutz}, {Naab}, {Arimoto}, {Arribas}, {Cimatti}, {Colina},
  {Daddi}, {Daigle}, {Erb}, {Hernandez}, {Kong}, {Mignoli}, {Onodera},
  {Renzini}, {Shapley}, \& {Steidel}}]{shapiro08}
{Shapiro}, K.~L., {Genzel}, R., {F{\"o}rster Schreiber}, N.~M., {Tacconi},
  L.~J., {Bouch{\'e}}, N., {Cresci}, G., {Davies}, R., {Eisenhauer}, F.,
  {Johansson}, P.~H., {Krajnovi{\'c}}, D., {Lutz}, D., {Naab}, T., {Arimoto},
  N., {Arribas}, S., {Cimatti}, A., {Colina}, L., {Daddi}, E., {Daigle}, O.,
  {Erb}, D., {Hernandez}, O., {Kong}, X., {Mignoli}, M., {Onodera}, M.,
  {Renzini}, A., {Shapley}, A., \& {Steidel}, C. 2008, \apj, 682, 231

\bibitem[{{Shapley} {et~al.}(2005){Shapley}, {Steidel}, {Erb}, {Reddy},
  {Adelberger}, {Pettini}, {Barmby}, \& {Huang}}]{shapley05}
{Shapley}, A.~E., {Steidel}, C.~C., {Erb}, D.~K., {Reddy}, N.~A., {Adelberger},
  K.~L., {Pettini}, M., {Barmby}, P., \& {Huang}, J. 2005, \apj, 626, 698

\bibitem[{{Sharon}(2013)}]{sharon13}
{Sharon}, C.~E. 2013, PhD thesis, Rutgers, the State University of New Jersey

\bibitem[{{Shetty} {et~al.}(2011){Shetty}, {Glover}, {Dullemond}, \&
  {Klessen}}]{shetty11}
{Shetty}, R., {Glover}, S.~C., {Dullemond}, C.~P., \& {Klessen}, R.~S. 2011,
  \mnras, 412, 1686

\bibitem[{{Shetty} {et~al.}(2013){Shetty}, {Kelly}, {Rahman}, {Bigiel},
  {Bolatto}, {Clark}, {Klessen}, \& {Konstandin}}]{shetty13}
{Shetty}, R., {Kelly}, B.~C., {Rahman}, N., {Bigiel}, F., {Bolatto}, A.~D.,
  {Clark}, P.~C., {Klessen}, R.~S., \& {Konstandin}, L.~K. 2013,
  ArXiv:1306.2951

\bibitem[{{Siana} {et~al.}(2008){Siana}, {Teplitz}, {Chary}, {Colbert}, \&
  {Frayer}}]{siana08}
{Siana}, B., {Teplitz}, H.~I., {Chary}, R.-R., {Colbert}, J., \& {Frayer},
  D.~T. 2008, \apj, 689, 59

\bibitem[{{Smail} {et~al.}(2007){Smail}, {Swinbank}, {Richard}, {Ebeling},
  {Kneib}, {Edge}, {Stark}, {Ellis}, {Dye}, {Smith}, \& {Mullis}}]{smail07}
{Smail}, I., {Swinbank}, A.~M., {Richard}, J., {Ebeling}, H., {Kneib}, J.-P.,
  {Edge}, A.~C., {Stark}, D., {Ellis}, R.~S., {Dye}, S., {Smith}, G.~P., \&
  {Mullis}, C. 2007, \apjl, 654, L33

\bibitem[{{Smith} {et~al.}(2009){Smith}, {Ebeling}, {Limousin}, {Kneib},
  {Swinbank}, {Ma}, {Jauzac}, {Richard}, {Jullo}, {Sand}, {Edge}, \&
  {Smail}}]{smith09}
{Smith}, G.~P., {Ebeling}, H., {Limousin}, M., {Kneib}, J.-P., {Swinbank},
  A.~M., {Ma}, C.-J., {Jauzac}, M., {Richard}, J., {Jullo}, E., {Sand}, D.~J.,
  {Edge}, A.~C., \& {Smail}, I. 2009, \apjl, 707, L163

\bibitem[{{Solomon} {et~al.}(1997){Solomon}, {Downes}, {Radford}, \&
  {Barrett}}]{solomon97}
{Solomon}, P.~M., {Downes}, D., {Radford}, S.~J.~E., \& {Barrett}, J.~W. 1997,
  \apj, 478, 144

\bibitem[{{Sommariva} {et~al.}(2012){Sommariva}, {Mannucci}, {Cresci},
  {Maiolino}, {Marconi}, {Nagao}, {Baroni}, \& {Grazian}}]{sommariva12}
{Sommariva}, V., {Mannucci}, F., {Cresci}, G., {Maiolino}, R., {Marconi}, A.,
  {Nagao}, T., {Baroni}, A., \& {Grazian}, A. 2012, \aap, 539, A136

\bibitem[{{Stark} {et~al.}(2009){Stark}, {Ellis}, {Bunker}, {Bundy}, {Targett},
  {Benson}, \& {Lacy}}]{stark09}
{Stark}, D.~P., {Ellis}, R.~S., {Bunker}, A., {Bundy}, K., {Targett}, T.,
  {Benson}, A., \& {Lacy}, M. 2009, \apj, 697, 1493

\bibitem[{{Stark} {et~al.}(2012){Stark}, {Schenker}, {Ellis}, {Robertson},
  {McLure}, \& {Dunlop}}]{stark12}
{Stark}, D.~P., {Schenker}, M.~A., {Ellis}, R.~S., {Robertson}, B., {McLure},
  R., \& {Dunlop}, J. 2012, arXiv:1208.3529

\bibitem[{{Sturm} {et~al.}(2011){Sturm}, {Gonz{\'a}lez-Alfonso}, {Veilleux},
  {Fischer}, {Graci{\'a}-Carpio}, {Hailey-Dunsheath}, {Contursi}, \&
  {Poglitsch}}]{sturm11}
{Sturm}, E., {Gonz{\'a}lez-Alfonso}, E., {Veilleux}, S., {Fischer}, J.,
  {Graci{\'a}-Carpio}, J., {Hailey-Dunsheath}, S., {Contursi}, A., \&
  {Poglitsch}, A. e.~a. 2011, \apjl, 733, L16

\bibitem[{{Swinbank} {et~al.}(2010){Swinbank}, {Smail}, {Longmore}, {Harris},
  {Baker}, {De Breuck}, {Richard}, {Edge}, \& et~al.}]{swinbank10}
{Swinbank}, A.~M., {Smail}, I., {Longmore}, S., {Harris}, A.~I., {Baker},
  A.~J., {De Breuck}, C., {Richard}, J., {Edge}, A.~C., \& et~al. 2010, \nat,
  464, 733

\bibitem[{{Symeonidis} {et~al.}(2013){Symeonidis}, {Vaccari}, {Berta}, {Page},
  {Lutz}, \& et~al.}]{symeonidis13}
{Symeonidis}, M., {Vaccari}, M., {Berta}, S., {Page}, M.~J., {Lutz}, D., \&
  et~al. 2013, \mnras, 431, 2317

\bibitem[{{Tacconi} {et~al.}(2010){Tacconi}, {Genzel}, {Neri}, {Cox}, {Cooper},
  {Shapiro}, {Bolatto}, {Bouch{\'e}}, \& et~al.}]{tacconi10}
{Tacconi}, L.~J., {Genzel}, R., {Neri}, R., {Cox}, P., {Cooper}, M.~C.,
  {Shapiro}, K., {Bolatto}, A., {Bouch{\'e}}, N., \& et~al. 2010, \nat, 463,
  781

\bibitem[{{Tacconi} {et~al.}(2008){Tacconi}, {Genzel}, {Smail}, {Neri},
  {Chapman}, \& et~al.}]{tacconi08}
{Tacconi}, L.~J., {Genzel}, R., {Smail}, I., {Neri}, R., {Chapman}, S.~C., \&
  et~al. 2008, \apj, 680, 246

\bibitem[{{Tacconi} {et~al.}(2013){Tacconi}, {Neri}, {Genzel}, {Combes},
  {Bolatto}, \& et~al.}]{tacconi13}
{Tacconi}, L.~J., {Neri}, R., {Genzel}, R., {Combes}, F., {Bolatto}, A., \&
  et~al. 2013, \apj, 768, 74

\bibitem[{{Teplitz} {et~al.}(2000){Teplitz}, {McLean}, {Becklin}, {Figer},
  {Gilbert}, {Graham}, {Larkin}, {Levenson}, \& {Wilcox}}]{teplitz00}
{Teplitz}, H.~I., {McLean}, I.~S., {Becklin}, E.~E., {Figer}, D.~F., {Gilbert},
  A.~M., {Graham}, J.~R., {Larkin}, J.~E., {Levenson}, N.~A., \& {Wilcox},
  M.~K. 2000, \apjl, 533, L65

\bibitem[{{Veilleux} {et~al.}(2002){Veilleux}, {Kim}, \&
  {Sanders}}]{veilleux02}
{Veilleux}, S., {Kim}, D.-C., \& {Sanders}, D.~B. 2002, \apjs, 143, 315

\bibitem[{{Volino} {et~al.}(2010){Volino}, {Wucknitz}, {McKean}, \&
  {Garrett}}]{volino10}
{Volino}, F., {Wucknitz}, O., {McKean}, J.~P., \& {Garrett}, M.~A. 2010, \aap,
  524, A79

\bibitem[{{Wardlow} {et~al.}(2013){Wardlow}, {Cooray}, {De Bernardis},
  {Amblard}, {Arumugam}, {Aussel}, {Baker}, {B{\'e}thermin}, \&
  et~al.}]{wardlow13}
{Wardlow}, J.~L., {Cooray}, A., {De Bernardis}, F., {Amblard}, A., {Arumugam},
  V., {Aussel}, H., {Baker}, A.~J., {B{\'e}thermin}, M., \& et~al. 2013, \apj,
  762, 59

\bibitem[{{Weinmann} {et~al.}(2011){Weinmann}, {Neistein}, \&
  {Dekel}}]{weinmann11}
{Weinmann}, S.~M., {Neistein}, E., \& {Dekel}, A. 2011, \mnras, 417, 2737

\bibitem[{{Weiss} {et~al.}(2007){Weiss}, {Downes}, {Walter}, \&
  {Henkel}}]{weiss07}
{Weiss}, A., {Downes}, D., {Walter}, F., \& {Henkel}, C. 2007, in Astronomical
  Society of the Pacific Conference Series, Vol. 375, From Z-Machines to ALMA:
  (Sub)Millimeter Spectroscopy of Galaxies, ed. {A.~J.~Baker, J.~Glenn,
  A.~I.~Harris, J.~G.~Mangum, \& M.~S.~Yun }, 25--+

\bibitem[{{Whitaker} {et~al.}(2012){Whitaker}, {van Dokkum}, {Brammer}, \&
  {Franx}}]{whitaker12}
{Whitaker}, K.~E., {van Dokkum}, P.~G., {Brammer}, G., \& {Franx}, M. 2012,
  \apjl, 754, L29

\bibitem[{{Wilson}(1995)}]{wilson95}
{Wilson}, C.~D. 1995, \apjl, 448, L97+

\bibitem[{{Wilson} {et~al.}(2008){Wilson}, {Petitpas}, {Iono}, {Baker}, {Peck},
  {Krips}, {Warren}, {Golding}, {Atkinson}, {Armus}, \& et~al.}]{wilson08}
{Wilson}, C.~D., {Petitpas}, G.~R., {Iono}, D., {Baker}, A.~J., {Peck}, A.~B.,
  {Krips}, M., {Warren}, B., {Golding}, J., {Atkinson}, A., {Armus}, L., \&
  et~al. 2008, \apjs, 178, 189

\bibitem[{{Wisnioski} {et~al.}(2011){Wisnioski}, {Glazebrook}, {Blake},
  {Wyder}, {Martin}, {Poole}, {Sharp}, {Couch}, {Kacprzak}, {Brough},
  {Colless}, {Contreras}, {Croom}, {Croton}, {Davis}, {Drinkwater}, {Forster},
  {Gilbank}, {Gladders}, {Jelliffe}, {Jurek}, {Li}, {Madore}, {Pimbblet},
  {Pracy}, {Woods}, \& {Yee}}]{wisnioski11}
{Wisnioski}, E., {Glazebrook}, K., {Blake}, C., {Wyder}, T., {Martin}, C.,
  {Poole}, G.~B., {Sharp}, R., {Couch}, W., {Kacprzak}, G.~G., {Brough}, S.,
  {Colless}, M., {Contreras}, C., {Croom}, S., {Croton}, D., {Davis}, T.,
  {Drinkwater}, M.~J., {Forster}, K., {Gilbank}, D.~G., {Gladders}, M.,
  {Jelliffe}, B., {Jurek}, R.~J., {Li}, I.-H., {Madore}, B., {Pimbblet}, K.,
  {Pracy}, M., {Woods}, D., \& {Yee}, H.~K.~C. 2011, \mnras, 417, 2601

\bibitem[{{Wright} {et~al.}(2007){Wright}, {Larkin}, {Barczys}, {Erb},
  {Iserlohe}, {Krabbe}, {Law}, {McElwain}, {Quirrenbach}, {Steidel}, \&
  {Weiss}}]{wright07}
{Wright}, S.~A., {Larkin}, J.~E., {Barczys}, M., {Erb}, D.~K., {Iserlohe}, C.,
  {Krabbe}, A., {Law}, D.~R., {McElwain}, M.~W., {Quirrenbach}, A., {Steidel},
  C.~C., \& {Weiss}, J. 2007, \apj, 658, 78

\bibitem[{{Wuyts} {et~al.}(2012{\natexlab{a}}){Wuyts}, {Rigby}, {Gladders},
  {Gilbank}, {Sharon}, {Gralla}, \& {Bayliss}}]{wuyts12}
{Wuyts}, E., {Rigby}, J.~R., {Gladders}, M.~D., {Gilbank}, D.~G., {Sharon}, K.,
  {Gralla}, M.~B., \& {Bayliss}, M.~B. 2012{\natexlab{a}}, \apj, 745, 86

\bibitem[{{Wuyts} {et~al.}(2012{\natexlab{b}}){Wuyts}, {Rigby}, {Sharon}, \&
  {Gladders}}]{wuyts12b}
{Wuyts}, E., {Rigby}, J.~R., {Sharon}, K., \& {Gladders}, M.~D.
  2012{\natexlab{b}}, \apj, 755, 73

\bibitem[{{Wuyts} {et~al.}(2011{\natexlab{a}}){Wuyts}, {F{\"o}rster Schreiber},
  {Lutz}, {Nordon}, {Berta}, {Altieri}, {Andreani}, \& et~al.}]{wuyts11sfr}
{Wuyts}, S., {F{\"o}rster Schreiber}, N.~M., {Lutz}, D., {Nordon}, R., {Berta},
  S., {Altieri}, B., {Andreani}, P., \& et~al. 2011{\natexlab{a}}, \apj, 738,
  106

\bibitem[{{Wuyts} {et~al.}(2011{\natexlab{b}}){Wuyts}, {F{\"o}rster Schreiber},
  {van der Wel}, {Magnelli}, {Guo}, {Genzel}, {Lutz}, {Aussel}, \&
  et~al.}]{wuyts11}
{Wuyts}, S., {F{\"o}rster Schreiber}, N.~M., {van der Wel}, A., {Magnelli}, B.,
  {Guo}, Y., {Genzel}, R., {Lutz}, D., {Aussel}, H., \& et~al.
  2011{\natexlab{b}}, \apj, 742, 96

\bibitem[{{Yuan} {et~al.}(2011){Yuan}, {Kewley}, {Swinbank}, {Riterchard}, \&
  {Livermore}}]{yuan11}
{Yuan}, T.-T., {Kewley}, L.~J., {Swinbank}, A.~M., {Riterchard}, J., \&
  {Livermore}, R.~C. 2011, \apjl, 732, L14

\end{thebibliography}

%===================================
% Beginning of the appendices
%===================================

\appendix

\section{SDSS J1137+4936 lens model}
\label{lensmodel}

\begin{figure}[h!tb]
\begin{center}$
\begin{array}{cc}
\includegraphics[width=0.48\textwidth]{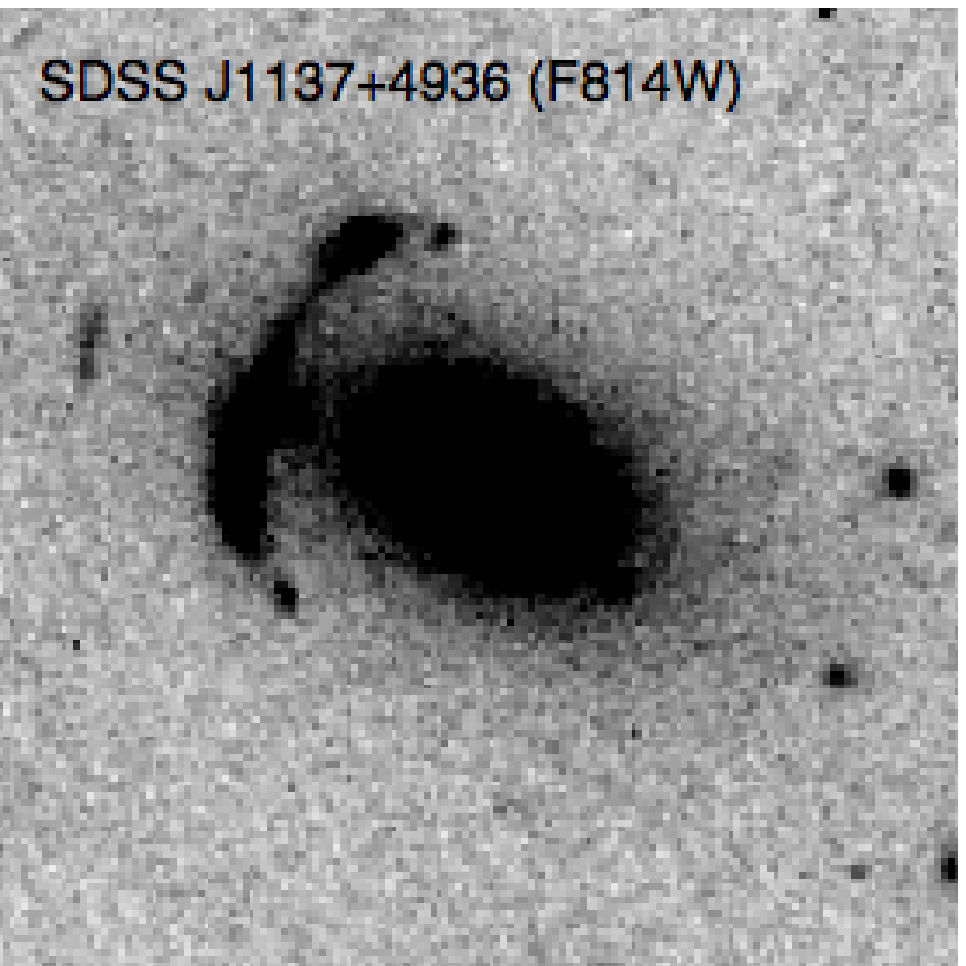} &
\includegraphics[width=0.48\textwidth]{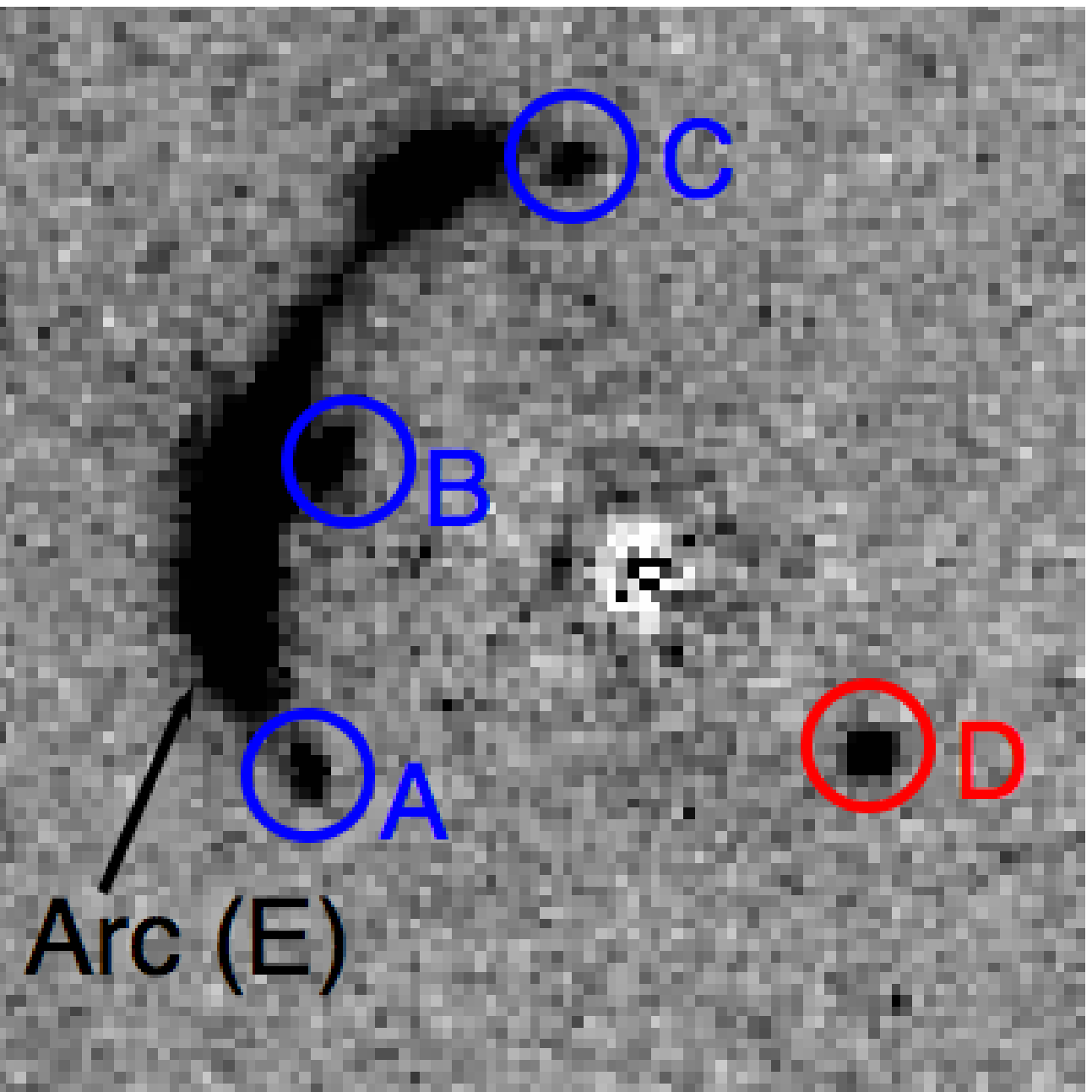}
\end{array}$
\end{center}
\caption{\footnotesize {\it HST}-WFPC2 observations of SDSS J1137+4936. {\it Left:} {\it HST}-WFPC2 image of SDSS J1137+4936 taken in the F814W filter ($\rm{7\farcs5 \times 7\farcs5}$). {\it Right:} The lens galaxy subtracted F814W image ($\rm{5\farcs0 \times 5\farcs0}$) showing the lensed features. The lens galaxy light profile was modeled using multiple $\rm{S\acute{e}rsic}$ \citep{sersic68} components. Images {\bf A}, {\bf B} and {\bf C} correspond to a single source on the source plane (at $z = 1.38$). Image {\bf D} corresponds to an extraneous feature that is commonly mistaken as a counter-image of the lensed arc. This feature was masked during the initial lens modeling. Arc {\bf E} corresponds to the primary source galaxy at $z = 1.41$.} \label{fig:sdssj1137+4936}
\end{figure}

J1137 is a galaxy-scale gravitational lens system, where a luminous red galaxy (i.e. the lens galaxy) at $z = 0.45$ is forming a bright blue arc of the background source (i.e. the source galaxy). Follow-up spectroscopy of the bright blue arc by \citet{kubo09} confirms a primary source galaxy redshift of $z = 1.41$ and shows evidence of a secondary source galaxy at $z = 1.38$. As shown in Figure 2 of \citet{kubo09}, SDSS imaging of the blue arc resembles two split knots and does not distinguish between the two background sources that are very close in redshift, forming nearly overlapping multiple images. However, multiwavelength {\it HST}-WFPC2 imaging of SDSS J1137 indicates that the bright blue arc is comprised of several images, as shown in Figure~\ref{fig:sdssj1137+4936}({\it left}). Figure~\ref{fig:sdssj1137+4936}({\it right}) demonstrates the subtraction of the lens galaxy light profile, showing the lensed features more clearly:
\begin{itemize}
\item features {\bf A} through {\bf C}, which resemble three small knots merging together to form a faint secondary arc, result from the lensing of the source galaxy at $z = 1.38$
\item feature {\bf D} is an extraneous, non-lensing object that is masked during the initial lens modeling process 
\item feature {\bf E} corresponds to the primary lensed arc, resulting from the lensing of the source galaxy at $z = 1.41$. 
\end{itemize}
Therefore, an accurate lens model must account for all images except {\bf D}.

We use LENSFIT \citep{peng06}, an extension of the galaxy decomposition software GALFIT \citep{peng02,pengc10} for strong gravitational lens analysis, to derive the lens model. We refer the reader to \citet{peng06} and Bandara et al. (2013, ApJ submitted), for a detailed overview of LENSFIT and its application for the analysis of galaxy-scale gravitational lenses discovered in the Sloan Lens ACS (SLACS) Survey, and summarize our methodology below. To describe the mass distribution of the lens galaxy, we assume a singular isothermal ellipsoid mass model \citep[SIE,][]{kormann94} with an external shear field to model the tidal effects by nearby objects. The projected mass density of the SIE model is,
\begin{eqnarray}
\kappa(x,y)\:=\:\frac{b}{2}\Big[\frac{2q^{2}}{1+q^{2}}(x^{2} + y^{2} / q^{2})\Big]^{-1/2}
\label{eq:SIE}
\end{eqnarray}
where $b$ is the mass scale and $q$ is the axis ratio of the mass model. The mass scale parameter ($b$) is approximately the Einstein radius of the lens (denoted as $b_{SIE}$); however, this relation is only exact at the $q\:=\:1$ limit \citep{kochanek01,peng06}. At this limit, $b_{SIE}$ is related to the physical quantities of the mass model by,
\begin{eqnarray}
b_{SIE}\:=\:4\pi\:\frac{{\sigma_{SIE}}^{2}}{c^{2}}\:\frac{D_{LS}}{D_{S}}
\label{eq:einstein_radius}
\end{eqnarray}
 where $\sigma_{SIE}$ is the velocity dispersion of the mass model and $D_{LS}$ and $D_{S}$ are angular diameter distances from the lens to the source galaxy and from the observer to the source galaxy respectively \citep{kochanek01}. The SIE mass model of LENSFIT is characterized by the following parameters: mass model centroid ($x_{SIE}, y_{SIE}$), Einstein radius ($b_{SIE}$), axis ratio of the mass model ($q_{SIE}$), position angle of the major axis measured E from N ($PA_{SIE}$), external shear ($\gamma_{SIE}$) and the position angle of the external shear component measured E from N ($PA_{\gamma}$). Furthermore, LENSFIT is a parametric lensing code that describes the {\it unlensed} source galaxy light profile through a set of parametric functions. To model the source galaxies of J1137, we use S$\rm{\acute{e}}$rsic profiles \citep{sersic68}, 
\begin{eqnarray}
\Sigma(r) = \Sigma_{e}\exp{(-\kappa({\it n})\:[(r/r_{hl})^{1/{\it n}} - 1]})
\label{eq:sersic}
\end{eqnarray}
where $\Sigma(r)$ is the surface brightness at a given radius $r$, $r_{hl}$ is the half-light radius (i.e. also referred to as the effective radius, $r_{e}$), $\Sigma_{e}$ is the pixel surface brightness at the effective radius and $n$ is the concentration parameter. The elliptically symmetric $\rm{S\acute{e}rsic}$ profile is characterized by the following parameters: position of the $\rm{S\acute{e}rsic}$ component ($x_{s},y_{s}$), half-light radius ($r_{hl}$), apparent magnitude ($m$), the $\rm{S\acute{e}rsic}$ index ({\it n}), axis ratio of the elliptical profile ($q$) and the position angle of the major axis measured E from N ($PA_{SIE}$).

We perform the lens modeling of J1137 in the {\it HST}-WFPC2 filters F814W, F606W and F450W.  Since the gravitational lensing phenomenon is achromatic, the mass-model should be identical across multiple filters within the systematic uncertainties. Therefore, when modeling a gravitational lens in multiple filters, the mass-model parameters are typically fixed to those determined by the highest signal-to-noise filter (in this case, the F814W filter). However, we allow the SIE mass model parameters to vary freely for the F606W filter (which also has high signal-to-noise), such that we can test whether the lens modeling process is sufficiently robust to converge to the same mass model parameters for different filters. We find that the SIE mass model parameters from F814W and F606W imaging are virtually indistinguishable, thus confirming that the lens model solution is robust. Since F450W imaging has a lower signal-to-noise, we initially fix the SIE mass model parameters to those constrained by F814W and F606W imaging. During the final iteration of the F450W lens model, we allow the mass model to vary freely and find that the fractional difference between F450W parameters and those obtained from F814W/F606W imaging is less than $\sim 2\%$.

\begin{figure}[h!tb]
\begin{center}$
\begin{array}{llll}
\includegraphics[width=0.23\textwidth]{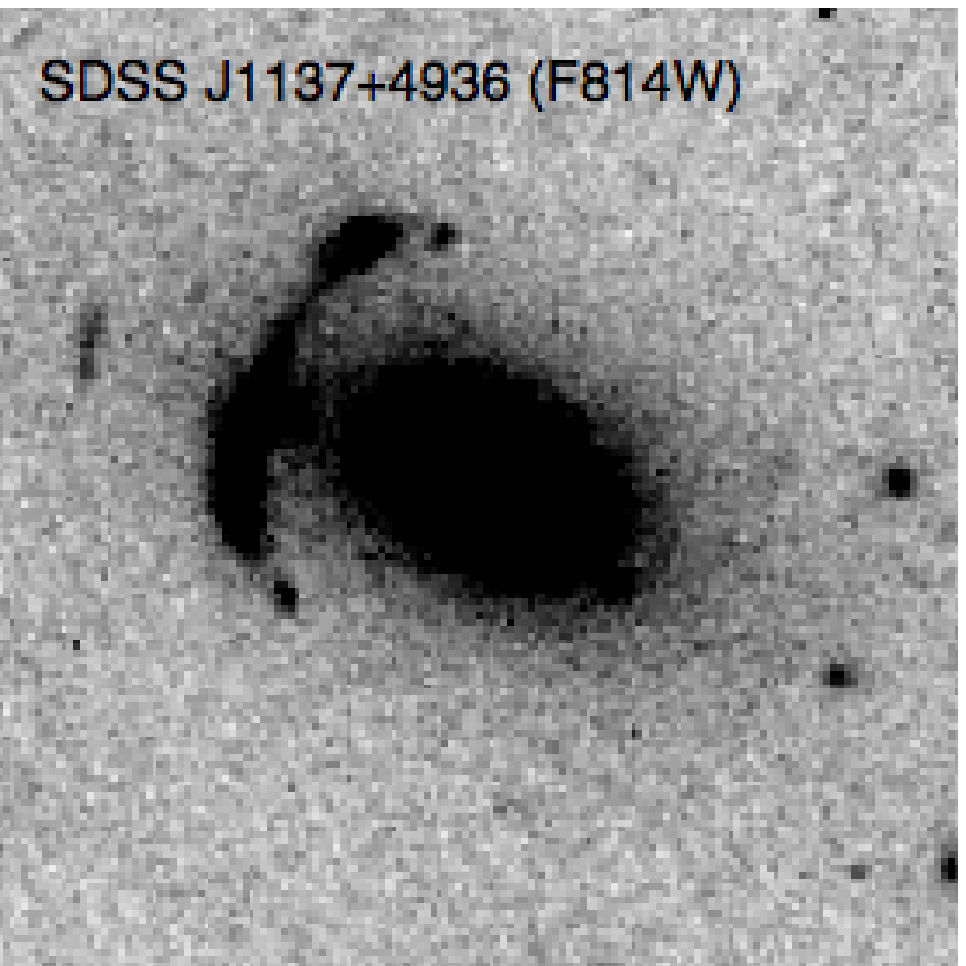} &
\includegraphics[width=0.23\textwidth]{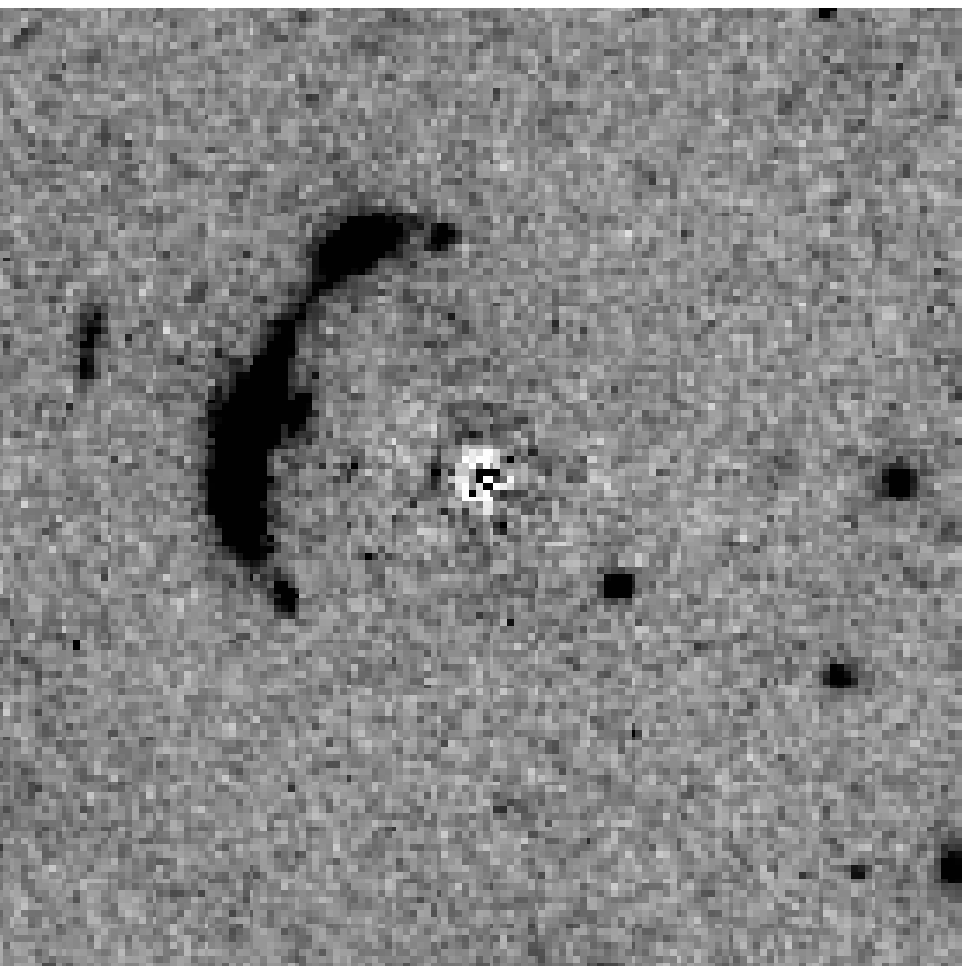} &
\includegraphics[width=0.23\textwidth]{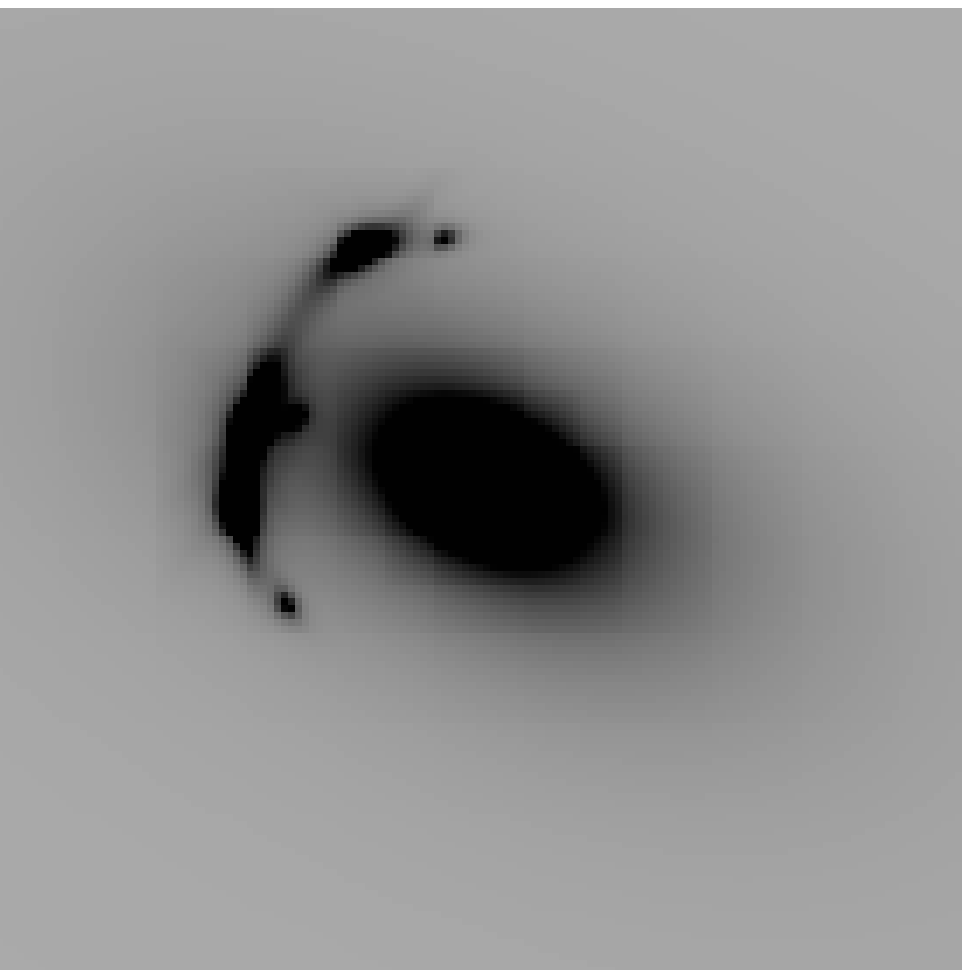} &
\includegraphics[width=0.23\textwidth]{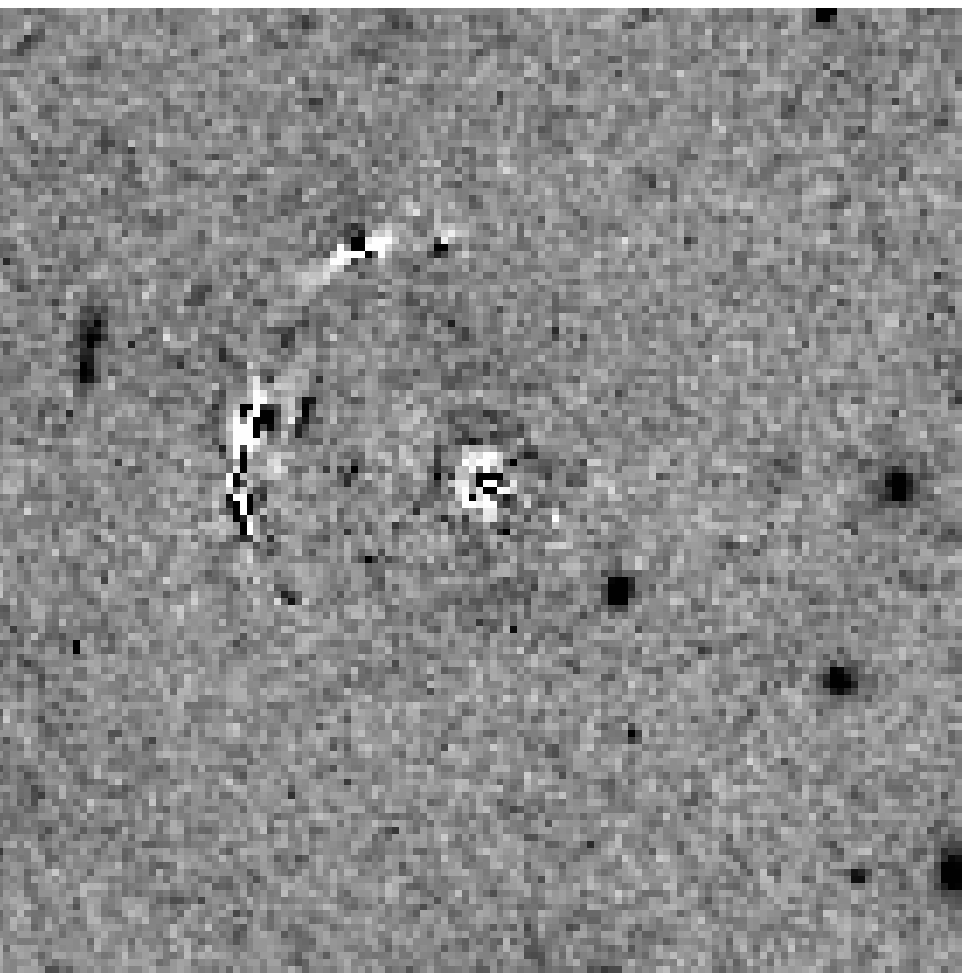}
\end{array}$
\end{center}
\caption{\footnotesize LENSFIT model of SDSS J1137+4936 in the F814W filter. {\it From left to right:} {\it HST}-WFPC2 image of SDSS J1137+4936; The lens galaxy subtracted image showing the lensed features; The complete lens model on the image plane including the light profile of the lens galaxy and the lensed features from the best-fit SIE mass-model; The ``double residual'' image after subtracting the PSF convolved lens model from the F814W image. All images are $\rm{7\farcs5 \times 7\farcs5}$ in size.} \label{fig:lensfit_breakdown1}
\end{figure}

The morphology of the $z = 1.41$ {\it unlensed} source galaxy (which forms the lensed arc {\bf E} on the image plane) is best described by three $\rm{S\acute{e}rsic}$ components, in the F814W and F606W filters, and two $\rm{S\acute{e}rsic}$ components in the F450W filter. In addition, the morphology of the $z = 1.38$ unlensed source galaxy (which forms features A through C on the image plane) is best described by a single $\rm{S\acute{e}rsic}$ component in all three filters. Figure~\ref{fig:lensfit_breakdown1} shows the results of lens modeling of SDSS J1137 in the F814W filter. As indicated by the ``double residual'' image of SDSS J1137, shown in the fourth panel of Figure~\ref{fig:lensfit_breakdown1}, the complete lens model predicts images {\bf A} through {\bf C} and {\bf E} but not {\bf D} (which was unmasked during the final steps of the lensing analysis). 

LENSFIT output parameters include the unlensed flux of each $\rm{S\acute{e}rsic}$ component of the source galaxy light profile. Due to the circular feedback mechanism between the mass model and the source galaxy light profile, the use of multiple $\rm{S\acute{e}rsic}$ components to describe the unlensed source galaxy yields the best-fit mass model. In other words, it is important to minimize the residuals of the source galaxy light profile to fully constrain the mass model, using multiple $\rm{S\acute{e}rsic}$ components if necessary. However, from the standpoint of computing the overall intrinsic properties of the source galaxy (e.g. for comparison with non-lensed galaxy samples), we require a simplified representation that mimics the analysis techniques of high-redshift studies which typically use a single $\rm{S\acute{e}rsic}$ component in galaxy fitting. Therefore, we compute the {\it global} properties of the unlensed $z = 1.41$ source galaxy (i.e. the flux of a {\it single} $\rm{S\acute{e}rsic}$ component that best minimizes the overall residuals) by performing an additional LENSFIT iteration using single $\rm{S\acute{e}rsic}$ components. The SIE mass model, which was fully constrained by the use of multiple $\rm{S\acute{e}rsic}$ components to define the $z = 1.41$ source galaxy, and the $z = 1.38$ source galaxy light profile parameters are fixed to those implied by the best-fit lens model. 

For this study, the quantity of interest is the magnification factor of the source galaxy at $z = 1.41$ resulting in lensed image {\bf E}, defined as
\begin{eqnarray}
\rm{magnification\:factor = \frac{flux\:of\:the\:lensed\:image}{flux\:of\:the\:unlensed\:galaxy}}.
\label{eq:magnification}
\end{eqnarray}
We use the SExtractor photometry package \citep{bertin96} to deblend and measure the flux within the lensed image {\bf E} in all three filters. We then combine the total flux of the lensed image, obtained from the SExtractor photometry, and the unlensed flux of the {\it global} source galaxy profile to derive the magnification factor, according to equation~\ref{eq:magnification}. The total magnification factor of SDSS J1137 is the mean value of F814W, F606W and F450W filters and corresponds to $\sim 17\times$.  To estimate the uncertainty on this measurement, we propagate the errors coming from the determination of both the unlensed and lensed galaxy fluxes, of which the magnification factor is the ratio.  The former includes the error on the slope of the mass model \citep{marshall07}, the error caused by the lens galaxy subtraction, and the error associated with the PSF model, while the lensed galaxy flux uncertainty comes only from the SExtractor photometry.   Including all these sources of uncertainty, we arrive at a total error of 18$\%$ for the magnification factor of J1137.  This also justifies our choice of a blanket 20\% uncertainty on the magnification factors taken from the literature when published without an uncertainty value (see Table \ref{Sampletab}). 

\section{IRAC photometry of the lensed galaxies}
\label{iracphoto}

\begin{figure}
\epsscale{1.0}
\plotone{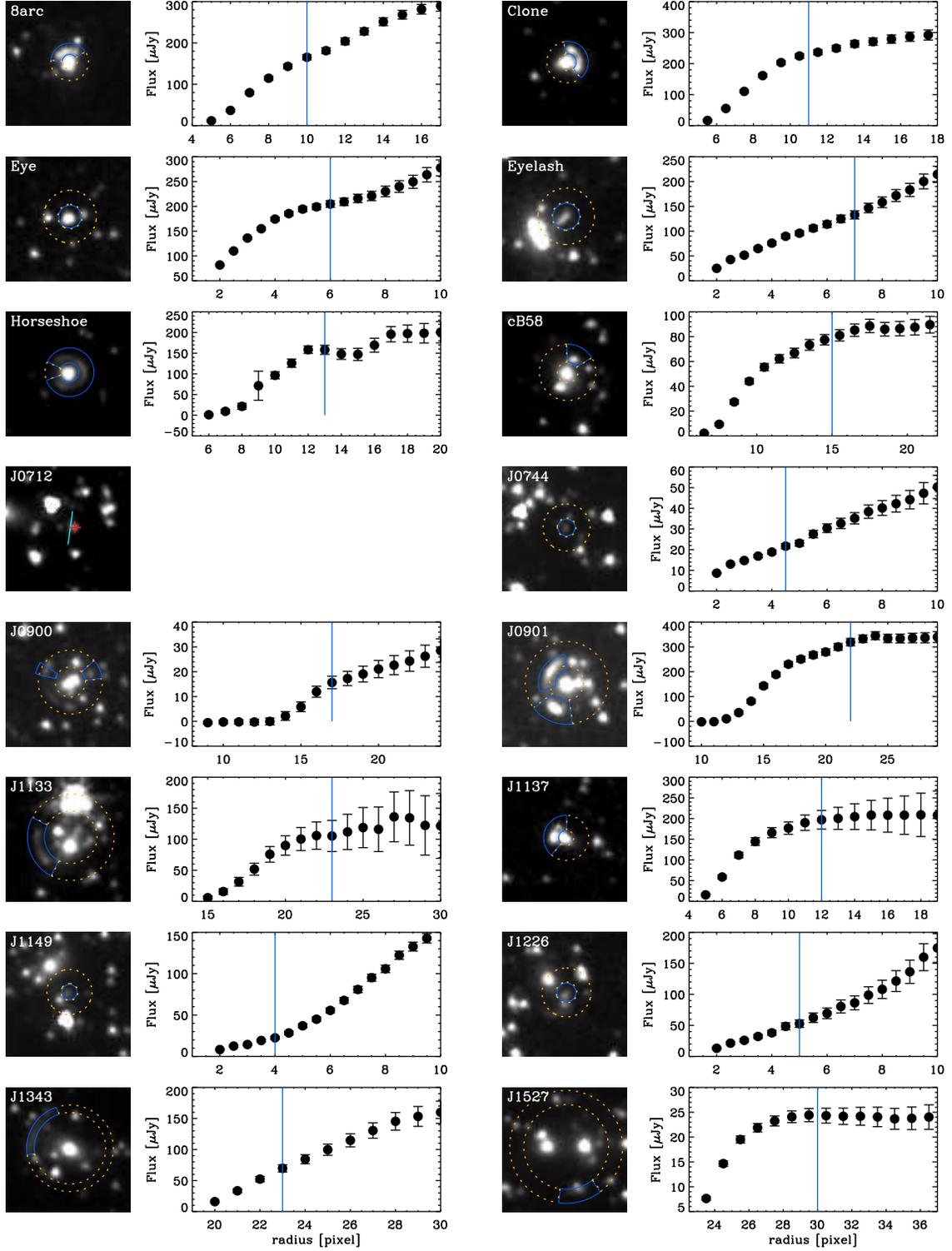}
\caption{IRAC 3.6$\mu$m images, each 40\arcsec$\times$40\arcsec, for all lensed galaxies in the sample with the exception of J1441 where no such data exist.  The apertures used to extract flux measurements are shown with solid blue lines, and the areas used for background estimation are delineated by dashed orange lines.  In the case of J0712, no reliable flux measurement could be made due to the severe blending between the foreground lens (red cross) and the faint arc (blue line).  For all other galaxies, the right panel shows the curve of growth, as was used to determine the maximum size of the aperture (vertical blue lines).   \label{stamps}} 
\end{figure}

The {\it Spitzer} archive was queried for observations of the lensed galaxies in the 3.6 and 4.5$\mu$m bands, and the calibrated images retrieved.  The only lensed galaxy in our sample without IRAC coverage is J1441.  The 3.6$\mu$m images of the other 16 galaxies are shown in Figure \ref{stamps}.  Most of the lensed galaxies appear as extended arcs in the IRAC images, and they are generally located within the wings of the foreground lenses, which are typically bright at these wavelengths.  For these two main reasons, standard photometry packages are not optimal and instead a custom IDL script was written to measure accurate fluxes.  

We define an annulus, typically centered on the foreground lens, that encompasses fully the arc.  The section of this annulus containing the arc is used to integrate the flux of the object, and the remaining section is used to estimate the background, as indicated n Figure \ref{stamps}. The advantages of this approach are twofold: (1) the shape of the aperture naturally matches well that of the lensed arcs, and (2) the background is estimated over a region with equivalent noise properties and contamination from the bright lens.  The exact parameters defining the aperture (inner and outer radii, position angle and opening angle) are fixed using curve of growth arguments.  In Figure \ref{stamps}, the curves of growth obtained by changing the outer radius of the apertures are shown.  The background level within the aperture is estimated by measuring the mode of the sky pixels in the ``background aperture" and multiplying it by the area of the aperture.  The mode is calculated with the MMM task in IDL (based of the DAOPHOT task of the same name), allowing an accurate measure of the sky properties even in the presence of bright point sources within the ``background aperture".   The error on the total flux of the lensed galaxy is obtained by adding in quadrature the uncertainty on the flux in the aperture from the IRAC error map, the scatter in the sky values, and the uncertainty in the mean sky brightness.  

Some of the lensed galaxies are however best modeled by a simple circular aperture than with an arc aperture.  This is the case of the Eye, the Eyelash, cB58, J0744, J1149 and J1226.  In those cases, a simple circular aperture is used, with a radius determined from the curve of growth (see Fig. \ref{stamps}).  The background is determined from an annulus around this aperture, using the same technique described above.   For the lensed galaxy J0712, neither technique produces reliable results, since the faint blue arc is heavily blended with the brighter lens, as shown in Fig. \ref{stamps}.  For that galaxy, we adopt the stellar mass of $\log M_{\ast}/M_{\odot}=10.23\pm0.48$ derived by \citet{richard11} from SED fitting to the HST photometry.  

The final, background subtracted fluxes at 3.6 and 4.5$\mu$m are summarized in Table \ref{masstab}, and our methodology to derive stellar masses from these fluxes is described in \S \ref{mstarsection}.

\section{Notes on individual objects}

\subsection{AGN contamination in J0901?}
\label{J0901agn}
\begin{figure}[h!]
\epsscale{0.6}
\plotone{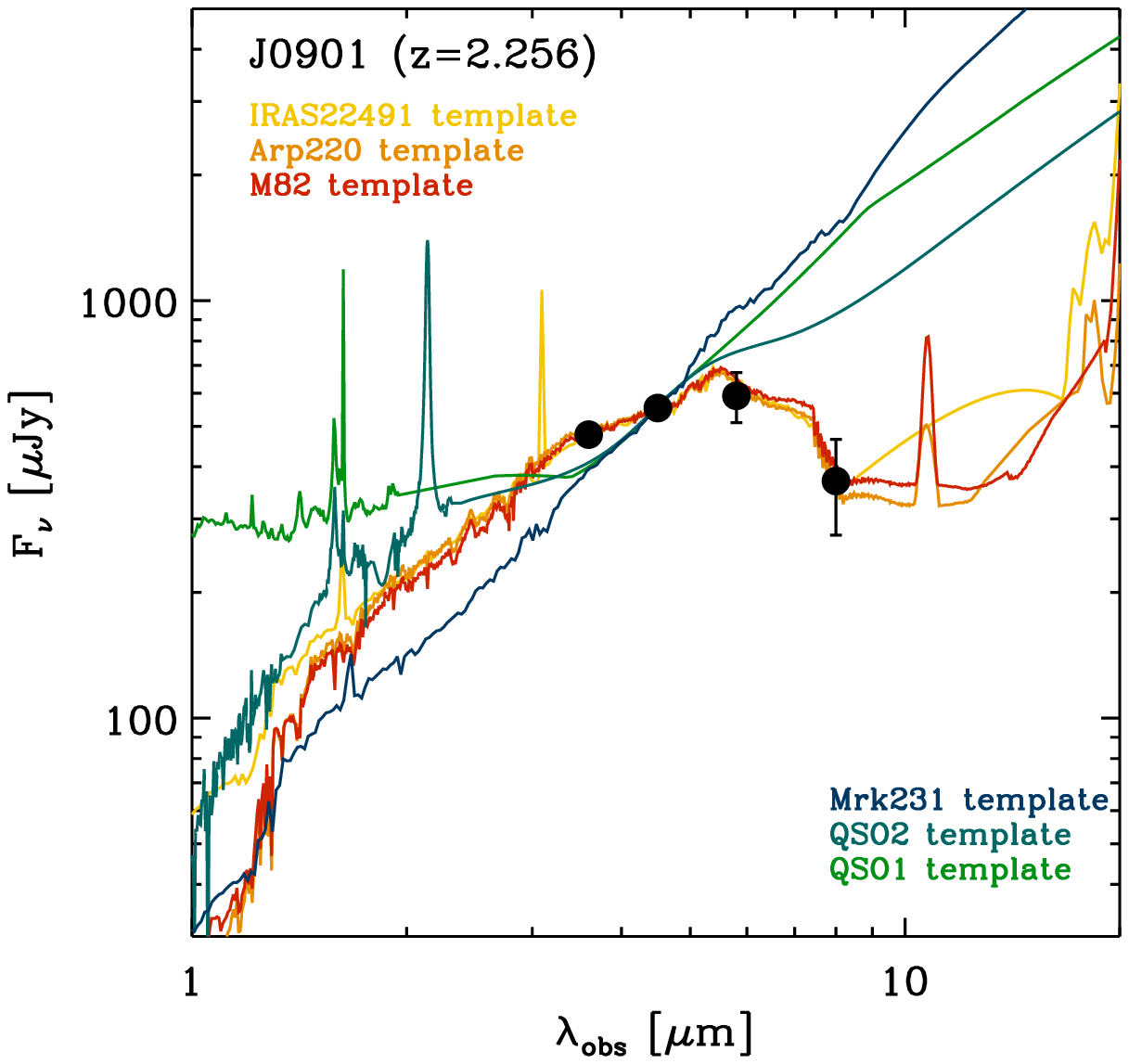}
\caption{Fluxes in the IRAC bands for J0901, compared with template SEDs from the library of \citet{polletta07} normalized to the observed wavelength of 4.5$\mu$m.  Shown are spectra of both starbursts (yellow, orange and red lines) and AGN-dominated objects (blue and green lines), confirming the conclusion of \citet{fadely10} that the AGN contribution to the mid-infrared emission is negligible.    \label{J0901SED}}
\end{figure}

The lensed galaxy J0901 has been reported by \citet{hainline09} to harbor an AGN, based on a high [NII]/H$\alpha$ ratio, above the range expected from star forming galaxies, suggesting excitation of the lines through AGN and/or shocks \citep[e.g.][]{kewley01,levesque10}.   Care must therefore be taken that the AGN emission does not affect our estimates of stellar mass and star formation rate.  Since our stellar masses are derived from the IRAC photometry at 3.6 and 4.5 $\mu$m, we compare in Figure \ref{J0901SED} the shape of the observed near-infrared spectral energy distribution to templates from the SWIRE library \citep{polletta07}.   The emission observed in the IRAC bands is fully consistent with the starburst templates, suggesting that the 3.6 and 4.5 $\mu$m fluxes (and therefore the stellar mass derived from them) are not affected by the presence of an AGN.  \citet{fadely10} also concluded from {\it Spitzer}/IRS spectroscopy that the contribution of the AGN to the mid-infrared emission is insignificant and that the system is instead starburst-driven.  In the absence of an AGN signature at near- and mid-infrared wavelengths, it can be safely concluded that there is no contamination at even longer wavelengths and therefore that the SFR and dust mass/temperature derived from the Herschel photometry are secure \citep[see][for a discussion of AGN contamination in the far-infrared]{rosario12}.   However, given the un-physically high [NII]/H$\alpha$ ratio for star-forming regions, we adopt for this galaxy the metallicity given by the MZ relation, $12+\log{\rm O/H}=8.91$.  This value is consistent with the lower limit of 1.3$Z_{\odot}$ reported by \citet{fadely10} based on Argon line fluxes from {\it Spitzer} mid-infrared spectroscopy.

\subsection{Metallicity measurement for J1226}
\label{J1226metal}
In the absence of a [NII]/H$\alpha$ ratio measurement, we computed the metallicity for J1226 using the $R_{23}$ and $O_{23}$ indicators, using H$\beta$,  [O{\sc III}] and [O{\sc II}] line fluxes from the spectrum of \citet{wuyts12}, further corrected for extinction (J.R. Rigby, private communication).  The values derived are $\log R_{23}=1.00\pm0.13$ and $\log O_{32}=0.09\pm0.18$.  To compute a metallicity from these line ratios, we use the calibration of \citet{mcgaugh91} as parametrized by \citet{denaray04}.  The measured values of $R_{23}$ and $O_{32}$ put this galaxy at the transition between the two branches of the relation, with 12$+\log$O/H$=$8.46 and 8.45 for the lower and upper branches, respectively.  We can therefore obtain a reliable metallicity estimate, even without [NII]/H$\alpha$ to discriminate between the two branches.  Finally, we use the relation of \citet{kewley08} to convert this metallicity on the \citet{denicolo02} scale, for consistency with the rest of the sample (Section \ref{metalsection}).  The final value we adopt for J1226 is therefore 12$+\log$O/H$=8.27\pm0.19$, where the uncertainty is the sum in quadrature of the propagated measurement errors, and the average scatter of 0.15 in the \citet{mcgaugh91} calibration. 

\end{document}